\RequirePackage{lineno}

\RequirePackage{setspace}
\documentclass[aps,prd,twocolumn,superscriptaddress,showkeys,showpacs,preprintnumbers,amsmath,amssymb,amsfonts]{revtex4}
\usepackage{graphicx} 
\usepackage{booktabs}
\usepackage{dcolumn}  
\usepackage{bigstrut}
\graphicspath{{ps}}

\usepackage[format=plain,justification=raggedright,singlelinecheck=off]{caption}
\usepackage[format=plain,justification=justified,singlelinecheck=off]{subfig}%
\usepackage{color}
\usepackage{multirow}
\usepackage{makecell}
\usepackage{pdflscape}
\usepackage[T1]{fontenc}
\usepackage{etoolbox} 
\usepackage{chngcntr}

\usepackage[capitalize]{cleveref}
\renewcommand{\arraystretch}{1.1}
\newcommand\extraclusters{{\operatorname{extra-clusters}}}

\AtBeginDocument{
\heavyrulewidth=.08em
\lightrulewidth=.05em
\cmidrulewidth=.03em
\belowrulesep=.65ex
\belowbottomsep=0pt
\aboverulesep=.4ex
\abovetopsep=0pt
\cmidrulesep=\doublerulesep
\cmidrulekern=.5em
\defaultaddspace=.5em
}

\makeatletter
\appto{\appendix}{%
  \@ifstar{\def\thesection{\unskip}\def\theequation@prefix{A.}}%
          {\def\thesection{\Alph {section}}}%
}
\makeatother

\begin{document}
\setlength\linenumbersep{0.2cm}

\preprint{\vbox{ \hbox{   }
			\hbox{Belle preprint 2020-06}
			\hbox{KEK preprint 2020-04}
}}

\title{\quad\\[1.0cm] Measurement of the Branching Fraction of the Decay
  $\boldsymbol{B^{+}\to\pi^{+}\pi^{-}\ell^{+}\nu_\ell}$ in Fully Reconstructed
  Events at Belle}


\noaffiliation
\affiliation{University of Bonn, 53115 Bonn}
\affiliation{Brookhaven National Laboratory, Upton, New York 11973}
\affiliation{Budker Institute of Nuclear Physics SB RAS, Novosibirsk 630090}
\affiliation{Faculty of Mathematics and Physics, Charles University, 121 16 Prague}
\affiliation{Chonnam National University, Gwangju 61186}
\affiliation{University of Cincinnati, Cincinnati, Ohio 45221}
\affiliation{Deutsches Elektronen--Synchrotron, 22607 Hamburg}
\affiliation{Key Laboratory of Nuclear Physics and Ion-beam Application (MOE) and Institute of Modern Physics, Fudan University, Shanghai 200443}
\affiliation{II. Physikalisches Institut, Georg-August-Universit\"at G\"ottingen, 37073 G\"ottingen}
\affiliation{SOKENDAI (The Graduate University for Advanced Studies), Hayama 240-0193}
\affiliation{Department of Physics and Institute of Natural Sciences, Hanyang University, Seoul 04763}
\affiliation{University of Hawaii, Honolulu, Hawaii 96822}
\affiliation{High Energy Accelerator Research Organization (KEK), Tsukuba 305-0801}
\affiliation{J-PARC Branch, KEK Theory Center, High Energy Accelerator Research Organization (KEK), Tsukuba 305-0801}
\affiliation{Higher School of Economics (HSE), Moscow 101000}
\affiliation{Forschungszentrum J\"{u}lich, 52425 J\"{u}lich}
\affiliation{Indian Institute of Science Education and Research Mohali, SAS Nagar, 140306}
\affiliation{Indian Institute of Technology Bhubaneswar, Satya Nagar 751007}
\affiliation{Indian Institute of Technology Hyderabad, Telangana 502285}
\affiliation{Indian Institute of Technology Madras, Chennai 600036}
\affiliation{Indiana University, Bloomington, Indiana 47408}
\affiliation{Institute of High Energy Physics, Chinese Academy of Sciences, Beijing 100049}
\affiliation{Institute of High Energy Physics, Vienna 1050}
\affiliation{Institute for High Energy Physics, Protvino 142281}
\affiliation{INFN - Sezione di Napoli, 80126 Napoli}
\affiliation{Advanced Science Research Center, Japan Atomic Energy Agency, Naka 319-1195}
\affiliation{J. Stefan Institute, 1000 Ljubljana}
\affiliation{Institut f\"ur Experimentelle Teilchenphysik, Karlsruher Institut f\"ur Technologie, 76131 Karlsruhe}
\affiliation{Kennesaw State University, Kennesaw, Georgia 30144}
\affiliation{Korea Institute of Science and Technology Information, Daejeon 34141}
\affiliation{Korea University, Seoul 02841}
\affiliation{Kyungpook National University, Daegu 41566}
\affiliation{P.N. Lebedev Physical Institute of the Russian Academy of Sciences, Moscow 119991}
\affiliation{Faculty of Mathematics and Physics, University of Ljubljana, 1000 Ljubljana}
\affiliation{Ludwig Maximilians University, 80539 Munich}
\affiliation{Luther College, Decorah, Iowa 52101}
\affiliation{Malaviya National Institute of Technology Jaipur, Jaipur 302017}
\affiliation{University of Maribor, 2000 Maribor}
\affiliation{Max-Planck-Institut f\"ur Physik, 80805 M\"unchen}
\affiliation{School of Physics, University of Melbourne, Victoria 3010}
\affiliation{University of Mississippi, University, Mississippi 38677}
\affiliation{University of Miyazaki, Miyazaki 889-2192}
\affiliation{Moscow Physical Engineering Institute, Moscow 115409}
\affiliation{Graduate School of Science, Nagoya University, Nagoya 464-8602}
\affiliation{Universit\`{a} di Napoli Federico II, 80055 Napoli}
\affiliation{Nara Women's University, Nara 630-8506}
\affiliation{National Central University, Chung-li 32054}
\affiliation{National United University, Miao Li 36003}
\affiliation{Department of Physics, National Taiwan University, Taipei 10617}
\affiliation{H. Niewodniczanski Institute of Nuclear Physics, Krakow 31-342}
\affiliation{Nippon Dental University, Niigata 951-8580}
\affiliation{Niigata University, Niigata 950-2181}
\affiliation{Novosibirsk State University, Novosibirsk 630090}
\affiliation{Osaka City University, Osaka 558-8585}
\affiliation{Pacific Northwest National Laboratory, Richland, Washington 99352}
\affiliation{Panjab University, Chandigarh 160014}
\affiliation{Peking University, Beijing 100871}
\affiliation{University of Pittsburgh, Pittsburgh, Pennsylvania 15260}
\affiliation{Punjab Agricultural University, Ludhiana 141004}
\affiliation{Theoretical Research Division, Nishina Center, RIKEN, Saitama 351-0198}
\affiliation{University of Science and Technology of China, Hefei 230026}
\affiliation{Seoul National University, Seoul 08826}
\affiliation{Showa Pharmaceutical University, Tokyo 194-8543}
\affiliation{Soochow University, Suzhou 215006}
\affiliation{Soongsil University, Seoul 06978}
\affiliation{University of South Carolina, Columbia, South Carolina 29208}
\affiliation{Sungkyunkwan University, Suwon 16419}
\affiliation{School of Physics, University of Sydney, New South Wales 2006}
\affiliation{Department of Physics, Faculty of Science, University of Tabuk, Tabuk 71451}
\affiliation{Tata Institute of Fundamental Research, Mumbai 400005}
\affiliation{Toho University, Funabashi 274-8510}
\affiliation{Department of Physics, Tohoku University, Sendai 980-8578}
\affiliation{Earthquake Research Institute, University of Tokyo, Tokyo 113-0032}
\affiliation{Department of Physics, University of Tokyo, Tokyo 113-0033}
\affiliation{Tokyo Institute of Technology, Tokyo 152-8550}
\affiliation{Tokyo Metropolitan University, Tokyo 192-0397}
\affiliation{Virginia Polytechnic Institute and State University, Blacksburg, Virginia 24061}
\affiliation{Wayne State University, Detroit, Michigan 48202}
\affiliation{Yamagata University, Yamagata 990-8560}
\affiliation{Yonsei University, Seoul 03722}
\author{C.~Bele\~{n}o}\affiliation{II. Physikalisches Institut, Georg-August-Universit\"at G\"ottingen, 37073 G\"ottingen} 
\author{A.~Frey}\affiliation{II. Physikalisches Institut, Georg-August-Universit\"at G\"ottingen, 37073 G\"ottingen} 
  \author{I.~Adachi}\affiliation{High Energy Accelerator Research Organization (KEK), Tsukuba 305-0801}\affiliation{SOKENDAI (The Graduate University for Advanced Studies), Hayama 240-0193} 
  \author{H.~Aihara}\affiliation{Department of Physics, University of Tokyo, Tokyo 113-0033} 
  \author{D.~M.~Asner}\affiliation{Brookhaven National Laboratory, Upton, New York 11973} 
  \author{H.~Atmacan}\affiliation{University of Cincinnati, Cincinnati, Ohio 45221} 
  \author{T.~Aushev}\affiliation{Higher School of Economics (HSE), Moscow 101000} 
  \author{R.~Ayad}\affiliation{Department of Physics, Faculty of Science, University of Tabuk, Tabuk 71451} 
  \author{P.~Behera}\affiliation{Indian Institute of Technology Madras, Chennai 600036} 
  \author{J.~Bennett}\affiliation{University of Mississippi, University, Mississippi 38677} 
% ...
\collaboration{The Belle Collaboration}


\begin{abstract}
We present an analysis of the exclusive $B^{+}\to\pi^{+}\pi^{-}\ell^{+}\nu_{\ell}$ decay, where $\ell$ represents an electron or a muon, with the assumption of charge-conjugation symmetry and lepton universality. The analysis uses the full $\Upsilon(4S)$ data sample collected by the Belle detector, corresponding to 711 fb$^{-1}$ of integrated luminosity. We select the events by fully reconstructing one $B$ meson in hadronic decay modes, subsequently determining the properties of the other $B$ meson. We extract the signal yields using a binned maximum-likelihood fit to the missing-mass squared distribution in bins of the invariant mass of the two pions or the momentum transfer squared. We measure a total branching fraction of ${{\cal B}(B^{+}\to \pi^{+}\pi^{-}\ell^{+}\nu_{\ell})= [22.7 ^{+1.9}_{-1.6} (\mathrm{stat}) \pm 3.5(\mathrm{syst}) ]\times 10^{-5}}$, where the uncertainties are statistical and systematic, respectively. This result is the first reported measurement of this decay. 
\end{abstract}

\pacs{12.15.-y, 13.20.He, 13.20.-v, 14.40.Nd}

\maketitle

\tighten

{\renewcommand{\thefootnote}{\fnsymbol{footnote}}}
\setcounter{footnote}{0}

\section{INTRODUCTION}

The reported measurements of exclusive semileptonic ${b\to u \ell \nu_{\ell}}$ decays, with $\ell$ either a muon or electron, do not saturate the inclusive charmless semileptonic ${b\to u \ell \nu_{\ell}}$ decay rate. Summing up all observed exclusive modes, only about 25\% of the inclusive rate can be accounted for~\cite{PDG}. The remaining modes pose a sizeable source of systematic uncertainty on inclusive and exclusive semileptonic ${b\to u \ell \nu_{\ell}}$ measurements or in decays in which such processes constitute important backgrounds. The absolute value of the Cabibbo-Kobayashi-Maskawa (CKM) matrix element $|V_{ub}|$~\cite{KM,cabibbo} can be precisely determined by combining measured branching fractions with predictions for the total rate. Three direct methods are considered as mature at the present time: first, combining the measured branching fraction of ${B\to \pi\ell\bar{\nu}_{\ell}}$ with lattice quantum chromodynamics (QCD) information to determine $|V_{ub}|$ and the non-perturbative form factors in a global fit~\cite{PDG,HFLAV}; second, measurement of the inclusive charmless semileptonic branching fraction, which is combined with calculations of the decay rate at NNLO in QCD plus non-perturbative parameters, determined in global fits to ${b\to c}$ semileptonic decays~\cite{PDG,HFLAV}; and last, combining the measured ratio of branching fractions of ${\Lambda_{b}\to p\ell\bar{\nu}_{\ell}}$ and ${\Lambda_{b}\to \Lambda_{c}\ell\bar{\nu}_{\ell}}$ with lattice QCD information to extract the ratio $|V_{ub}|/|V_{cb}|$~\cite{LHCbVubVcbR}. The determinations from the exclusive and inclusive approaches are only marginally compatible, resulting in a difference more than two standard deviations~\cite{ReviewVub}. A fourth method is the indirect determination of $|V_{ub}|$ with combining angles and other measurements characterizing the unitarity triangle. This indirect method is carried out by such groups as CKMfitter~\cite{CKMfitter} and UTfit~\cite{UTfit}. The values determined in these fits favor the exclusive result.

In this paper, we present the first measurement of the branching fraction of the exclusive channel ${B^{+}\to\pi^{+}\pi^{-}\ell^{+}\nu_{\ell}}$, where $\ell$ represents electrons and muons, and charge-conjugation symmetry and lepton universality are assumed. This channel is of particular interest, as the $\pi^{+}\pi^{-}$ system receives contributions from nonresonant and various resonant states, giving rise to a rich spectroscopy of the system. In this manner, it can serve as a probe to inspect the internal structure of light mesons decaying to a charged-pion pair, given that in semileptonic decays the hadronic and leptonic currents can be treated independently because the latter are not affected by the strong force~\cite{Dto2pilnu}. Measurements of branching fractions of this decay will improve the calculation of the ${B\to\pi\pi}$ form factors, which are an essential hadronic input for other processes such as the rare flavor-changing-neutral-current decay ${B\to\pi\pi\ell^{+}\ell^{-}}$ and to hadronic decays such as ${B\to\pi\pi\pi}$~\cite{FFB2pipi,BL4}. The resonant channel ${B^{+}\to \rho^{0}\ell^{+}\nu_{\ell}}$, which contributes to the   ${B^{+}\to\pi^{+}\pi^{-}\ell^{+}\nu_{\ell}}$ final state, has been measured by the CLEO~\cite{CLEOrho0}, Belle~\cite{BELLErho0I,Sibidanov}, and BaBar~\cite{BaBarUncFSR1} collaborations. All these results focus on reconstructing the resonant $\rho^{0}$ final state and do not measure the full $\pi^{+}\pi^{-}$ invariant-mass spectrum. The exclusive measurement of the ${B^{+}\to\pi^{+}\pi^{-}\ell^{+}\nu_{\ell}}$ decay presented in this paper extends these previous studies. Furthermore, more precise knowledge of the nonresonant $\pi^{+}\pi^{-}$ contributions will help improve future measurements of the $\rho^{0}$ final state~\cite{ObservablesBl4}. With the rapid progress of lattice QCD, we are hopeful that the measured ${B^{+}\to\pi^{+}\pi^{-}\ell^{+}\nu_{\ell}}$ branching fraction and future measurements at Belle II will provide a new avenue to determine $|V_{ub}|$, which is expected to reach a precision at the 2\% level~\cite{B2TIP}.

\section{DETECTOR, DATA SET, AND MONTE CARLO SIMULATION }

The Belle detector is a large-solid-angle magnetic spectrometer consisting of a silicon vertex detector (SVD), a 50-layer central drift
chamber (CDC), an array of aerogel threshold Cherenkov counters (ACC),
a barrel-like arrangement of time-of-flight scintillation counters
(TOF), and an electromagnetic calorimeter comprised of CsI(Tl) crystals (ECL) located inside a superconducting solenoid coil that provides a 1.5~T magnetic field.  An iron flux-return located outside of the coil is instrumented to detect $K_L^0$ mesons and to identify muons (KLM). The detector is described in detail
elsewhere~\cite{Belle}.

We use the entire Belle $\Upsilon(4S)$ data sample of 711~fb$^{-1}$ collected at the KEKB asymmetric-energy $e^+e^-$ collider~\cite{KEKB}. The sample contains ${(772\pm 11)\times 10^6}$ ${e^+e^-\to\Upsilon(4S)\to B\bar B}$ events. The Belle detector used two inner detector configurations in the course of the experiment. The first arrangement consisted of a 2.0-cm-radius beampipe, and a three-layer silicon vertex detector used to collect a sample of $152 \times 10^6$ $B\bar B$~pairs, while the second comprised a 1.5-cm-radius beampipe, a four-layer silicon detector, and a small-cell inner drift chamber employed to record the remaining ${620\times 10^6}$ $B\bar B$ pairs~\cite{svd2}.

Monte Carlo (MC) simulated samples are generated using the EvtGen~\cite{EVTGEN} package, and the response of the detector is modeled using GEANT3~\cite{GEANT3}. We account for final-state radiation (FSR) effects from charged particles by using the PHOTOS package~\cite{PHOTOS1,PHOTOS2}. A sample of ${\Upsilon(4S)\to B\bar B}$ events, where the $B$ meson decays entirely via the dominating quark-level transition $b\to cW$ (generic $B$ decays), was generated with a size equivalent to ten times the integrated luminosity of the data sample. Continuum events of the form $e^+e^-\to q\bar q$, where $q$ denotes $u$, $d$, $s$, or $c$ quarks, were simulated using PYTHIA6.4~\cite{PYTHIA} in a sample containing six times the integrated luminosity of the data sample. Charmless rare $B$ decays, occurring among others via loop transitions such as $b\to s$ quark transition or via radiative decays, are generated with a sample size corresponding to 50 times the integrated luminosity of the data sample. 

The signal ${B^{+}\to\pi^{+}\pi^{-}\ell^{+}\nu_{\ell}}$ sample is produced with the phase-space (PHSP) model of EvtGen, to make sure that every point in phase space is populated, independent of whether or not it can be reached by an intermediate resonance. Given that branching fraction estimations for the ${B^{+}\to\pi^{+}\pi^{-}\ell^{+}\nu_{\ell}}$ decay in the entire phase space are not available from either lattice QCD or QCD sum-rule calculations, we assumed a branching fraction of $31.7\times 10^{-5}$ according to reference~\cite{Multipion} using ${|V_{ub}|/|V_{cb}|=0.083\pm 0.006}$~\cite{LHCbVubVcbR}. We generate 100 million $B\bar B$ events, with one $B$ meson decaying generically and the other through the ${B^{+}\to\pi^{+}\pi^{-}\ell^{+}\nu_{\ell}}$ channel. Various exclusive semileptonic decays proceeding through the Cabibbo-suppressed transition ${b\to u \ell \nu_{\ell}}$ at quark level were produced with a sample size equivalent to 20 times the integrated luminosity of the data. This sample contains the following decays: $B^{+}\to\pi^{0}\ell^{+}\nu_{\ell}$, $B^{+}\to\eta\ell^{+}\nu_{\ell}$, $B^{+}\to\eta^{\prime}\ell^{+}\nu_{\ell}$, $B^{+}\to\omega\ell^{+}\nu_{\ell}$, $B^{+}\to a_{0}(980)^{0}\ell^{+}\nu_{\ell}$, $B^{+}\to a_{1}(1260)^{0}\ell^{+}\nu_{\ell}$, $B^{+}\to a_{2}(1320)^{0}\ell^{+}\nu_{\ell}$, $B^{+}\to b_{1}(1235)^{0}\ell^{+}\nu_{\ell}$, $B^{+}\to f_{1}(1285)\ell^{+}\nu_{\ell}$, $B^{+}\to f_{2}^{\prime}(1525)\ell^{+}\nu_{\ell}$, $B^{0}\to\rho^{-}\ell^{+}\nu_{\ell}$, $B^{0}\to\pi^{-}\ell^{+}\nu_{\ell}$, $B^{0}\to a_{0}(980)^{-}\ell^{+}\nu_{\ell}$, $B^{0}\to a_{1}(1260)^{-}\ell^{+}\nu_{\ell}$, $B^{0}\to a_{2}(1320)^{-}\ell^{+}\nu_{\ell}$, and $B^{0}\to b_{1}(1235)^{-}\ell^{+}\nu_{\ell}$. These decays are generated using form factor calculations from ISGW2~\cite{ISGW2} and light-cone sum rules (LCSR)~\cite{Ball98}. We do not consider an inclusive component since the $V_{ub}$ generator~\cite{DeFazioNeubert}, used to model this contribution, incorrectly describes nonresonant states in the entire phase space. High-multiplicity mass terms that can contribute to the nonresonant component come from decays such as ${B^{+}\to\pi^{+}\pi^{-}\pi^{0}\ell^{+}\nu_{\ell}}$ and ${B^{+}\to\pi^{+}\pi^{-}\pi^{0}\pi^{0}\ell^{+}\nu_{\ell}}$. However, after simulating these processes with the PHSP generator and examining their contributions after the full selection, they are found to be negligible and thus are not considered further in this analysis.

We set the branching fractions of the decays $B\to D \ell \nu_{\ell}$, $B\to D^{*}\ell\nu_{\ell}$, $B\to D_{1}\ell\nu_{\ell}$, $B\to D_{1}^{\prime}\ell\nu_{\ell}$, $B\to D_{2}^{*}\ell\nu_{\ell}$, $B\to D_{0}^{*}\ell\nu_{\ell}$, and of the known exclusive charmless semileptonic $B$ decays to the latest experimental averages~\cite{PDG}. We reweight the Caprini-Lellouch-Neubert (CLN)-based form factors~\cite{CLN} of the decays $B\to D^{(*)}\ell\nu_{\ell}$ to the recent world-average values~\cite{HFLAV}, and the form factors of the $B\to D^{**}\ell\nu_{\ell}$ decay according to the model of Leibovich-Ligeti-Stewart-Wise (LLSW)~\cite{LLSW}. We also correct the MC for the efficiency of particle identification of charged tracks, derived from studies using control samples for known processes, as described later in the section about systematic uncertainties associated to the detector simulation. These corrections depend on the kinematics of the particles involved.  

\section{EVENT SELECTION}

This analysis employs a full reconstruction technique~\cite{Feindt} based on the NeuroBayes neural-network package~\cite{Feindt2}, in which we reconstruct one $B$ meson ($B_{\mathrm{tag}}$) stemming from the $\Upsilon(4S)$ resonance in 1104 hadronic modes. This tagging technique allows one to determine the properties of the other $B$ meson ($B_{\mathrm{sig}}$) from kinematic constraints via conservation laws. Subsequently, we reconstruct the $B_{\mathrm{sig}}$ using the rest of the event, except for the neutrino, which is invisible to the detector. 

To filter $B\bar B$ events from non-hadronic background such as two-photon, radiative Bhabha, and $\tau^{+}\tau^{-}$ processes, we implement a selection based on charged-track multiplicity and the total visible energy~\cite{gas_beam}. Afterward, to reject continuum events, we add 18 modified Fox-Wolfram~\cite{SFWM} moment variables to the NeuroBayes neural network used in the reconstruction of the $B_{\mathrm{tag}}$. The output classifier $o_{\mathrm{tag}}^{\mathrm{cs}}$ of the algorithm ranges from zero to unity, with higher values indicating a higher probability of correctly reconstructing a $B$ meson with low contamination of continuum events. We retain candidates with $\ln o_{\mathrm{tag}}^{\mathrm{cs}} >-4.9$ to ensure good quality of the $B_{\mathrm{tag}}$ candidate. This requirement is optimized using a figure-of-merit $N_{S}/\sqrt{N_{S}+N_{B}}$, where $N_{S}$ and $N_{B}$ are the expected number of events from MC for signal and background, respectively. With this selection criterion, we attain a tag-side efficiency of $0.1\%$ and a tag-side purity of around $23\%$ for charged $B$ mesons reconstructed with the full hadronic tagging algorithm. Differences in the tagging efficiency between data and MC have been evaluated in reference ~\cite{Sibidanov}; they depend on the value of the network output and the $B_{\mathrm{tag}}$ reconstructed channel. We take an event-by-event correction factor from this study, derived from a control sample of $B\to D^{(*)}\ell\nu$ decays on the signal side, to account for these discrepancies.

We require the beam-constrained mass, ${M_{\mathrm{bc}}=\sqrt{E_{\mathrm{beam}}^{2}-\left| \vec{p}_{B_{\mathrm{tag}}}\right|^{2}}}$, to be greater than 5.27 GeV~\cite{natural-units}. Here, $E_{\mathrm{beam}}$ and $\vec{p}_{B_{\mathrm{tag}}}$ are the beam energy and the three-momentum of the $B_{\mathrm{tag}}$  candidate in the $\Upsilon(4S)$ frame, respectively. We select only charged $B_{\mathrm{tag}}$ candidates since the signal mode only involves charged $B$ mesons.

The charged particles and neutral clusters in the event not associated with the $B_{\mathrm{tag}}$ candidate are used in the reconstruction of the $B_{\mathrm{sig}}$ candidate. Due to the magnetic field inside the detector, charged particles with low momenta spiral inside the CDC and may lead to multiple track candidates for the same particle. A pair of tracks is regarded as duplicated if they have momenta transverse to the beam direction below 275 MeV, with a small momentum difference (below 100 MeV) and an opening angle either below $15^\circ$ (same charges) or above $165^\circ$ (opposite charges). Once such a pair is identified, the track with the smaller value of the quantity $\left(5\times|\mathrm{d}r|\right)^{2}+|\mathrm{d}z|^{2}$ is kept, with $|\mathrm{d}r|$ and $|\mathrm{d}z|$ denoting the distance of closest approach of a given track to the interaction point (IP) in the plane perpendicular to the beam direction, or along the beam direction, respectively. This criterion was optimized using simulated tracks. In addition, we impose that all selected tracks satisfy $|\mathrm{d}r|<0.4$ cm and $|\mathrm{d}z|<2.0$ cm.

We identify charged hadrons using the ionization energy loss $\mathrm{d}E/\mathrm{d}x$ in the CDC, the time-of-flight in the TOF, and the Cherenkov light in the ACC~\cite{Nakano:2002jw}. The selection of charged pions in this analysis has an identification efficiency of 85\% and a kaon misidentification rate of 13\%. 

In this analysis, we only consider events with a single charged-lepton candidate on the signal side. Electron candidates are identified based on the ratio of the ECL energy to that of the CDC track, the ECL shower shape, the position matching between the CDC track and the ECL cluster, the energy loss in the CDC, and the response of the ACC ~\cite{hanagaki}. Furthermore, we require electrons to have a  minimum momentum of 0.3 GeV in the laboratory frame. Muon candidates are selected using their penetration range and transverse scattering in the KLM~\cite{abashian}, and requiring a minimum momentum of 0.6 GeV in the laboratory frame. In the momentum region relevant to this analysis, the average electron (muon) identification efficiency is about 87\% (89\%), and the probability of misidentifying a pion as an electron (muon) is 0.15\% (1.3\%). We veto charged leptons from photon conversion in the detector material and from $J/\psi$ and $\psi(2S)$ decays if the lepton candidate, when combined with an oppositely charged particle, gives an invariant mass $M_{\ell\ell}$  satisfying the following conditions: $M_{\ell\ell}<0.1$ GeV, $M_{\ell\ell} \in [3.00,3.15]$ GeV, or $M_{\ell\ell}\in [3.60,3.75]$ GeV. 

We reconstruct photons as clusters in the ECL not linked to a track in the CDC. To reject low-energy photons originating from background caused by the beam circulation, we require a minimum energy of 50 MeV, 100 MeV, and 150 MeV in the barrel, the forward endcap, and the backward endcap of the ECL, respectively. We reconstruct neutral pions from pairs of photons with an invariant mass in the range 120-150 MeV. The photons forming a neutral pion candidate are rejected from the one to be linked to a charged track. In electron events, we take into account possible Bremsstrahlung from the electron by searching for low-energy photons ($E_{\gamma}<1\,\text{GeV}$) within a $5^\circ$ cone around the lepton direction. If such a photon is found, it is merged with the electron and the sum of the momenta is taken to be the lepton momentum. If there is more than one photon candidate, only the nearest photon is merged with the electron.

\section{SIGNAL SELECTION AND BACKGROUND SUPPRESSION}

After applying the above criteria, we reconstruct the signal decay $B^{+}\to \pi^{+}\pi^{-}\ell^{+}\nu_{\ell}$ from the tracks not associated with $B^{-}_{\mathrm{tag}}$. In this manner, we require exactly three tracks on the signal side, the two charged pions and the lepton. Given that the neutrino is invisible to the detector, we infer its four-momentum from the missing momentum of the event, defined as 

\begin{equation}
P_{\mathrm{miss}} = P_{\Upsilon(4S)}-P_{B_{\mathrm{tag}}^{\pm}}-P_{\ell^{\mp}}-P_{\pi^{+}}-P_{\pi^{-}}, 
\end{equation}
where $P_{i}$ is the four-momentum of particle $i={\Upsilon(4S), B_{\mathrm{tag}}^{-}, \ell, \pi^{+}, \pi^{-}}$. We determine the missing-mass squared, $M_{\mathrm{miss}}^{2}=P_{\mathrm{miss}}^{2}$, to separate semileptonic decays from other processes. For correctly reconstructed semileptonic decays, $M_{\mathrm{miss}}^{2}$ sharply peaks at $0$, whereas other processes have a shoulder typically at positive values.

At this point in the reconstruction, the dominant background processes come from semileptonic $B$ decays to charmed mesons whose kinematic distributions resemble those of the signal. To suppress this background, we train a boosted decision tree (BDT) to recognize $B^{+}\to \pi^{+}\pi^{-}\ell^{+}\nu_{\ell}$ decays and identify $B$-meson decays into other final states. This BDT is coincidentially also effective against other backgrounds such as continuum, rare and charmless semileptonic $B$ decays. A statistically independent two sets of MC samples for signal and background are prepared. One set is used to train BDT with the stochastic gradient boosting approach from the TMVA software package ~\cite{TMVA}, which combines the baggings and boosting algorithms. Another one set is used for validation of the training. The following input variables are used:

\begin{figure*}[htpb]
	\subfloat{
		\includegraphics[width=0.3\textwidth ]{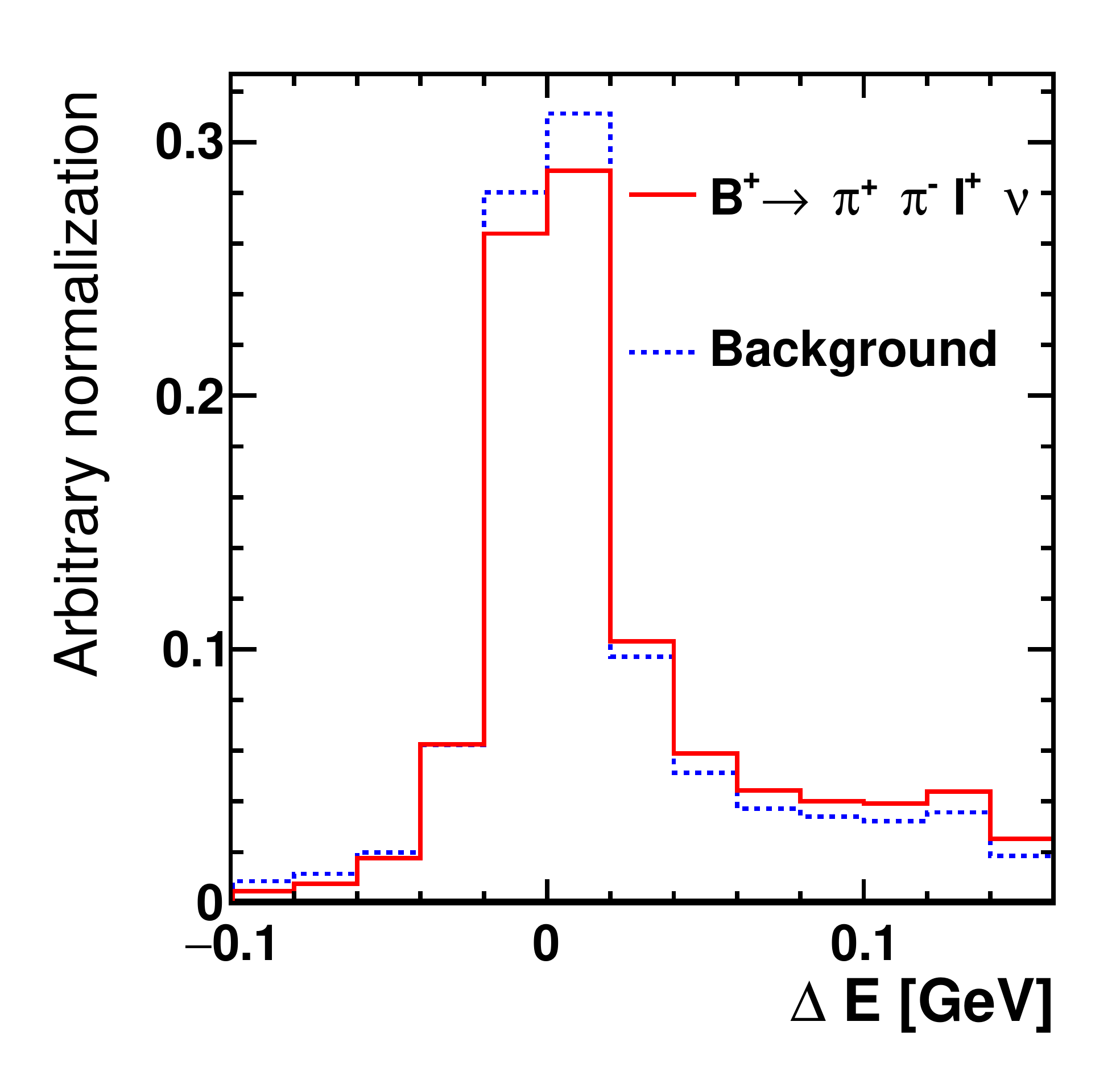}
	}
	\subfloat{
		\includegraphics[width=0.3\textwidth ]{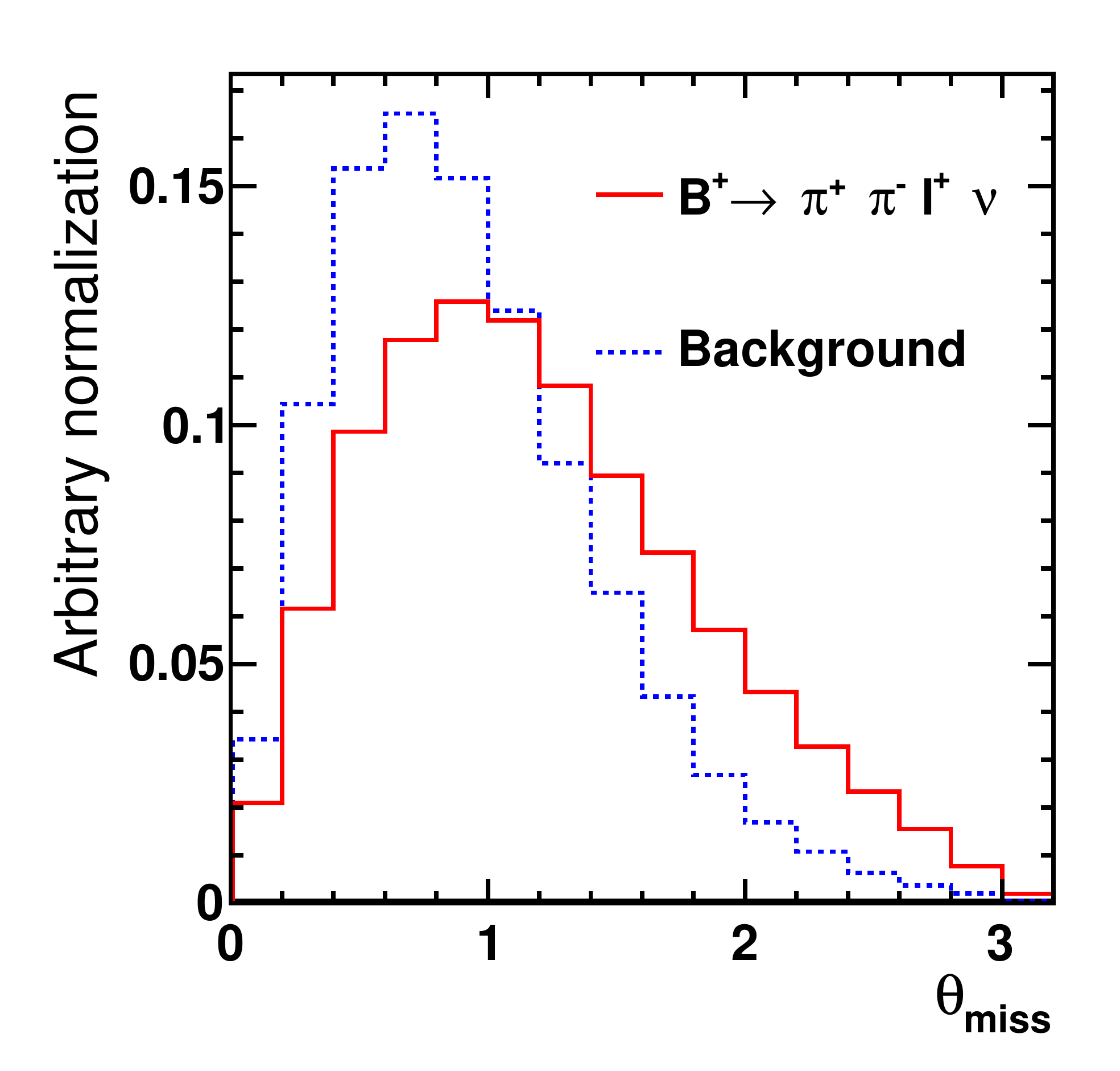}
	}
	\subfloat{
		\includegraphics[width=0.3\textwidth ]{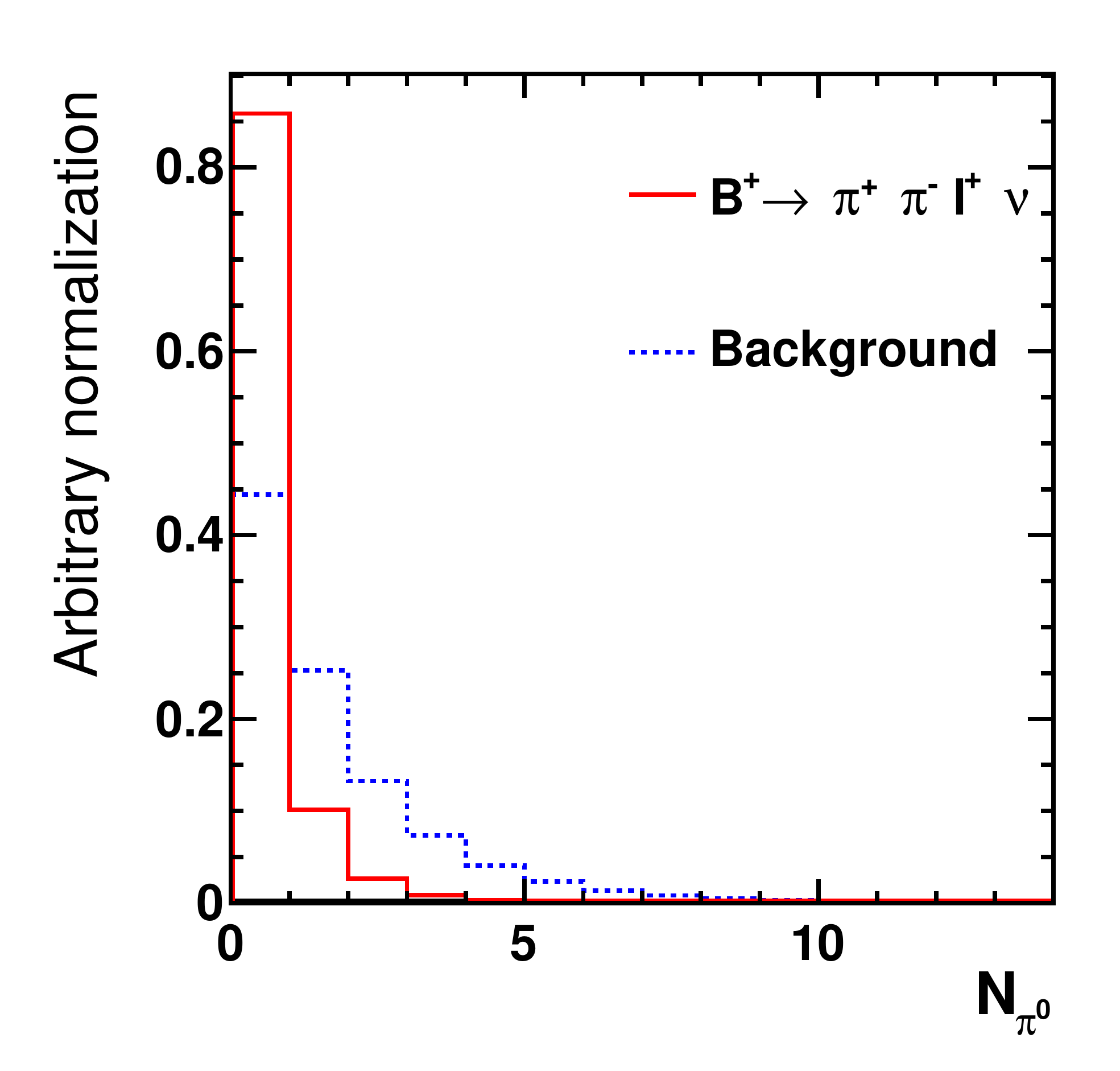}
	}\\
	\subfloat{
		\includegraphics[width=0.3\textwidth ]{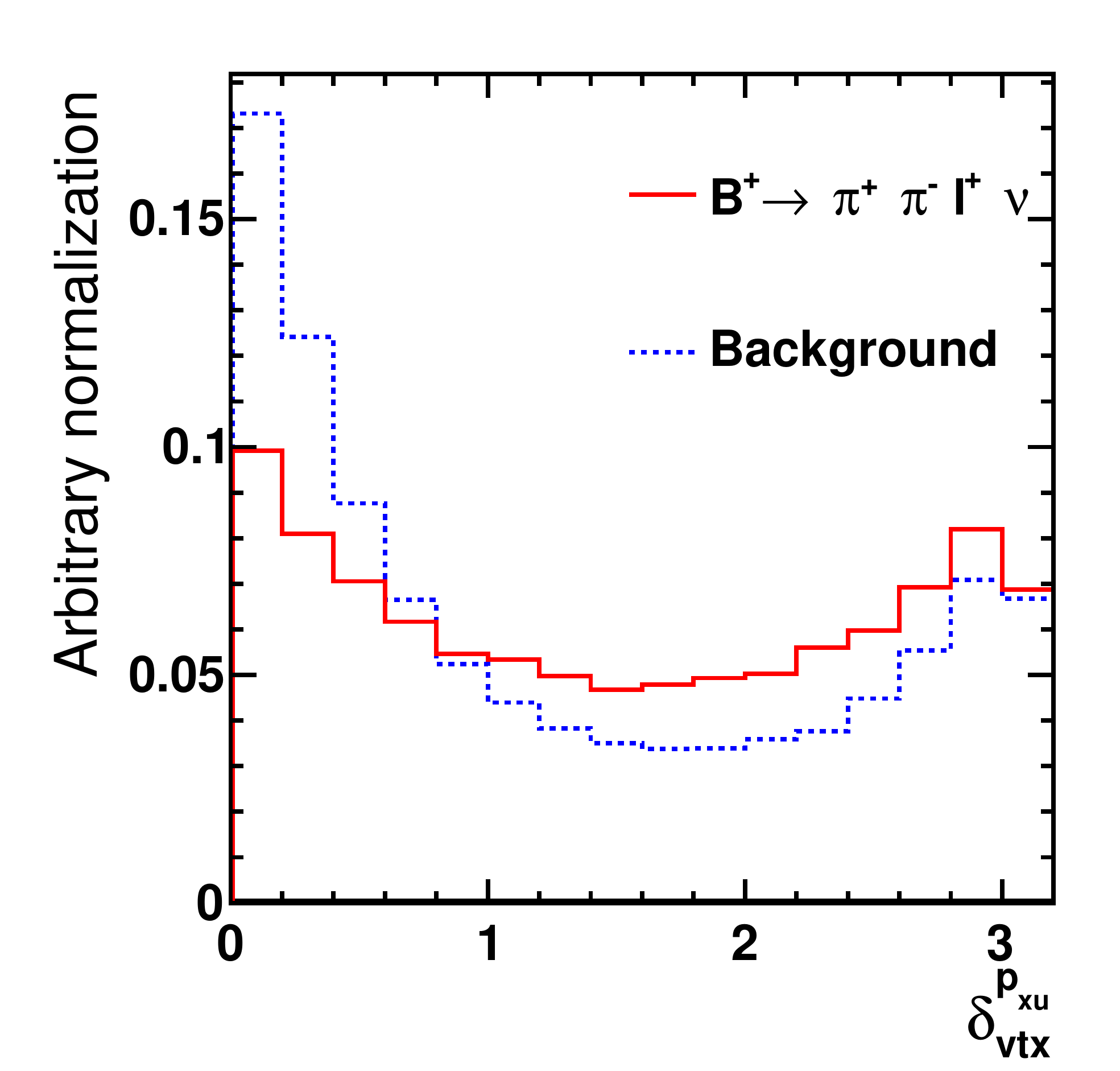}
	}
	\subfloat{
		\includegraphics[width=0.3\textwidth ]{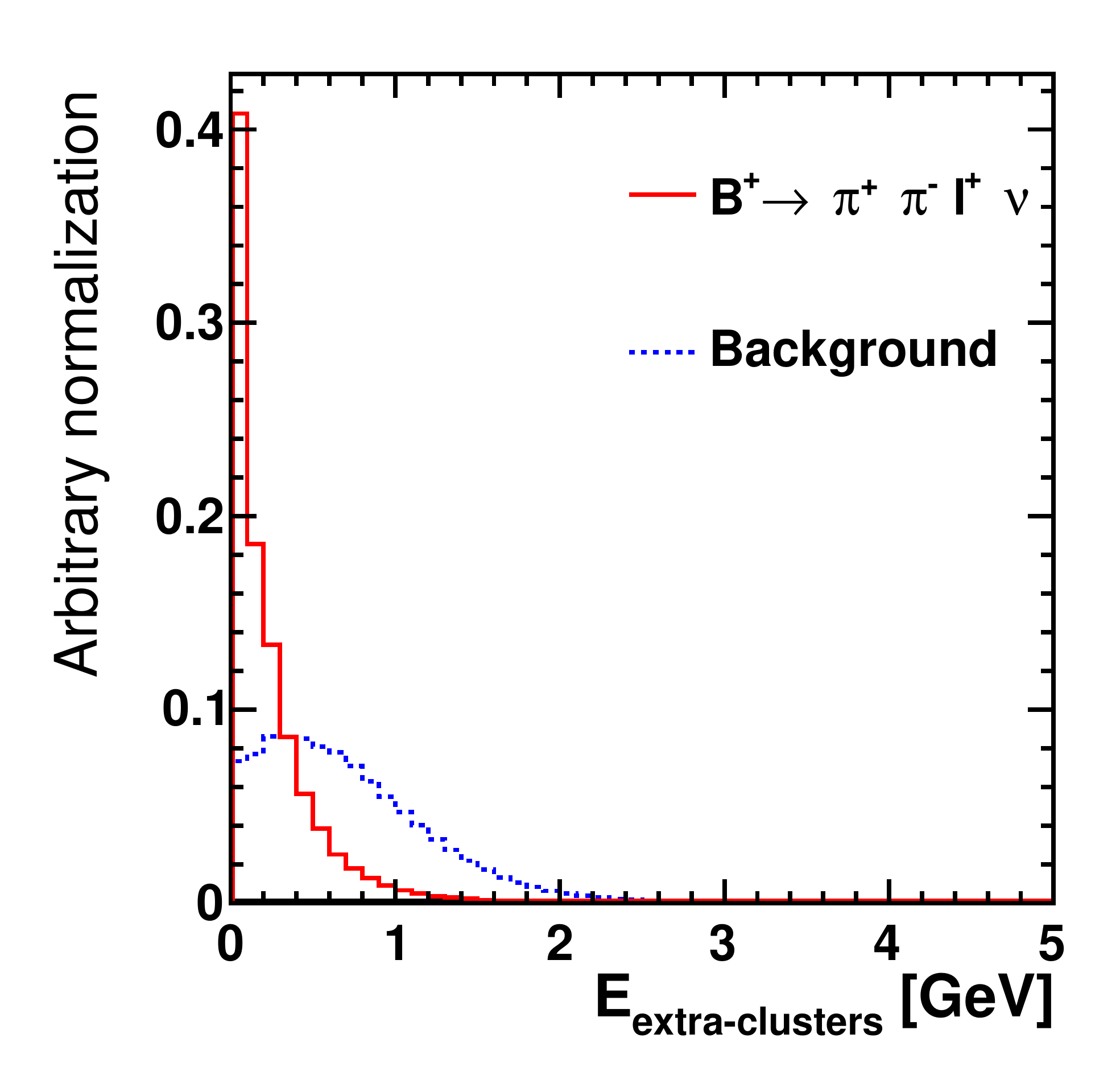}
	}
	\subfloat{
		\includegraphics[width=0.3\textwidth ]{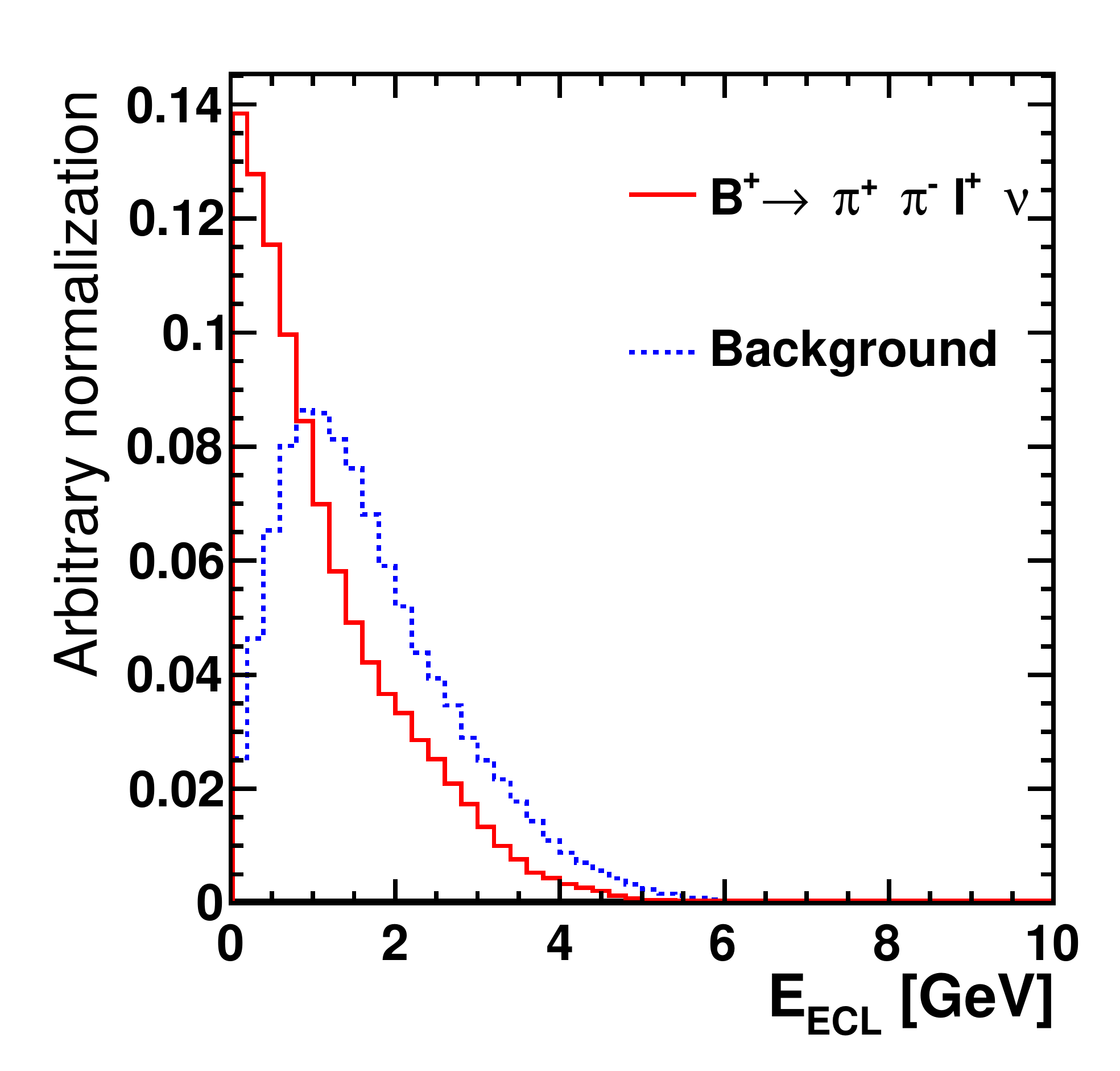}
	}
	\caption{Shape comparison of the input variables of the BDT before the selection on $\mathcal{O}_{\mathrm{BDT}}$ for simulated signal and background events.}
	\label{fig:InputVariablesShapes}
\end{figure*} 
\begin{enumerate}
 \item $\Delta E_{\mathrm{sig}}$: the difference between the beam and the $B_{\mathrm{sig}}$ meson energies in the center-of-mass system (c.m.), which is calculated using the $B_{\mathrm{tag}}$ meson, ${\Delta E_{\mathrm{sig}}=-\Delta E_{\mathrm{tag}}=-(E_{\mathrm{beam}}-E_{B_{\mathrm{tag}}})}$.
 \item $\theta_{\mathrm{miss}}$: the polar angle of the missing momentum in the laboratory frame.
 \item $N_{\pi^{0}}$: the multiplicity of $\pi^{0}$ candidates on the signal side.
 \item $\delta_{\mathrm{vtx}}^{p_{\mathrm{xu}}}$: the angle between the signal-side $\pi^{+}\pi^{-}$ momentum and the vector connecting the IP and the $\pi^{+}\pi^{-}$ decay vertex calculated in the laboratory frame. The distance of the $\pi^{+}\pi^{-}$-system to the IP for charmless intermediate states is smaller than that for two-track pairs associated with $D^{0}$ and $K_{S}^{0}$ mesons. Thus the angle $\delta_{\mathrm{vtx}}^{p_{\mathrm{xu}}}$ is useful in reducing these background processes.  
 \item $E_{\extraclusters}$: the total c.m. energy of photons within the barrel region not associated with either the $B_{\mathrm{tag}}$ or $B_{\mathrm{sig}}$ candidates.
 \item $E_{\mathrm{ECL}}$: the sum of the clusters in the ECL from the whole event not matching tracks and that pass the energy thresholds for photons. This calculation also includes ECL clusters made by photons that were incorrectly associated with a track and that satisfy $E9/E25>0.94$. The $E9/E25$ variable quantifies the transverse shower shape in the ECL, defined as the ratio of energy deposited in the $3\times3$ array of crystals centered on the track to that in the corresponding $5\times5$ array of crystals. This variable is suitable to separate overlapping hits in the ECL crystals caused by hadronic interaction with charged tracks and photons. For photons $E9/E25$ peaks at one, whereas for charged tracks it tends to have lower values.
\end{enumerate}
Distributions of the above variables for signal and background (with arbitrary normalizations) are shown in Fig.~\ref{fig:InputVariablesShapes}. 

\begin{figure}[tpb]
\hspace{-0.5cm}\includegraphics[width=0.5\textwidth ]{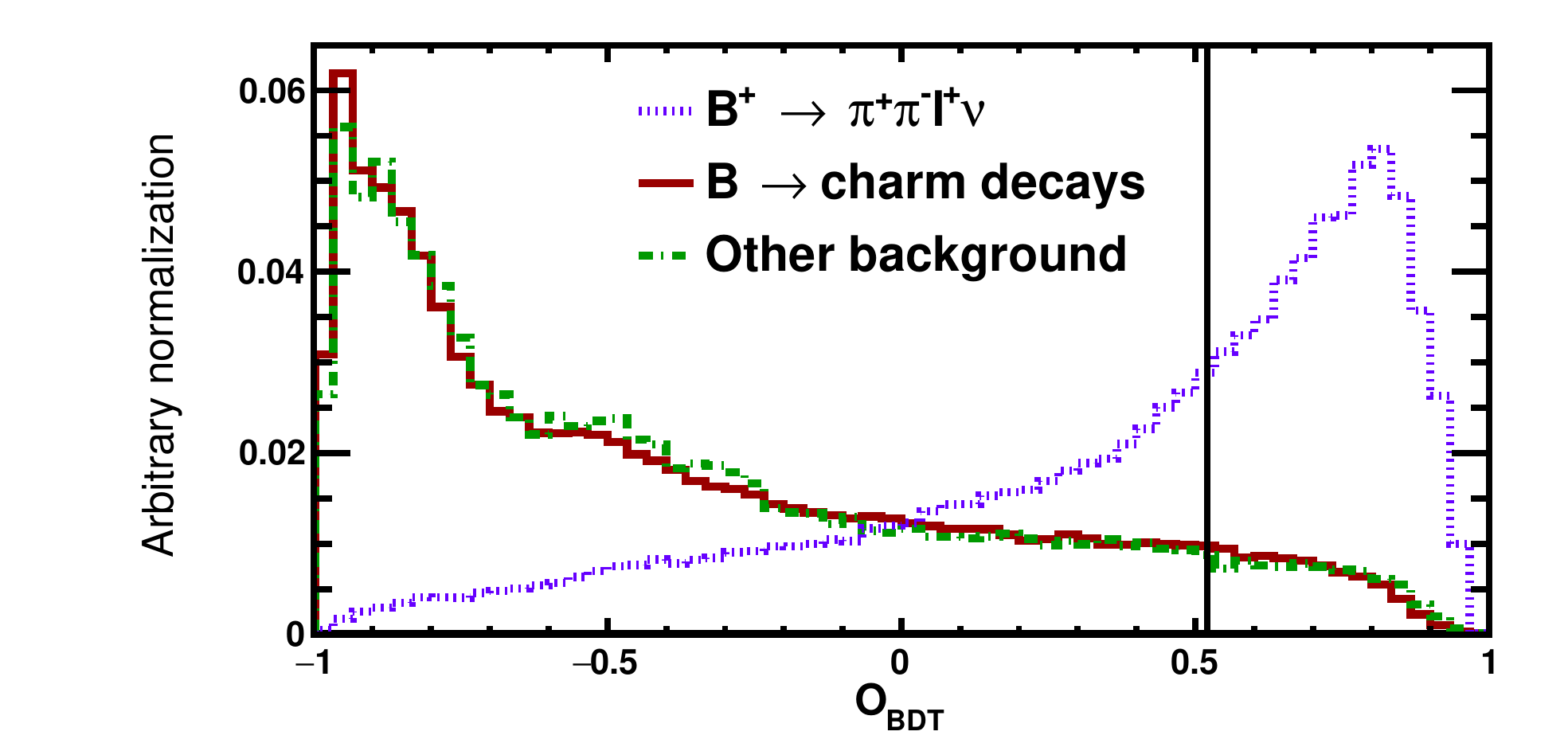}
\caption{Shapes of the BDT output for the signal and the major background processes, as predicted by MC. The vertical line shows the minimum requirement on this variable, obtained from optimizing a figure-of-merit $N_{S}/\sqrt{N_{S}+N_{B}}$.}
\label{fig:BDToutput}
\end{figure}

\begin{figure*}[htpb]
	\subfloat{
		\includegraphics[width=0.3\textwidth ]{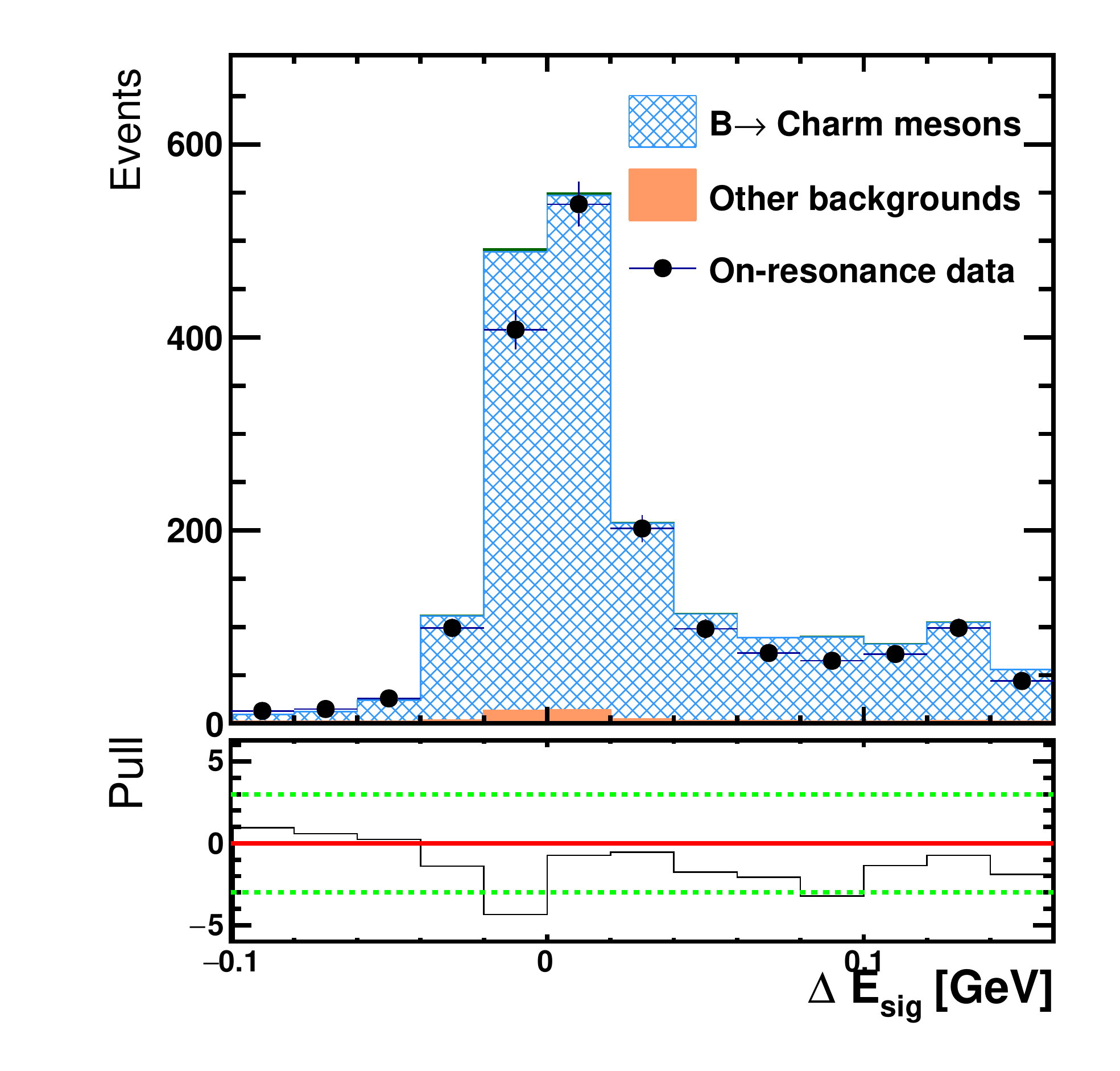}
	}
	\subfloat{
		\includegraphics[width=0.3\textwidth ]{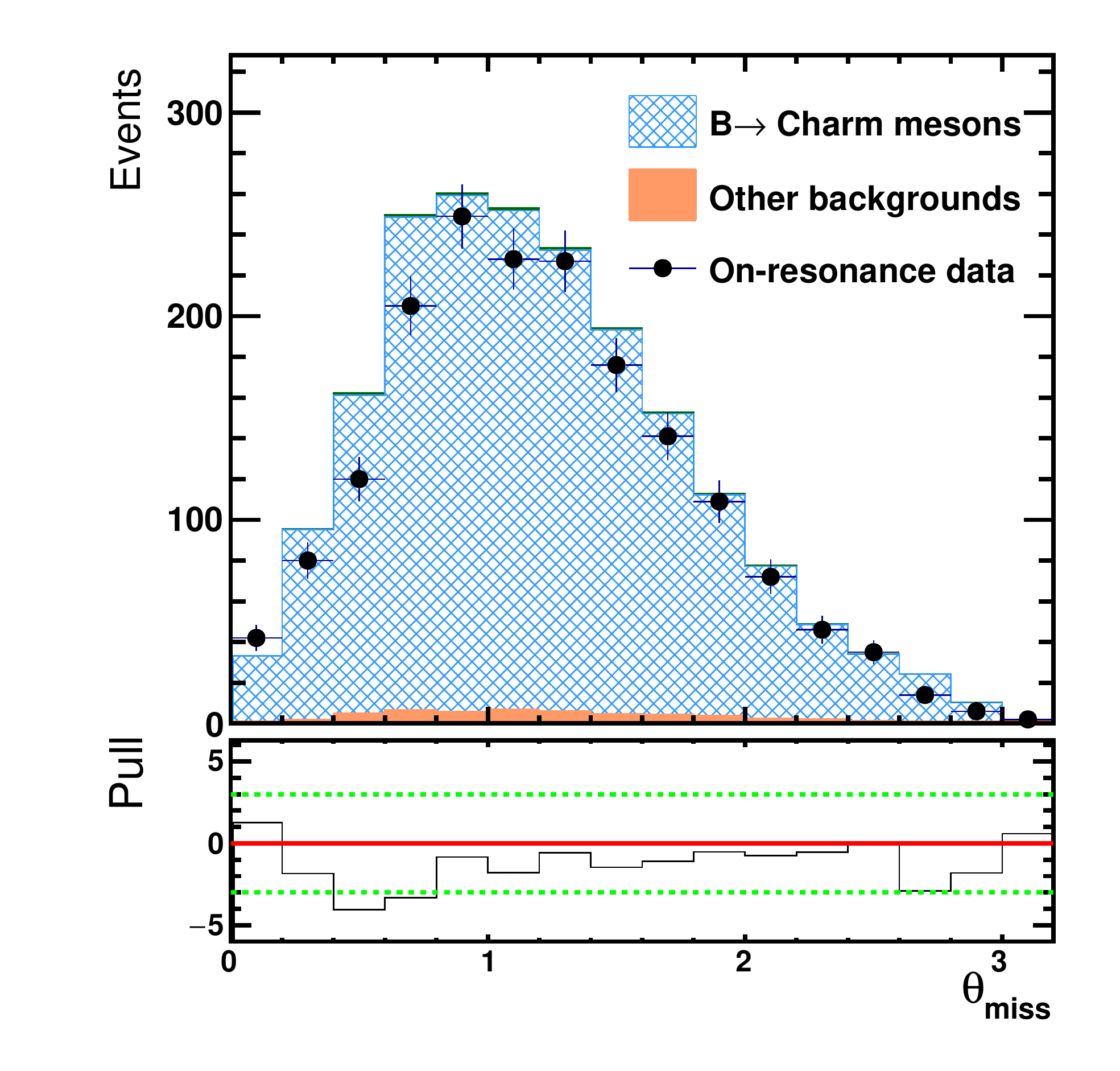}
	}
	\subfloat{
		\includegraphics[width=0.3\textwidth ]{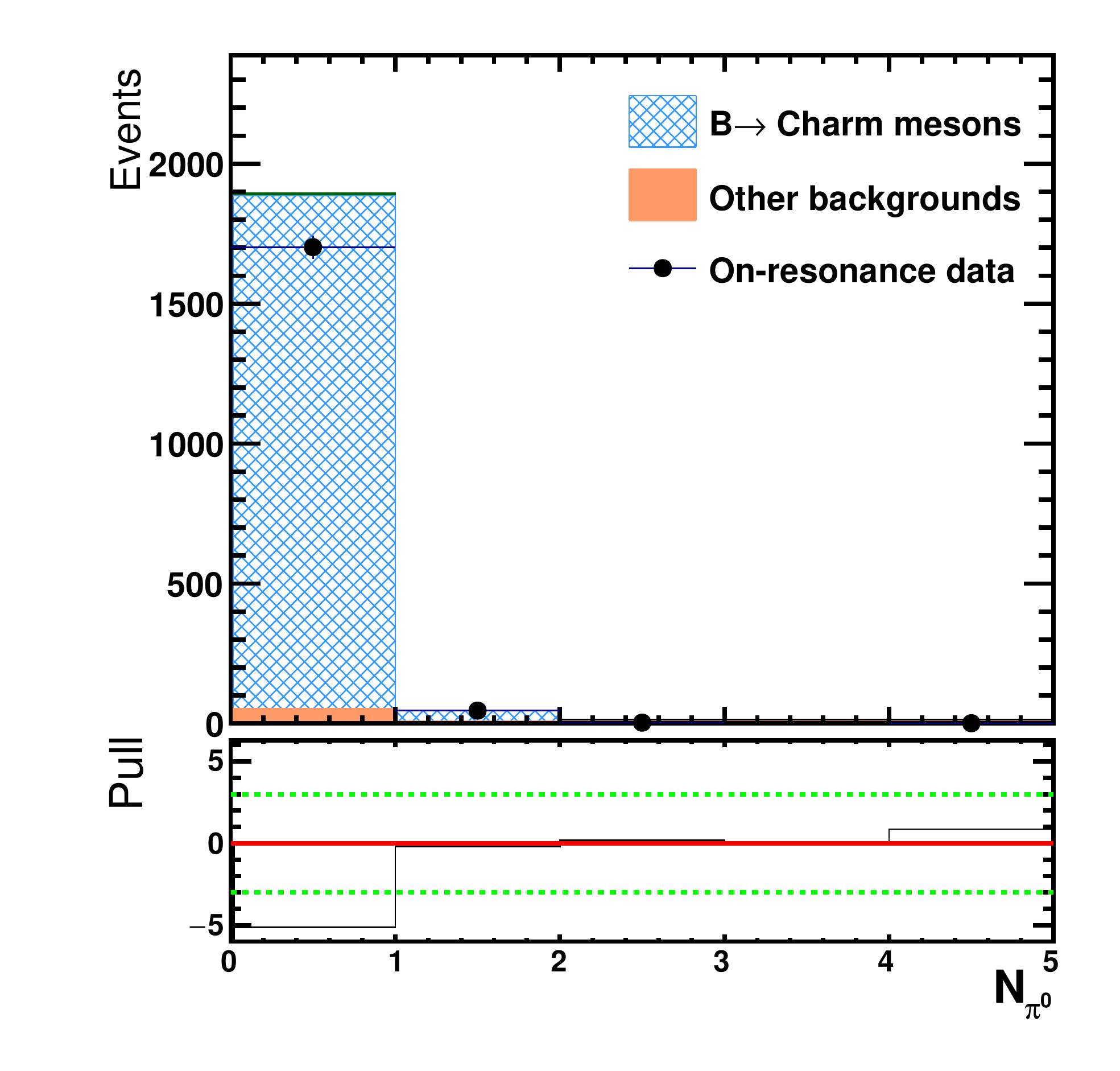}
	}\\
	\subfloat{
		\includegraphics[width=0.3\textwidth ]{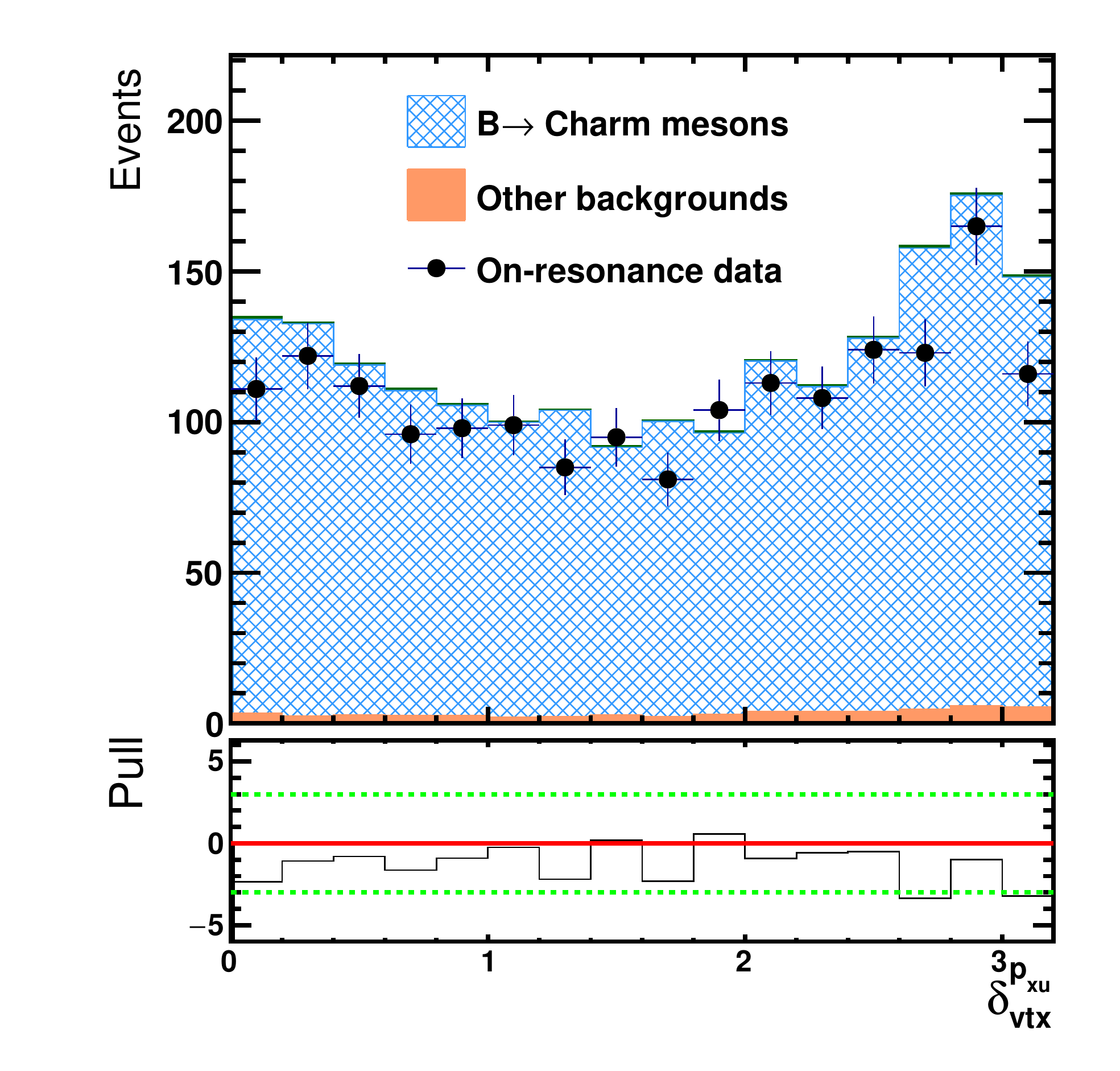}
	}
	\subfloat{
		\includegraphics[width=0.3\textwidth ]{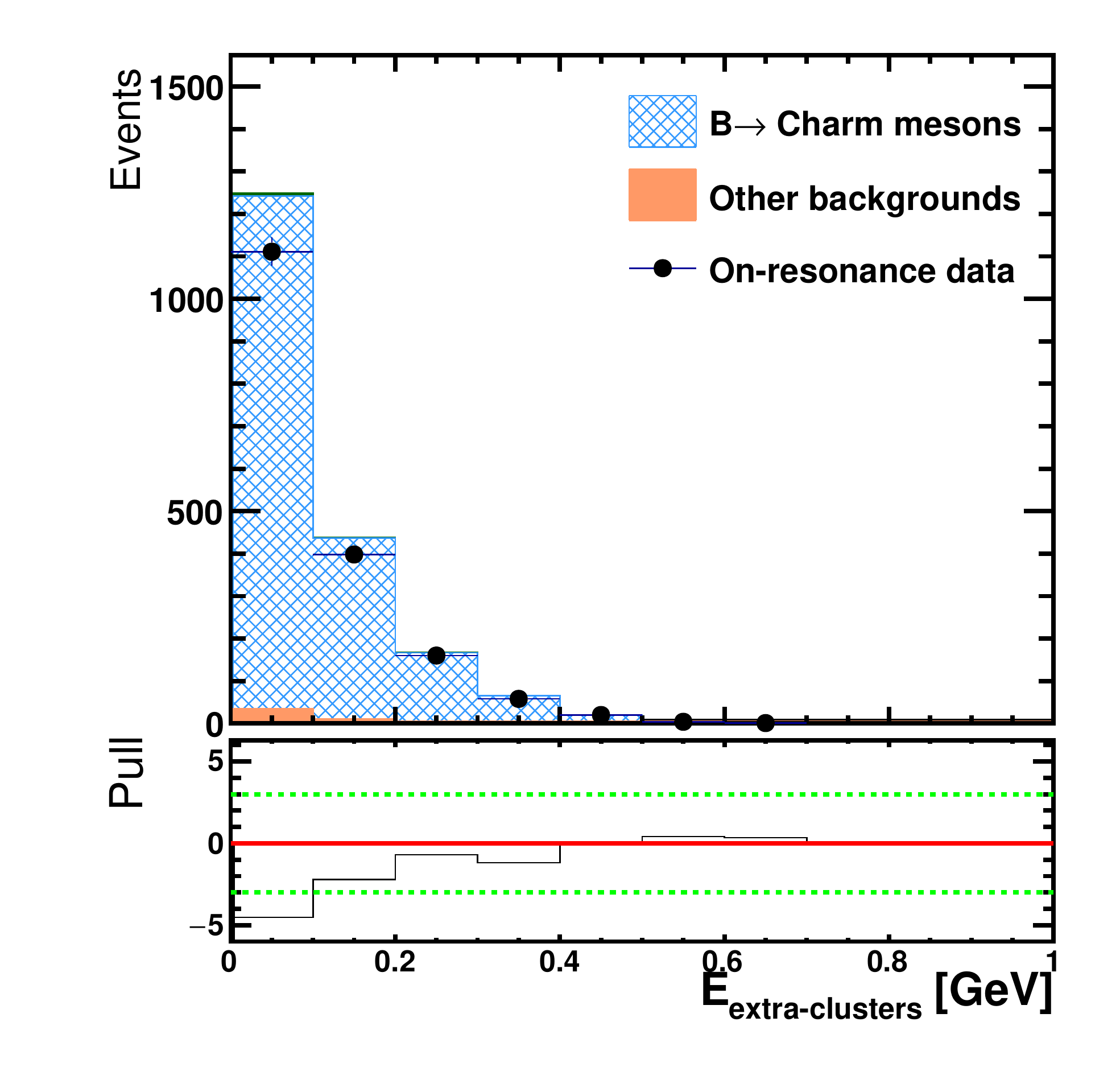}
	}
	\subfloat{
		\includegraphics[width=0.3\textwidth ]{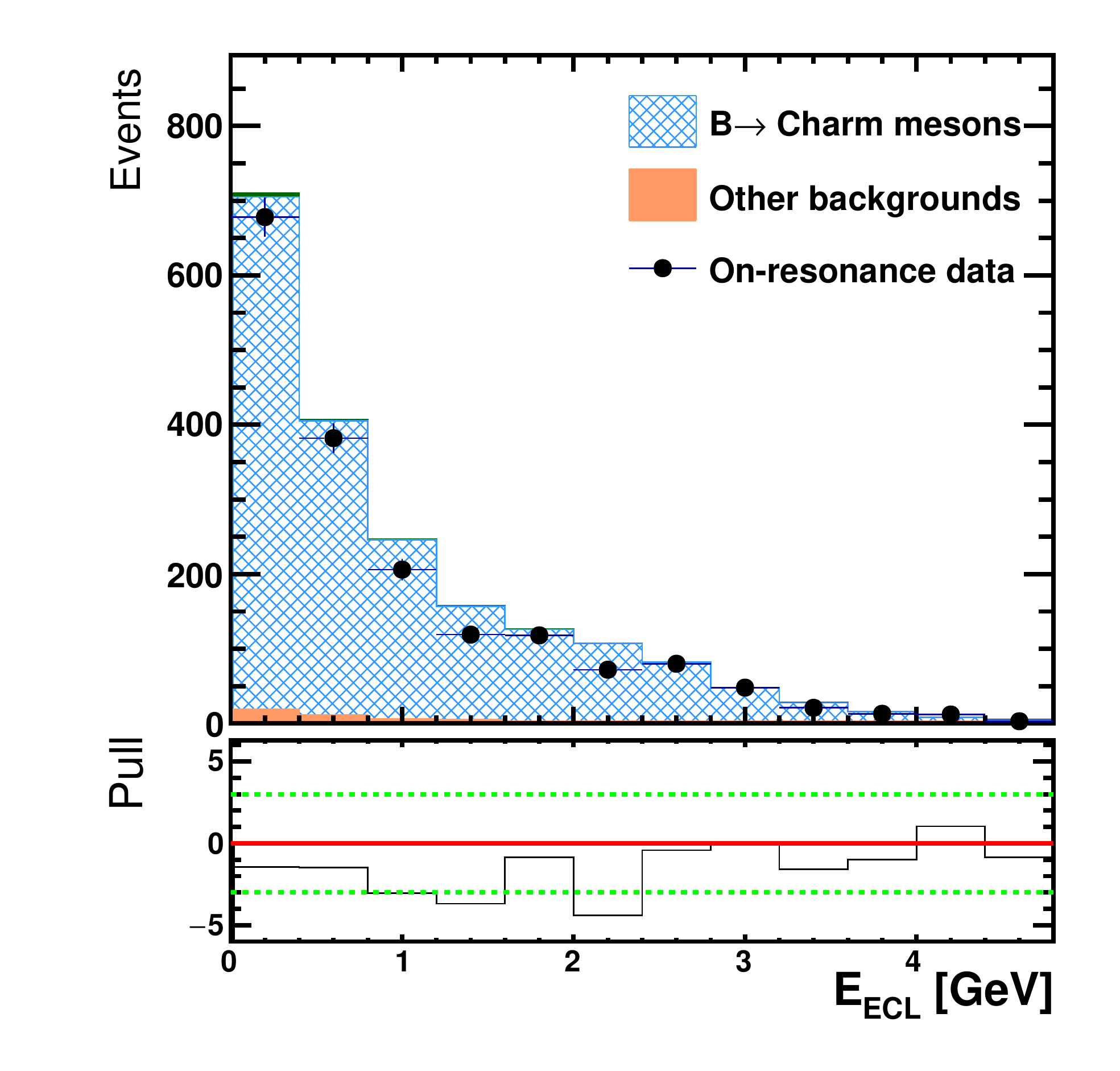}
	}
	\caption{Distributions of the input variables of the BDT in the sidebands of the missing-mass squared, after the selection on $\mathcal{O}_{\mathrm{BDT}}$. The shaded histogram shows the contribution from $B$ decays to charm mesons, while the solid histogram shows the contributions from other processes. The pull values are presented underneath each plot to display the difference of the data relative to the MC. The MC are normalized to the corresponding integrated luminosity.}
	\label{fig:InputVariables}
\end{figure*}

We choose a selection criterion on the BDT output classifier by optimizing a figure-of-merit $N_{S}/\sqrt{N_{S}+N_{B}}$. The distributions of the BDT classifier $\mathcal{O}_{\mathrm{BDT}}$ for the signal, $B$-meson decays to charm mesons and other backgrounds, as well as the selection criterion ($\mathcal{O}_{\mathrm{BDT}}>0.52$), are shown in Fig.~\ref{fig:BDToutput}. We validate the description of the variables used in the BDT using the sideband of the missing-mass squared distribution, defined as $M_{\mathrm{miss}}^{2}>2\,\text{GeV}^{2}$. These distributions are shown in Fig.~\ref{fig:InputVariables}.

\section{Signal extraction}

We perform a binned extended maximum-likelihood fit to the $M_{\mathrm{miss}}^{2}$ spectrum using histogram templates derived from MC simulation to determine the signal yields. We use a bin width of 0.2 $ \text{GeV}^{2}$ in the range $[-1.0,6.0]\; \text{GeV}^{2}$. Because of the negligible contribution of the continuum, $b\to u\ell\nu$, and rare $b\to s$ decay processes, we combine these into a single component and fix their event yields to the MC expectation (referred to as fixed background in the following). We thus distinguish among three components in our fit:
\begin{enumerate}
	\item the signal $B^{+}\to \pi^{+}\pi^{-}\ell^{+}\nu_{\ell}$,
	\item $B\to X_{c} \ell \nu$, where $X_{c}$ is a charm meson, and
	\item the fixed background,
\end{enumerate}
where yields of the first two components are floated in the fit.

\begin{figure*}[htpb]
\centering
\subfloat{
\includegraphics[width=0.5\textwidth ]{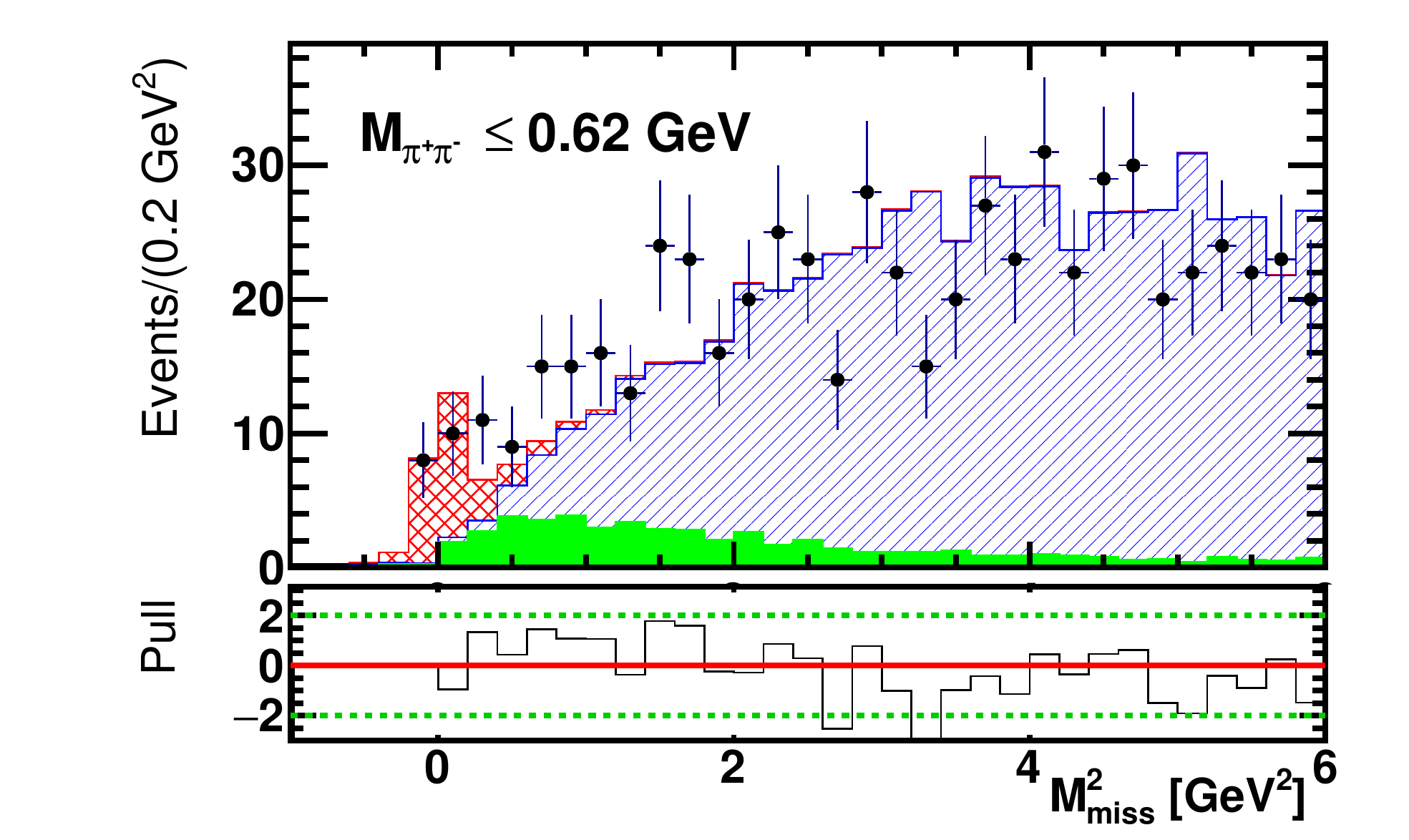}
}
\subfloat{
\includegraphics[width=0.5\textwidth ]{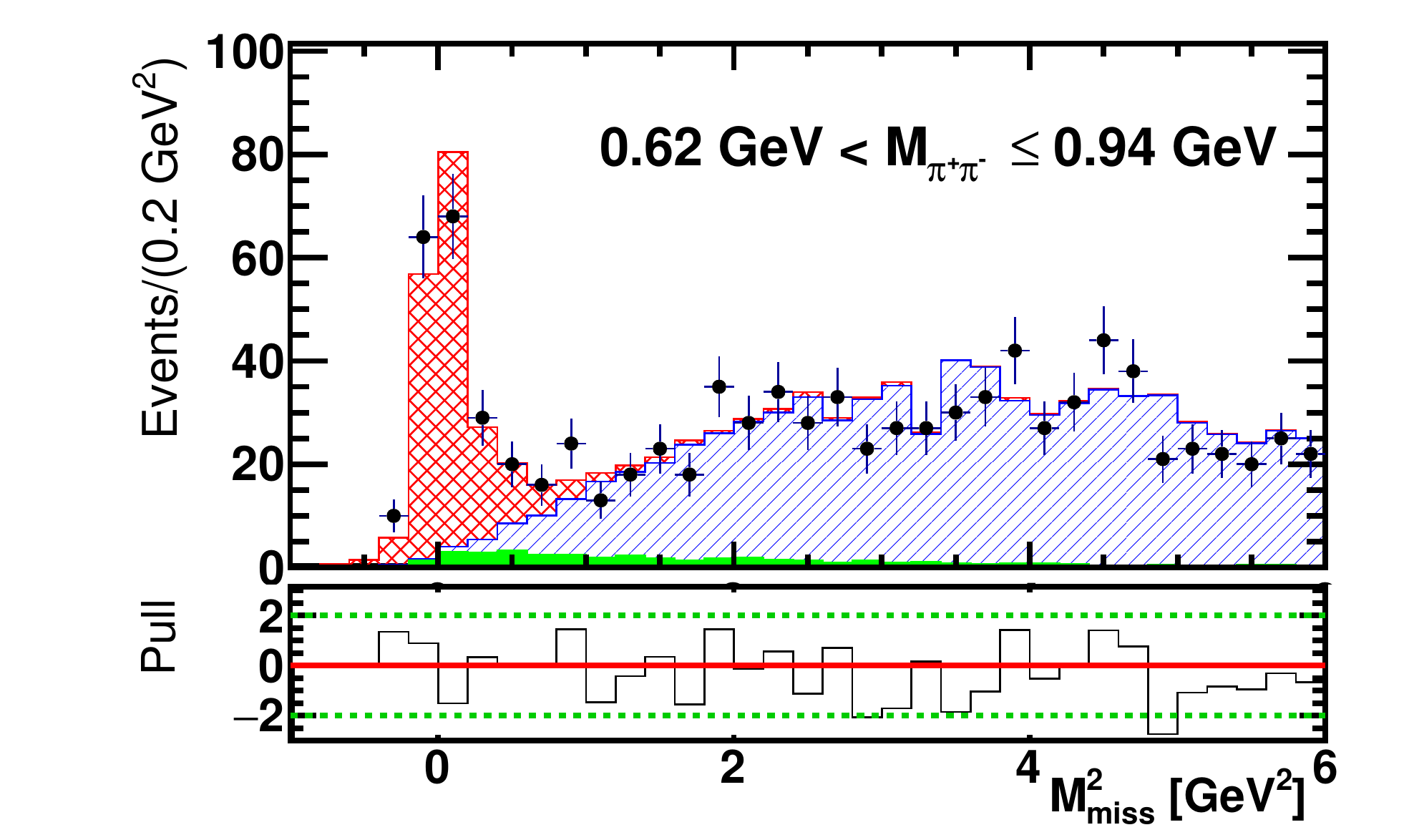}
}\\
\subfloat{
\includegraphics[width=0.5\textwidth ]{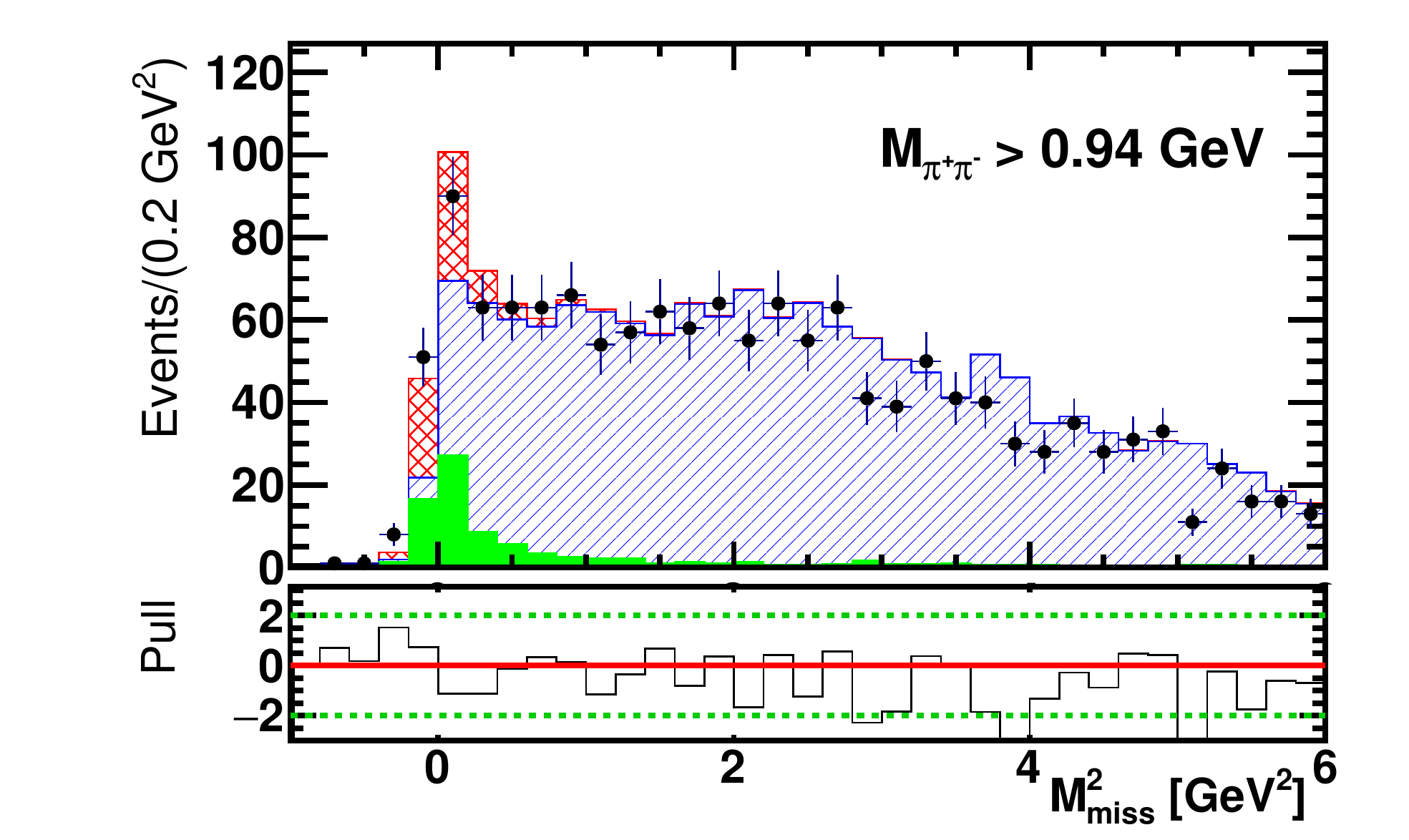}
}
\subfloat{
\includegraphics[width=0.5\textwidth]{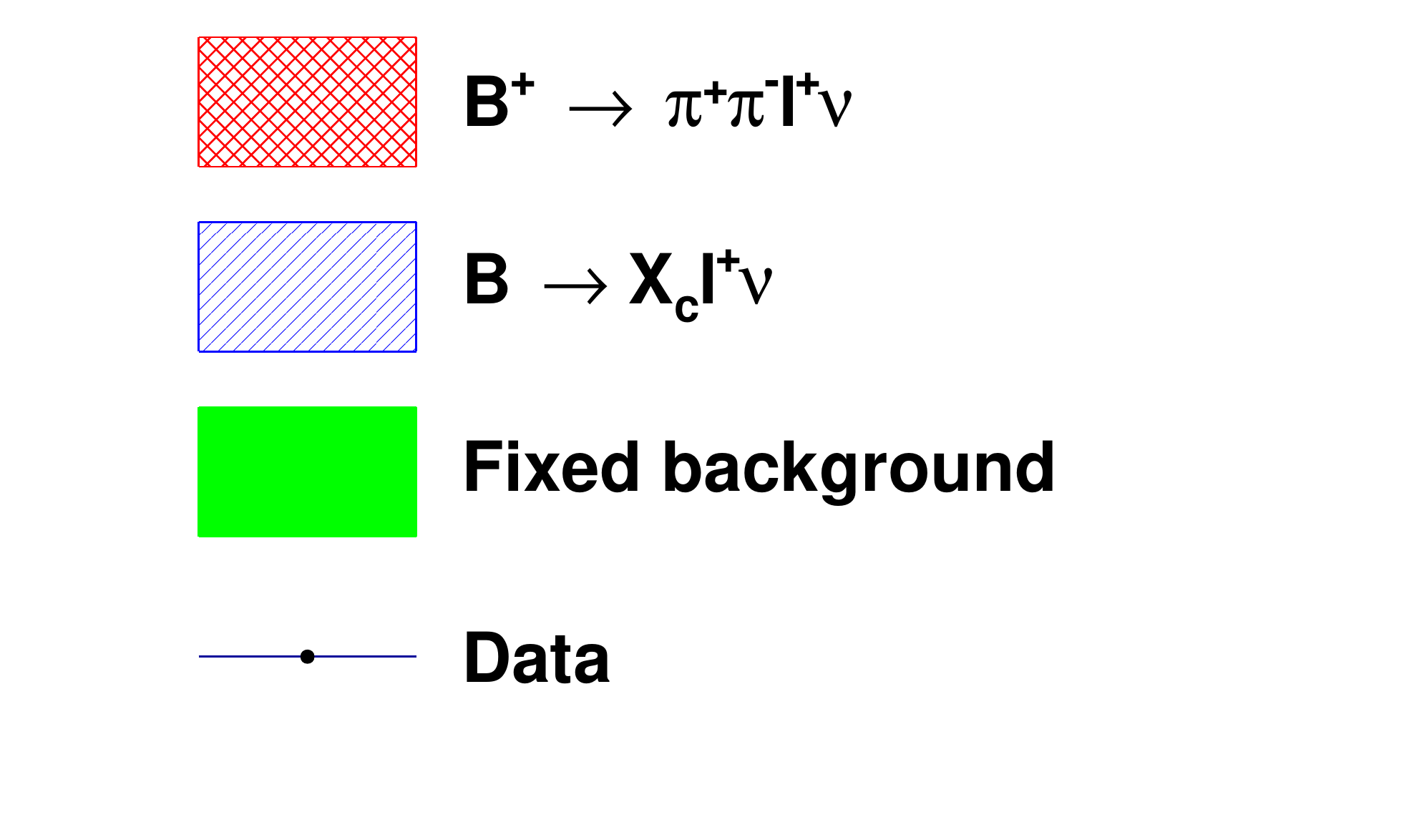}
}
\caption{Projection of the 1D$(M_{\pi\pi})$ configuration fit results in the $M^2_\mathrm{miss}$ distribution (points with error bars) in three regions of the dipion mass as labeled: (upper left) low-mass region ($M_{\pi^{+}\pi^{-}}\leq 0.62 \; \mathrm{GeV}$), (upper right) around the $\rho^{0}$ meson ($ 0.62\;  \mathrm{GeV}<M_{\pi^{+}\pi^{-}}  \leq 0.94\;  \mathrm{GeV}$) and (lower left) high-mass region ($M_{\pi^{+}\pi^{-}}> 0.94 \; \mathrm{GeV}$). The fit components are shown as the colored histograms as given in the lower right. The pull values are presented underneath each plot to display the accuracy of the fit relative to the data. The peaking structure in the fixed background around the signal region in the high dipion mass range is due to the $B^{+}\to \bar{D}^{0}(\pi^{+}\pi^{-})\ell^{+}\nu_{\ell}$ decay.}
\label{fig:fitmmiss2}
\end{figure*} 

To allow for a  $B^{+}\to \pi^{+}\pi^{-}\ell^{+}\nu_{\ell}$ decay-model-independent interpretation of the result, we analyze the measured yields in bins of $M_{\pi\pi}=\sqrt{(P_{\pi^{+}}+P_{\pi^{-}})^{2}}$ and $q^{2}=(P_{\ell}+P_{\nu_{\ell}})^{2}$ using three fit configurations. The minimum value for $M_{\pi\pi}$ corresponds to twice the mass of a charged pion, that is $0.28 \; \mathrm{GeV}$, whereas the maximum value is about the mass of the $B^{\pm}$ meson, which is approximately $5.28 \; \mathrm{GeV}$. On the other hand, $q^{2}$ ranges from $0 \; \mathrm{GeV}^{2}$ to approximately $25 \; \mathrm{GeV}^{2}$. The first configuration employs a fit of the dipion invariant-mass spectrum, referred to as 1D$(M_{\pi\pi})$ in the following. In the second configuration, abbreviated as 2D, we carry out a two-dimensional analysis and measure partial branching fractions in bins of $M_{\pi\pi}$ and $q^{2}$. Finally, in the third configuration we perform the measurement in bins of $q^{2}$, and denote this configuration as 1D$(q^{2})$. We use 13 bins in the 1D$(M_{\pi\pi})$ configuration, consisting of 11 bins with a uniform width in the dipion mass of 80 MeV, and two additional bins corresponding to the low dipion mass ($M_{\pi^{+}\pi^{-}}<0.46 \; \mathrm{GeV}$) and the high dipion mass ($M_{\pi^{+}\pi^{-}}>1.34 \; \mathrm{GeV}$) regions. In the 1D$(q^{2})$ configuration, we employ 17 bins with a uniform width of $1$ GeV$^{2}$ and an additional bin accounting for the region $q^{2}>17$ GeV$^{2}$. In the 2D configuration, we employ five bins of 300 MeV in the dipion mass and, depending on the size of the data sample for these regions, we split the $q^{2}$ spectrum into either two or three bins. Hence, for $M_{\pi^{+}\pi^{-}}<0.6 \; \mathrm{GeV}$ we use  $q^{2}\leq  \; 8\;\mathrm{GeV}^{2}$ and $q^{2}>8\; \mathrm{GeV}^{2}$; for $M_{\pi^{+}\pi^{-}}>1.5 \; \mathrm{GeV}$ we use  $q^{2}\leq  \; 4\;\mathrm{GeV}^{2}$ and $q^{2}>4\; \mathrm{GeV}^{2}$. For the remaining $M_{\pi^{+}\pi^{-}}$ bins, we separate $q^{2}$ into three regions: $q^{2}\leq  \; 4\;\mathrm{GeV}^{2}$,  $4<q^{2} \; \lbrack \mathrm{GeV}^{2}\rbrack \leq 8$, and $q^{2}>8\; \mathrm{GeV}^{2}$. For the highest bin in the 1D$(M_{\pi\pi})$ configuration ($M_{\pi^{+}\pi^{-}}>1.34 \; \text{GeV}$), we separate the $B\to X_{c}\ell\nu$ background into two components: one containing $B$ meson decays to $D^{0}$ mesons as a cross-feed ($B\to D^{0} \ell \nu$), and another involving the remaining charmed mesons (rest of $B\to X_{c}\ell\nu$). The decay $B^{+}\to \bar{D}^{0}\ell^{+}\nu_{\ell}$ with $D^{0}\to\pi^{+}\pi^{-}$ also peaks at $M_{\mathrm{miss}}^{2}\approx 0$ GeV$^{2}$ in the dipion mass ($M_{\pi^{+}\pi^{-}}$) region from 1.85 GeV to 1.88 GeV, with relatively small contamination from other processes. In this mass window, we measure ${\cal B}(B^{+}\to \bar{D}^{0}\ell^{+}\nu_{\ell})=(2.83\pm 0.54)\%$, where the uncertainty is only statistical, and the result is compatible with the world average  ${\cal B}(B^{+}\to \bar{D}^{0}\ell^{+}\nu_{\ell})_{\textrm{PDG}}=(2.33\pm 0.10)\%$~\cite{PDG}. We fix this component in MC according to the measured event yield in data and add it to the fixed background shape and yield. The detector resolution for the dipion mass and $q^{2}$ are about 4 MeV and $5\times 10^{-2}$ $\mathrm{GeV}^{2}$, respectively. These values are significantly smaller than the bin sizes used in our measurement, and hence no additional corrections to account for migrations between the reconstructed bins are applied.   

Figure~\ref{fig:fitmmiss2} shows the projection of the fit results in the 1D$(M_{\pi\pi})$ configuration in three regions of the dipion mass: a low-mass region ($M_{\pi^{+}\pi^{-}}\leq 0.62 \; \mathrm{GeV}$), an intermediate-mass region dominated by the $\rho^{0}$ meson (${ 0.62<M_{\pi^{+}\pi^{-}} \; \lbrack \mathrm{GeV} \rbrack \leq 0.94}$), and a high-mass region (${M_{\pi^{+}\pi^{-}}> 0.94 \; \mathrm{GeV}}$) where we can also observe contributions from the $B^{+}\to \bar{D}^{0}(\pi^{+}\pi^{-})\ell^{+}\nu_{\ell}$ decay. Tables~\ref{tab:fit1D},~\ref{tab:fit1Dq2}, and~\ref{tab:fit2D} present the fit results for every bin in the three configurations. In these tables, we provide the $\chi^{2}$ value and number of degrees of freedom to verify the goodness of fit, following the $\chi^{2}$ calculation of Baker and Cousins~\cite{BakerCousins}, which applies to fits derived from a maximum-likelihood method where the data obey Poisson statistics. The fit procedure was validated by generating an ensemble of pseudoexperiments using the fitted number of signal and background events in each of the bins. No bias in the coverage of the reported uncertainties was observed. The recovered central values show a small bias, which we include into the systematic uncertainties (discussed in the next section). To validate our measurement, we used control samples following a selection procedure similar to that implemented for the signal. For that purpose, we study four channels in the $B^{+}\to \bar{D}^{0}\ell^{+}\nu_{\ell}$ decay, with the $D^{0}$ meson reconstructed as a combination of two charged hadrons and the possibility to include a neutral pion: $K^{-}\pi^{+}$, $K^{-}K^{+}$, $\pi^{+}\pi^{-}\pi^{0}$ and $K^{-}\pi^{+}\pi^{0}$. The measured branching fractions are in agreement with the world averages~\cite{PDG}.

\begin{table*}[tpb]
	\caption{Event yields for the signal and background processes in the $B^+ \to \pi^{+} \pi^{-}  \ell^+ \nu$ decay obtained from an extended binned maximum-likelihood fit to the $M_{\mathrm{miss}}^{2}$ distribution in bins of $M_{\pi^{+} \pi^{-}}$. The $\chi^{2}$, the number of degrees of freedom (NDF) and the probability of the fit (Prob.) are provided. The $\chi^{2}$ calculation is based on the Baker-Cousins method~\cite{BakerCousins}.}
	\label{tab:fit1D}
	\begin{center}
		\small{
			\begin{tabular}{c c c D{,}{\pm}{-1} D{,}{\pm}{-1} c c c c c}
				\toprule
				Bin &  $M_{\pi\pi}$ [GeV] & \multicolumn{1}{c}{Signal}  &  \multicolumn{2}{c}{\makecell[c]{ $B^{+} \to X_{c} \ell \nu$}} & \makecell[c]{Fixed\\ background} & \makecell[c]{Total\\ MC} & Data & $\chi^{2}/\mathrm{NDF}$ & Prob.  \\
				\midrule
				1 & $\phantom{0.46<}M_{\pi\pi}<0.46$  & 7.1$^{+4.1}_{-3.2}$ & \multicolumn{2}{D{,}{\pm}{-1}}{195.0 , 14.6} & 20.2 & 222.3 & 225 & 27.5$/33$ & 73.7  \\
				2 & $0.46\leq M_{\pi\pi}<0.54$            & 10.0$^{+4.4}_{-3.5}$ & \multicolumn{2}{D{,}{\pm}{-1}}{146.7 , 12.7} & 17.1 & 173.8 & 179 & 30.2$/33$ & 60.7  \\
				3 & $0.54\leq M_{\pi\pi}<0.62$            & 10.6$^{+4.3}_{-3.5}$ & \multicolumn{2}{D{,}{\pm}{-1}}{190.1 , 14.2} & 14.8 & 215.5 & 216 & 38.3$/33$ & 24.3  \\
				4 & $0.62\leq M_{\pi\pi}<0.70$            & 23.3$^{+6.2}_{-5.4}$ & \multicolumn{2}{D{,}{\pm}{-1}}{185.4 , 14.1} & 9.3 & 218.0 & 220 & 27.9$/33$ & 71.7  \\
				5 & $0.70\leq M_{\pi\pi}<0.78$            & 90.3$^{+10.7}_{-10.0}$ & \multicolumn{2}{D{,}{\pm}{-1}}{234.8 , 16.0} & 12.4 & 337.5 & 337 & 45.9$/33$ & 6.8  \\
				6 & $0.78\leq M_{\pi\pi}<0.86$            & 50.5$^{+8.1}_{-7.4}$ & \multicolumn{2}{D{,}{\pm}{-1}}{151.6 , 12.8} & 12.3 & 214.4 & 214 & 30.2$/33$ & 60.8  \\
				7 & $0.86\leq M_{\pi\pi}<0.94$            & 29.6$^{+6.4}_{-5.7}$ & \multicolumn{2}{D{,}{\pm}{-1}}{108.5 , 10.8} & 7.8 & 145.9 & 146 & 43.6$/33$ & 10.4  \\
				8 & $0.94\leq M_{\pi\pi}<1.02$            & 10.2$^{+4.2}_{-3.4}$ & \multicolumn{2}{D{,}{\pm}{-1}}{102.4 , 10.3} & 6.1 & 118.7 & 119 & 15.2$/33$ & 99.7  \\
				9 & $1.02\leq M_{\pi\pi}<1.10$            & 8.9$^{+3.7}_{-3.0}$ & \multicolumn{2}{D{,}{\pm}{-1}}{127.6 , 11.3} & 4.0 & 140.5 & 140 & 26.3$/33$ & 78.9  \\
				10 & $1.10\leq M_{\pi\pi}<1.18$           & 5.7$^{+3.1}_{-2.4}$ & \multicolumn{2}{D{,}{\pm}{-1}}{149.2 , 12.1} & 2.9 & 157.8 & 158 & 40.0$/33$ & 18.8  \\
				11 & $1.18\leq M_{\pi\pi}<1.26$           & 15.7$^{+5.0}_{-4.2}$ & \multicolumn{2}{D{,}{\pm}{-1}}{186.6 , 13.8} & 3.0 & 205.3 & 205 & 41.2$/33$ & 15.5  \\
				12 & $1.26\leq M_{\pi\pi}<1.34$           & 11.8$^{+4.2}_{-3.4}$ & \multicolumn{2}{D{,}{\pm}{-1}}{221.4 , 14.9} & 3.1 & 236.3 & 236 & 30.5$/33$ & 59.3  \\
				\midrule
				&&&\multicolumn{1}{c}{$B^{+} \to \bar{D}^{0} \ell \nu$} &\multicolumn{1}{c}{\makecell[c]{Rest of \\ $B^{+} \to X_{c} \ell \nu$}} &  & & & &   \\
				\midrule
				13 & $\phantom{1.34<}M_{\pi\pi}\geq 1.34$ & 23.4$^{+14.7}_{-13.6}$ & 175.3 , 49.4 & 289.3 , 44.0 & 68.0 & 556.0 & 556 & 28.3$/32$ & 65.4  \\
				\bottomrule
		\end{tabular}}
	\end{center}
\end{table*}

\begin{table*}[tpb]
	\caption{Event yields for the signal and background processes in the $B^+ \to \pi^{+} \pi^{-}  \ell^+ \nu$ decay obtained from an extended binned maximum-likelihood fit to the $M_{\mathrm{miss}}^{2}$ distribution in bins of $q^{2}$. The $\chi^{2}$, the number of degrees of freedom (NDF) and the probability of the fit (Prob.) are provided. The $\chi^{2}$ calculation is based on the Baker-Cousins method~\cite{BakerCousins}.}
	\label{tab:fit1Dq2}
	\begin{center}
		\small{
			\begin{tabular}{c c c D{,}{\pm}{-1} c c c c c}
				\toprule
				Bin &  $q^{2}$ [GeV$^{2}$] & \multicolumn{1}{c}{Signal}  &  \multicolumn{1}{c}{\makecell[c]{ $B^{+} \to X_{c} \ell \nu$}} & \makecell[c]{Fixed\\ background} & \makecell[c]{Total\\ MC} & Data & $\chi^{2}/\mathrm{NDF}$ & Prob.  \\
				\midrule
				1 & $\phantom{1<}q^{2}<1$        & 16.5$^{+6.8}_{-6.1}$ & 126.4 , 12.1 & 20.6 & 163.5 & 163 & 32.1$/33.0$ & 51.2  \\
				2 & $1\leq q^{2}<2$              & 11.4$^{+6.0}_{-5.1}$ & 150.5 , 12.9 & 15.2 & 177.1 & 176 & 34.9$/33.0$ & 37.9  \\
				3 & $2\leq q^{2}<3$              & 13.0$^{+5.6}_{-4.7}$ & 166.3 , 13.4 & 12.8 & 192.1 & 192 & 40.9$/33.0$ & 16.4  \\
				4 & $3\leq q^{2}<4$              & 16.0$^{+6.6}_{-5.8}$ & 180.2 , 14.1 & 13.5 & 209.7 & 210 & 32.2$/33.0$ & 50.9  \\
				5 & $4\leq q^{2}<5$              & 24.3$^{+7.1}_{-6.3}$ & 224.9 , 15.6 & 13.8 & 263.0 & 263 & 41.9$/33.0$ & 13.7  \\
				6 & $5\leq q^{2}<6$              & 12.2$^{+5.7}_{-4.9}$ & 212.3 , 15.0 & 14.0 & 238.5 & 238 & 17.4$/33.0$ & 98.8  \\
				7 & $6\leq q^{2}<7$              & 10.8$^{+5.1}_{-4.1}$ & 235.6 , 15.7 & 10.8 & 257.2 & 257 & 54.2$/33.0$ & 1.1  \\
				8 & $7\leq q^{2}<8$              & 21.4$^{+6.5}_{-5.7}$ & 220.5 , 15.3 & 10.5 & 252.4 & 253 & 36.0$/33.0$ & 32.9  \\
				9 & $8\leq q^{2}<9$              & 9.6$^{+4.7}_{-3.9}$ & 220.5 , 15.1 & 9.5 & 239.6 & 239 & 34.1$/33.0$ & 41.6  \\
				10 & $9\leq q^{2}<10$            & 30.8$^{+6.7}_{-6.0}$ & 199.0 , 14.6 & 9.3 & 239.1 & 239 & 36.5$/33.0$ & 30.9  \\
				11 & $10\leq q^{2}<11$           & 11.6$^{+5.0}_{-4.1}$ & 159.4 , 13.0 & 9.2 & 180.2 & 181 & 19.2$/33.0$ & 97.3  \\
				12 & $11\leq q^{2}<12$           & 16.3$^{+4.9}_{-4.1}$ & 122.1 , 11.4 & 7.2 & 145.6 & 146 & 35.4$/33.0$ & 35.4  \\
				13 & $12\leq q^{2}<13$           & 19.4$^{+5.3}_{-4.5}$ & 93.7 , 10.0 & 6.1 & 119.2 & 119 & 16.5$/33.0$ & 99.3  \\
				14 & $13\leq q^{2}<14$           & 15.4$^{+4.7}_{-4.0}$ & 66.1 , 8.5 & 5.7 &  87.2 & 87 & 21.5$/33.0$ & 93.8  \\
				15 & $14\leq q^{2}<15$           & 15.1$^{+4.9}_{-4.1}$ & 37.1 , 6.6 & 5.9 & 58.1 & 61 & 24.3$/28.0$ & 66.3  \\
				16 & $15\leq q^{2}<16$           & 10.8$^{+4.0}_{-3.2}$ & 24.1 , 5.3 & 4.8 & 39.7 & 41 & 17.4$/23.0$ & 79.1  \\
				17 & $16\leq q^{2}<17$           & 12.3$^{+4.4}_{-3.7}$ & 18.5 , 5.0 & 4.8 & 35.6 & 36 & 13.0$/23.0$ & 95.2  \\
				18 & $\phantom{<15}q^{2}\geq 17$ & 32.3$^{+6.8}_{-6.1}$ & 7.8 , 4.1 & 7.4 & 47.5 & 50 & 14.5$/18.0$ & 69.4  \\
				\bottomrule
		\end{tabular}}
	\end{center}
\end{table*}

\begin{table*}[tb]
\caption{Event yields for the signal and background processes in the $B^+ \to \pi^{+} \pi^{-}  \ell^+ \nu$ decay obtained from an extended binned maximum-likelihood fit to the $M_{\mathrm{miss}}^{2}$ distribution in bins of $M_{\pi^{+} \pi^{-}}$ and $q^{2}$. The $\chi^{2}$, the number of degrees of freedom (NDF) and and the probability of the fit (Prob.) are provided. The $\chi^{2}$ calculation is based on the Baker-Cousins method~\cite{BakerCousins}.}
\label{tab:fit2D}
\begin{center}
\small{
\begin{tabular}{c c c c D{,}{\pm}{-1} D{,}{\pm}{-1} c c c c c}
\toprule
 Bin & $M_{\pi\pi}$ [GeV] & $q^{2}$ [GeV$^{2}$] & \multicolumn{1}{c}{Signal}  &  \multicolumn{2}{c}{\makecell[c]{$B^{+} \to X_{c} \ell \nu$}} & \makecell[c]{Fixed\\ background}  & \makecell[c]{Total\\ MC} & Data & $\chi^{2}/\mathrm{NDF}$ & Prob.[\%]  \\
\midrule
1 & $\phantom{0.6<}M_{\pi\pi}\leq 0.6$   &   $\phantom{4<}q^{2}\leq 8$  &  9.8$^{+4.6}_{-3.7}$  &  \multicolumn{2}{D{,}{\pm}{-1}}{218.2  ,  15.3}  &  15.6  &  243.5  & 249 & 50.1$/33$ & 2.8  \\
2 & $\phantom{0.6<}M_{\pi\pi}\leq 0.6$   &   $8<q^{2}\phantom{\leq 8}$  &  15.8$^{+5.5}_{-4.6}$  &  \multicolumn{2}{D{,}{\pm}{-1}}{275.9  ,  17.4}  &  33.1  &  324.8 &  329 & 30.6$/33$ & 58.5  \\
3 & $0.6<M_{\pi\pi}\leq 0.9$   &   $\phantom{4<}q^{2}\leq 4$  &  29.5$^{+6.4}_{-5.7}$  &  \multicolumn{2}{D{,}{\pm}{-1}}{134.9  ,  12.1}  &  9.8  &  174.2  &  175 & 31.4$/33$ & 54.7  \\
4 & $0.6<M_{\pi\pi}\leq 0.9$   &   $4<q^{2}\leq 8$  &  34.8$^{+7.0}_{-6.2}$  &  \multicolumn{2}{D{,}{\pm}{-1}}{216.9  ,  15.1}  &  9.9  &  261.5  &  262 & 24.4$/33$ & 86.0  \\
5 & $0.6<M_{\pi\pi}\leq 0.9$   &   $8<q^{2}\phantom{\leq 8}$  &  116.2$^{+12.2}_{-11.5}$  &  \multicolumn{2}{D{,}{\pm}{-1}}{318.8  ,  18.7}  &  22.5  &  457.6  &  457 & 39.0$/33$ & 21.8  \\
6 & $0.9<M_{\pi\pi}\leq 1.2$   &   $\phantom{4<}q^{2}\leq 4$  &  8.0$^{+3.7}_{-2.9}$  &  \multicolumn{2}{D{,}{\pm}{-1}}{110.8  ,  10.6}  &  5.8  &  124.6  &  124 & 20.4$/33$ & 95.8  \\
7 & $0.9<M_{\pi\pi}\leq 1.2$   &   $4<q^{2}\leq 8$  &  9.2$^{+4.0}_{-3.2}$  &  \multicolumn{2}{D{,}{\pm}{-1}}{190.2  ,  13.9}  &  4.9  &  204.3  &  204 & 32.3$/33$ & 50.4  \\
8 & $0.9<M_{\pi\pi}\leq 1.2$   &   $8<q^{2}\phantom{\leq 8}$  &  27.6$^{+6.4}_{-5.6}$  &  \multicolumn{2}{D{,}{\pm}{-1}}{169.8  ,  13.4}  &  6.6  &  204.1  &  204 & 39.5$/33$ & 20.3  \\
9 & $1.2<M_{\pi\pi}\leq 1.5$   &   $\phantom{4<}q^{2}\leq 4$  &  11.3$^{+4.3}_{-3.5}$  &  \multicolumn{2}{D{,}{\pm}{-1}}{142.4  ,  12.1}  &  4.1  &  157.8  &  158 & 36.7$/33$ & 30.0  \\
10 & $1.2<M_{\pi\pi}\leq 1.5$   &   $4<q^{2}\leq 8$  &  9.7$^{+4.3}_{-3.5}$  &  \multicolumn{2}{D{,}{\pm}{-1}}{227.1  ,  15.1}  &  2.5  &  239.2  &  239 & 25.6$/33\
$ & 81.8  \\
11 & $1.2<M_{\pi\pi}\leq 1.5$   &   $8<q^{2}\phantom{\leq 8}$  &  13.2$^{+4.4}_{-3.7}$  &  \multicolumn{2}{D{,}{\pm}{-1}}{132.5  ,  11.6}  &  2.4  &  148.1  &  148 & 27.2$/33$ & 75.0  \\
\midrule
 &&&&\multicolumn{1}{c}{$B^{+} \to D^{0} \ell \nu$}&\multicolumn{1}{c}{\makecell[c]{Rest of \\ $B^{+} \to X_{c} \ell \nu$}} & & & & &   \\
\midrule 
  12 & $\phantom{1.2<}M_{\pi\pi}>1.5$   &   $\phantom{4<}q^{2}\leq 4$  &  8.5$^{+10.1}_{-9.0}$  &  72.1  ,  17.3  &  63.1  ,  13.2  &  36.2  &  179.8  &  180 & 25.4$/32$ & 79.1  \\
13 & $\phantom{1.2<}M_{\pi\pi}>1.5$   &   $4<q^{2}\phantom{\leq 8}$  &  7.6$^{+7.2}_{-4.9}$  &  96.4  ,  22.4  &  93.2  ,  20.2  &  27.8  &  225.1  &  222 & 27.8$/32$ & 68.1  \\
\bottomrule
\end{tabular}
}
\end{center}
\end{table*}
%

\section{Systematic uncertainties}
\label{sec:systematics}
The sources of systematic uncertainties considered in this analysis fall into three categories: those related to detector performance, those due to the modeling of the signal and background processes, and uncertainties associated with the fitting procedure. In most cases, we estimate the systematic uncertainties by varying each fixed parameter in the simulation by one standard deviation up and down ($\pm 1\sigma$) and repeating the fit to the $M_{\mathrm{miss}} ^{2}$ distribution. We then take the relative difference between the signal yield from the nominal fit and that with the parameter varied as the $\pm 1\sigma$ systematic uncertainty. We calculate these uncertainties separately for each bin in our measurement. 

\subsection{Signal and background modeling}

The sources of uncertainties related to the modeling of physical processes include the lack of precise knowledge of hadronic form factors that describe a specific decay, and the relative contributions of background processes. To assess the systematic uncertainty arising from the signal modeling, we compare the signal reconstruction efficiency calculated for each bin in $M_{\pi\pi}$, $q^{2}$, or ($M_{\pi\pi},q^{2}$), using the phase space $B^{+}\to \pi^{+}\pi^{-}\ell^{+}\nu_{\ell}$ and other $B$ semileptonic channels with an intermediate resonance decaying to a $\pi^{+}\pi^{-}$ pair. As these channels simulate the same final state, the resulting efficiencies should be similar. Nonetheless, resonances do not span as much of the domain in the phase space as an inclusive simulation since they have a finite width; hence their coverage in the dipion mass is essentially limited to the interval $[M_{\mathrm R}-2\Gamma_{\mathrm R}, M_{\mathrm R}+2\Gamma_{\mathrm R}]$, with $M_{\mathrm R}$ the nominal mass of the resonance and $\Gamma_{\mathrm R}$ its decay width. The range of $q^{2}$ varies with the resonant state as the maximum value depends on the mass of the resonance through $q^{2}_{\mathrm max}=(M_{B}-M_{\mathrm R})^{2}$, where $M_{B}$ is the mass of the $B$ meson. We thus simulate semileptonic $B$ decays with four intermediate resonances covering the phase space of the $B^{+}\to \pi^{+}\pi^{-}\ell^{+}\nu_{\ell}$ decay, namely $f_{0}(500)$, $\rho^{0}$, $f_{2}(1270)$, and $\rho^{0}(1450)$, and produce these with the phase space and ISGW2~\cite{ISGW2} models. Furthermore, we  use form factors from light-cone sum rule (LCSR) calculations for the $B^{+}\to \rho^{0}\ell^{+}\nu_{\ell}$  and the $B^{+}\to f_{2}(1270)\ell^{+}\nu_{\ell}$ decays according to references~\cite{Ball98} and~\cite{f2ffLCSR}, respectively. We calculate the root mean square error between the nominal efficiency (phase space $B^{+}\to \pi^{+}\pi^{-}\ell^{+}\nu_{\ell}$) and the resonant models valid for a given bin as the systematic uncertainty due to signal modeling. In addition, we also consider the finite size of the sample used to estimate the signal reconstruction efficiency. We include this (statistics-based) error in the systematic uncertainty due to reconstruction efficiency. The values of the efficiencies used for this assessment are presented in the appendix in Tables~\ref{tab:SigModEff1D},~\ref{tab:SigModEff1Dq2}, and~\ref{tab:SigModEff2D} for the 1D$(M_{\pi\pi})$, 1D$(q^{2})$, and 2D fit binning configurations, respectively.

Given that the continuum background is almost negligible after the selection, we compare the continuum MC with the off-resonance data using a loose selection to assign the uncertainty due to the description of this process. Consequently, we determine an asymmetric variation in the continuum normalization ($^{+50\%}_{-20\%}$)  and repeat the fit with these changes. Contributions from rare decays are also very small. To evaluate their effect on our measurement, we carried out 1000 pseudoexperiments (using the same prescription described in the previous section) with and without this component. The systematic uncertainty is then derived from the difference in mean values from both ensembles for each bin. To assess the impact of the background shape on the calculation of the branching fraction, we reweight a specific decay in the MC with another model. Specifically, we adjust the CLN-based form factors~\cite{CLN} of the $B\to D^{*}\ell\nu_{\ell}$ decays in the MC to the new world-average values~\cite{HFLAV}. Similarly, we reweight the form factors for the $B\to D^{**}\ell\nu_{\ell}$ decays from the ISGW2~\cite{ISGW2} to the LLSW model~\cite{LLSW}. In both cases, we add in quadrature the change in the branching fraction due to variation of each form factor to obtain a total uncertainty associated with these sources. The $B\to\pi\ell\nu_{\ell}$ and $B\to\omega\ell\nu_{\ell}$ were generated in the MC with LCSR form factors taken from reference~\cite{Ball98}. We reweight the $B\to\omega\ell\nu_{\ell}$ form factors to the calculation of ~\cite{Ball05} and use the difference in efficiencies compared to the nominal sample as the uncertainty. The $B\to\pi\ell\nu_{\ell}$ form factors are reweighted to the Bourrely-Caprini-Lellouch model~\cite{BCL}, which combines information from the measured spectra, light-cone sum rules (valid at low $q^{2}$) and lattice QCD (valid at high $q^{2}$), and the same procedure to calculate the uncertainty is used. We also reweight the form factors of the $B\to\eta^{(\prime)}\ell\nu_{\ell}$ decay from the ISGW2~\cite{ISGW2} and LCSR models according to~\cite{Ball07}. Other exclusive charmless semileptonic $B$ decays considered in this analysis were generated with the ISGW2 model. As they do not have well-established form factors derived from QCD calculations, we compare their shapes with those produced using the phase space and FLATQ2 generators~\cite{EVTGEN,Cote}. 

We correct the branching fractions of the ${B\to (D^{(*,**)},\pi,\eta^{(\prime)},\omega)\ell\nu_{\ell}}$ decay modes according to the world-averages~\cite{PDG} and vary these values within their measured uncertainties as presented in Table~\ref{tab:BRcorChannels}. For the unmeasured exclusive charmless semileptonic $B$ decays, we assign a $\pm100\%$ uncertainty in the variation of the branching fraction. We modify the contribution of the secondary leptons relative to the total uncertainty in the measurement of the branching fraction of the decay chain $B^{+}\to X_{\bar{c}}\ell^{+}\nu_{\ell}$ with $X_{\bar{c}}\to \ell^{-}+\mathrm{anything}$. To consider the effect of the BDT selection on our result, we evaluate the difference in efficiency in data and MC and find it to be negligible as compared to the statistical error.  

To assess the effect of inclusive charmless semileptonic $B$ decays we include an additional component to the fixed background simulated with the De Fazio-Neubert model~\cite{DeFazioNeubert}. The differences respect to the nominal fit are taken as a systematic uncertainty. 

\begin{table*}[tb]
\caption{Decay channels that were corrected in the MC, with their branching fractions in the MC, world-average, and their respective correction (weight). }
\label{tab:BRcorChannels}
\begin{center}
\bgroup
\begin{tabular}{l l l l l}
\toprule
Decay mode                                                                   & MC                   &  World-average ${\cal B}$~\cite{PDG,HFLAV}                                 & Weight  \\
\midrule                                   
$B^{-}\to D^{0}\ell^{-}\bar{\nu}_{\ell}$                                     & $2.31\times 10^{-2}$ & $(2.33\pm 0.10)\times 10^{-2}$             &1.01   \\
$B^{-}\to D^{*0}\ell^{-}\bar{\nu}_{\ell}$                                    & $5.79\times 10^{-2}$ & $(5.59\pm 0.19)\times 10^{-2}$             &0.97   \\
$B^{-}\to D_{1}^{0}\ell^{-}\bar{\nu}_{\ell}$, $D_{1}^{0}\to D^{*+}\pi^{-}$   & $5.40\times 10^{-3}$  & $(2.8\pm 0.1\pm 1.5)\times 10^{-3}$        &0.52   \\
$B^{-}\to D_{2}^{*0}\ell^{-}\bar{\nu}_{\ell}$, $D_{2}^{*0}\to D^{*+}\pi^{-}$ & $8.20\times 10^{-4}$  & $(7.7\pm 0.6\pm 0.4)\times 10^{-4}$        &0.94   \\
$B^{-}\to D_{1}^{'0}\ell^{-}\bar{\nu}_{\ell}$, $D_{1}^{'0}\to D^{*+}\pi^{-}$ & $5.40\times 10^{-3}$  & $(1.3\pm 0.3\pm 0.2)\times 10^{-3}$        &0.24   \\
$B^{-}\to D_{0}^{*0}\ell^{-}\bar{\nu}_{\ell}$, $D_{0}^{*0}\to D^{+}\pi^{-}$  & $6.10\times 10^{-3}$  & $(2.8\pm 0.3\pm 0.4)\times 10^{-3}$        &0.46   \\                                   
$\bar{B}^{0}\to D^{0}\ell^{-}\bar{\nu}_{\ell}$                               & $2.13\times 10^{-3}$ & $(2.20\pm 0.10)\times 10^{-2}$             &1.03   \\
$\bar{B}^{0}\to D^{*+}\ell^{-}\bar{\nu}_{\ell}$                              & $5.33\times 10^{-3}$ & $(4.88\pm 0.10)\times 10^{-2}$             &0.92   \\                                            
$B^{-}\to \pi^{0}\ell^{-}\bar{\nu}_{\ell}$                                   & $7.80\times 10^{-5}$  & $(7.80\pm 0.27)\times 10^{-5}$             &1.07   \\                                  
$B^{-}\to \rho^{0}\ell^{-}\bar{\nu}_{\ell}$                                  & $1.49\times 10^{-4}$ & $(1.58\pm 0.11)\times 10^{-4}$             &1.06   \\                                  
$B^{-}\to \omega\ell^{-}\bar{\nu}_{\ell}$                                    & $1.15\times 10^{-4}$ & $(1.19\pm 0.09)\times 10^{-4}$             &1.04   \\                                  
$B^{-}\to \eta\ell^{-}\bar{\nu}_{\ell}$                                      & $8.40\times 10^{-5}$  & $(3.8\pm 0.6)\times 10^{-5}$               &0.45   \\                                  
$B^{-}\to \eta^{\prime}\ell^{-}\bar{\nu}_{\ell}$                             & $3.30\times 10^{-5}$  & $(2.3\pm 0.8)\times 10^{-5}$               &0.70   \\                                  
$\bar{B}^{0}\to \pi^{+}\ell^{-}\bar{\nu}_{\ell}$                             & $1.36\times 10^{-4}$ & $(1.45\pm 0.05)\times 10^{-4}$             &1.07   \\                                  
$\bar{B}^{0}\to \rho^{+}\ell^{-}\bar{\nu}_{\ell}$                            & $2.77\times 10^{-4}$ & $(2.94\pm 0.21)\times 10^{-4}$             &1.06   \\                     
\bottomrule
\end{tabular}
\egroup
\end{center}
\end{table*}

\subsection{Detector simulation}

Since the analysis relies extensively on MC simulation, the detection of final-state particles affects the reconstruction of the signal and background decays and the subsequent extraction of the signal yields used in the measurement of the branching fractions. The efficiency for detecting these particles in data usually differs from that in MC, for which we applied a correction derived from independent control samples. We take the total uncertainty associated with this correction as a systematic uncertainty. These uncertainties include those related to charged-lepton and charged-pion identification efficiencies. Analogously to secondary leptons, charged tracks misidentified as leptons, i.e., fake leptons, can also originate from the continuum and from charmed semileptonic $B$ decays. To inspect their effect, we estimate the fake rate in data and MC using an enriched hadronic sample corresponding to the $D^{*+}\to D^{0}(K^{-}\pi^{+})\pi^{+}$ decay and determine a weight factor for each lepton type. We then correct the contribution of fake leptons in the MC and vary the central value by its error. We assign the relative difference between the fit results as the uncertainties associated with fake leptons.

To assess the size of the FSR photons uncertainty, we prepare histogram templates normalized to the fit results in data using two versions of the signal component: one where the signal was generated with the PHOTOS package (as in the nominal fit) and another without it. We then carry out 1000 pseudoexperiments for each case and take 20\% of the mean difference in the signal yields from these two scenarios~\cite{Richter,BaBarUncFSR1,BaBarUncFSR2}. 

\subsection{Fit procedure}
 
We perform 5000 pseudoexperiments to validate our fit procedure, obtaining pull distributions that takes into account the asymmetric statistical uncertainties. These distributions exhibit Gaussian behavior with a slight deviation from zero in the mean, which in most cases is an effect at the $1\%$ level. We do not correct the signal yields or their uncertainties; instead we assign a systematic uncertainty due to the fit procedure. The size of the systematic uncertainty is estimated by the difference between the mean signal yield of the ensemble of pseudoexperiments and the signal yield used as the central value in the generation of the ensemble. This is evaluated separately for each bin. 

Tables~\ref{tab:syst1D},~\ref{tab:systq2}, and~\ref{tab:syst2D}, in the appendix, list the systematic uncertainties for the 1D$(M_{\pi\pi})$, 1D$(q^{2})$, and 2D configurations, respectively. 

To estimate correlations among the systematic uncertainties of the values of the partial branching fractions for each bin, we consider two scenarios. The first scenario corresponds to uncertainties derived from the variation of one parameter in the MC simulation, such as those involving normalization of a background component or branching fractions of some decay processes. We characterize each component by a Gaussian distribution with a width equal to the systematic uncertainty investigated, draw a random variable for each parameter, and repeat the entire analysis procedure 1000 times. For each systematic uncertainty, we associate a correlation matrix $\mathrm{COR}_{ij}$ calculated as:

\begin{equation}
 \mathrm{COR}_{ij}=\frac{\langle(\Delta{\cal B}^{i}-\overline{\Delta{\cal B}^{i}})(\Delta{\cal B}^{j}-\overline{\Delta{\cal B}^{j}})\rangle}{\sigma_{i}\sigma_{j}}
\end{equation}
where the indices $i,j$ run over the bins in the sample, $\overline{\Delta{\cal B}^{i}}$ is the mean of the randomly generated partial branching fractions for the $i$-th bin, $\sigma_{i}$ is its standard deviation, and $\langle\rangle$ denotes an average over the 1000 iterations. We then compute the associated covariance matrix as

\begin{equation}
\mathrm{COV}_{ij}=\mathrm{COR}_{ij}\sigma_{i}\sigma_{j}.
\label{eq:CovI}
\end{equation}
The second scenario applies to systematic uncertainties assessed under a different procedure, e.g., signal model dependence or final-state radiation, among others. In this case, we evaluate the effect of a particular systematic uncertainty $k$ in the $i$-th bin, $\sigma_{i}^{k}$, on $\Delta{\cal B}^{i}$ and assign it to $\xi_{i}^{k}=\sigma_{i}^{k}\Delta{\cal B}^{i}$ and from this quantity, determine the corresponding covariant matrix as 

\begin{equation}
\mathrm{COV}_{ij}^{k}=\xi_{i}^{k}\xi_{j}^{k}.
\label{eq:CovII}
\end{equation}
We provide the total systematic correlation matrices for the different fit scenarios in the appendix in Tables~\ref{tab:SystCorr1D},~\ref{tab:SystCorr1Dq2}, and~\ref{tab:SystCorr2D}, respectively.

\subsection{Normalization uncertainties}

The uncertainty on the measurement of the number of $B$-meson pairs produced at Belle is 1.4\%, while that on the branching fraction of $\Upsilon(4S)\to B^{+}B^{-}$ is 1.17\%. We assume an 0.35\% uncertainty on the track-finding efficiency for each charged particle reconstructed on the signal side and add each contribution linearly. Finally, we take the uncertainty due to the tagging efficiency correction, which originates from incorrect assumptions of the hadronic branching fractions on the tag side, as 4.2\%~\cite{Sibidanov}. These uncertainties are assumed to be 100\% correlated across all bins and are also included in the correlation matrices of Tables~\ref{tab:SystCorr1D},~\ref{tab:SystCorr1Dq2}, and~\ref{tab:SystCorr2D}, respectively.

\section{Results and discussion}

\begin{table*}[tb]
\caption{Signal yields ($Y_{\mathrm{signal}}^{i}$), signal reconstruction efficiency ($\epsilon^{i}$), and partial branching fractions ($\Delta {\cal B}^{i} $) for each bin $i$ in the 1D($M_{\pi\pi}$) and 2D configurations, respectively, with the bin number convention as defined in Tables~\ref{tab:fit1D} and~\ref{tab:fit2D}, respectively. The first quoted uncertainty is statistical, and the second is systematic.}
\label{tab:BR_results}
\begin{center}
\def\arraystretch{1.4}
\begin{tabular}{c c c c c c c }
\toprule                      
    &    \multicolumn{3}{c}{\makecell[c]{1D($M_{\pi\pi}$) Configuration }}    &    \multicolumn{3}{c}{2D Configuration} \bigstrut \\
Bin    &    \multicolumn{1}{c}{\makecell[c]{$Y_{\mathrm{signal}}^{i}$}}    &    \multicolumn{1}{c}{$\epsilon^{i}\: [10^{-4}]$}&    $\Delta {\cal B}^{i} \: [10^{-5}]$  &    \multicolumn{1}{c}{\makecell[c]{$Y_{\mathrm{signal}}^{i}$}}    &    \multicolumn{1}{c}{$\epsilon^{i}\: [10^{-4}]$}&    $\Delta {\cal B}^{i} \: [10^{-5}]$ \\
\midrule     
1	&$	7.1	^{+	4.1	}_{-	3.2	}$	&$	7.92	\pm	0.66	$&$	0.57	^{+	0.33	}_{-	0.25	}\pm	0.09	$	    &$	9.8	^{+	4.6	}_{-	3.7	}$	&$	7.39	\pm	0.57	$&$	0.84	^{+	0.39	}_{-	0.32	}\pm	0.18	$	\\
2	&$	10.0	^{+	4.4	}_{-	3.5	}$	&$	8.20	\pm	0.77	$&$	0.77	^{+	0.34	}_{-	0.27	}\pm	0.16	$	&$	15.8	^{+	5.5	}_{-	4.6	}$	&$	8.45	\pm	0.63	$&$	1.18	^{+	0.41	}_{-	0.34	}\pm	0.20	$	\\
3	&$	10.6	^{+	4.3	}_{-	3.5	}$	&$	7.75	\pm	0.68	$&$	0.86	^{+	0.35	}_{-	0.28	}\pm	0.24	$	&$	29.5	^{+	6.4	}_{-	5.7	}$	&$	8.61	\pm	0.60	$&$	2.16	^{+	0.47	}_{-	0.42	}\pm	0.23	$	\\
4	&$	23.3	^{+	6.2	}_{-	5.4	}$	&$	7.82	\pm	0.64	$&$	1.88	^{+	0.50	}_{-	0.44	}\pm	0.41	$	&$	34.8	^{+	7.0	}_{-	6.2	}$	&$	8.35	\pm	0.63	$&$	2.63	^{+	0.53	}_{-	0.47	}\pm	0.32	$	\\
5	&$	90.3	^{+	10.7	}_{-	10.0	}$	&$	9.32	\pm	0.66	$&$	6.11	^{+	0.72	}_{-	0.68}\pm	1.12$	&$	116.2	^{+	12.2	}_{-	11.5	}$	&$	7.98	\pm	0.48	$&$	9.18	^{+	0.96	}_{-	0.91	}\pm	1.02	$	\\
6	&$	50.5	^{+	8.1	}_{-	7.4	}$	&$	7.76	\pm	0.58	$&$	4.10	^{+	0.70	}_{-	0.66	}\pm	0.74	$	&$	8.0	^{+	3.7	}_{-	2.9	}$	&$	7.20	\pm	0.46	$&$	0.70	^{+	0.32	}_{-	0.25	}\pm	0.20	$	\\
7	&$	29.6	^{+	6.4	}_{-	5.7	}$	&$	8.18	\pm	0.57	$&$	2.28	^{+	0.49	}_{-	0.44	}\pm	0.36	$	&$	9.2	^{+	4.0	}_{-	3.2	}$	&$	9.07	\pm	0.56	$&$	0.64	^{+	0.28	}_{-	0.22	}\pm	0.11	$	\\
8	&$	10.2	^{+	4.2	}_{-	3.4	}$	&$	8.47	\pm	0.57	$&$	0.76	^{+	0.31	}_{-	0.25	}\pm	0.12	$	&$	27.6	^{+	6.4	}_{-	5.6	}$	&$	9.78	\pm	0.50	$&$	1.78	^{+	0.41	}_{-	0.36	}\pm	0.25	$	\\
9	&$	8.9	^{+	3.7	}_{-	3.0	}$	&$	8.79	\pm	0.56	$&$	0.64	^{+	0.27	}_{-	0.22	}\pm	0.11	    $	&$	11.3	^{+	4.3	}_{-	3.5	}$	&$	7.82	\pm	0.43	$&$	0.91	^{+	0.35	}_{-	0.28	}\pm	0.12	$	\\
10	&$	5.7	^{+	3.1	}_{-	2.4	}$	&$	8.98	\pm	0.56	$&$	0.40	^{+	0.22	}_{-	0.17	}\pm	0.07	$	    &$	9.7	^{+	4.3	}_{-	3.5	}$	&$	8.45	\pm	0.49	$&$	0.72	^{+	0.32	}_{-	0.26	}\pm	0.08	$	\\
11	&$	15.7	^{+	5.0	}_{-	4.2	}$	&$	9.04	\pm	0.55	$&$	1.09	^{+	0.35	}_{-	0.29	}\pm	0.12	$	&$	13.2	^{+	4.4	}_{-	3.7	}$	&$	8.97	\pm	0.49	$&$	0.93	^{+	0.31	}_{-	0.26	}\pm	0.11	$	\\
12	&$	11.8	^{+	4.2	}_{-	3.4	}$	&$	8.20	\pm	0.52	$&$	0.91	^{+	0.32	}_{-	0.26	}\pm	0.14	$	&$	8.5	^{+	10.1	}_{-	9.0	}$	&$	6.77	\pm	0.12	$&$	0.79	^{+	0.94	}_{-	0.84	}\pm	0.26	$	\\
13	&$	23.4	^{+	14.7	}_{-	13.6	}$	&$	7.45	\pm	0.10	$&$	1.98	^{+	1.24	}_{-	1.15	}\pm 0.46$	&$	7.6	^{+	7.2	}_{-	4.9	}$	&$	8.55	\pm	0.20	$&$	0.56	^{+	0.53	}_{-	0.36	}\pm	0.08	$	\\
\bottomrule
\end{tabular}
\end{center}
\end{table*}

\begin{table*}[tb]
\caption{Signal yields ($Y_{\mathrm{signal}}^{i}$), signal reconstruction efficiency ($\epsilon^{i}$), and partial branching fractions ($\Delta {\cal B}^{i} $) for each bin $i$ in the 1D($q^{2}$) configuration with the bin number convention defined according to Table~\ref{tab:fit1Dq2}. The first quoted uncertainty is statistical, and the second is systematic.}
\label{tab:BR_resultsq2}
\begin{center}
\def\arraystretch{1.4}
\begin{tabular}{c c c c c c c c }
\toprule
Bin    &    \multicolumn{1}{c}{\makecell[c]{$Y_{\mathrm{signal}}^{i}$}}    &    \multicolumn{1}{c}{$\epsilon^{i}\: [10^{-4}]$}&    $\Delta {\cal B}^{i} \: [10^{-5}]$  &  Bin &  \multicolumn{1}{c}{\makecell[c]{$Y_{\mathrm{signal}}^{i}$}}    &    \multicolumn{1}{c}{$\epsilon^{i}\: [10^{-4}]$}&    $\Delta {\cal B}^{i} \: [10^{-5}]$ \\
\midrule     
1	&$	16.5	^{+	6.8	}_{-	6.1	}$&$	6.27	\pm	0.19	$&$	1.66	^{+	0.68	}_{-	0.61	} \pm	0.53	$&	10	&$	30.8	^{+	6.7	}_{-	6.0	}$&$	8.96	\pm	0.53	$&$	2.17	^{+	0.47	}_{-	0.42} \pm	0.23	$\\
2	&$	11.4	^{+	6.0	}_{-	5.1	}$&$	6.97	\pm	0.22	$&$	1.03	^{+	0.54	}_{-	0.46	} \pm	0.25	$&	11	&$	11.6	^{+	5.0	}_{-	4.1	}$&$	9.52	\pm	0.60	$&$	0.77	^{+	0.33	}_{-	0.27	} \pm	0.09	$\\
3	&$	13.0	^{+	5.6	}_{-	4.7	}$&$	7.45	\pm	0.25	$&$	1.10	^{+	0.47	}_{-	0.40	} \pm	0.20	$&	12	&$	16.3	^{+	4.9	}_{-	4.1	}$&$	9.14	\pm	0.66	$&$	1.12	^{+	0.34	}_{-	0.28	} \pm	0.14	$\\
4	&$	16.0	^{+	6.6	}_{-	5.8	}$&$	7.56	\pm	0.27	$&$	1.33	^{+	0.55	}_{-	0.48	} \pm	0.26	$&	13	&$	19.4	^{+	5.3	}_{-	4.5	}$&$	8.62	\pm	0.72	$&$	1.42	^{+	0.39	}_{-	0.33	} \pm	0.17	$\\
5	&$	24.3	^{+	7.1	}_{-	6.3	}$&$	8.13	\pm	0.31	$&$	1.88	^{+	0.55	}_{-	0.49	} \pm	0.29	$&	14	&$	15.4	^{+	4.7	}_{-	4.0	}$&$	10.1	\pm	0.88	$&$	0.96	^{+	0.29	}_{-	0.25	} \pm	0.15	$\\
6	&$	12.2	^{+	5.7	}_{-	4.9	}$&$	8.60	\pm	0.35	$&$	0.89	^{+	0.42	}_{-	0.36	} \pm	0.10	$&	15	&$	15.1	^{+	4.9	}_{-	4.1	}$&$	8.65	\pm	0.93	$&$	1.10	^{+	0.36	}_{-	0.30	} \pm	0.17	$\\
7	&$	10.8	^{+	5.1	}_{-	4.1	}$&$	8.43	\pm	0.38	$&$	0.81	^{+	0.38	}_{-	0.31	} \pm	0.13	$&	16	&$	10.8	^{+	4.0	}_{-	3.2	}$&$	9.31	\pm	1.12	$&$	0.73	^{+	0.27	}_{-	0.22	} \pm	0.12	$\\
8	&$	21.4	^{+	6.5	}_{-	5.7	}$&$	9.17	\pm	0.44	$&$	1.47	^{+	0.45	}_{-	0.39	} \pm	0.17	$&	17	&$	12.3	^{+	4.4	}_{-	3.7	}$&$	8.94	\pm	1.28	$&$	0.87	^{+	0.31	}_{-	0.26	} \pm	0.15	$\\
9	&$	9.6	^{+	4.7	}_{-	3.9	}$&$	8.03	\pm	0.45	$&$	0.75	^{+	0.37	}_{-	0.31	} \pm	0.13	    $&	18	&$	32.3	^{+	6.8	}_{-	6.1	}$&$	7.85	\pm	0.88	$&$	2.59	^{+	0.55	}_{-	0.49	} \pm	0.47	$\\
\bottomrule
\end{tabular}
\end{center}
\end{table*} 

The main result of this analysis is the measurement of the total branching fraction for the ${B^{+}\to \pi^{+}\pi^{-}\ell^{+}\nu_{\ell}}$ decay. Since we carried out this measurement in bins of the kinematic variables $M_{\pi\pi}$ or $q^{2}$, we calculate the total branching fraction as the sum over all bins of the partial branching fractions, ${{\cal B}(B^{+}\to \pi^{+}\pi^{-}\ell^{+}\nu_{\ell})=\sum_{i} \Delta{\cal B}^{i}}$ with

\begin{equation}
 \Delta{\cal B}^{i}=\frac{1}{4}\frac{Y^{i}_{\mathrm{signal}}}{\epsilon_{i}}\frac{1}{{\cal B}(\Upsilon(4S)\to B^{+}B^{-})N_{B\bar{B}}}.
\label{eq:BF}
\end{equation}
Here, $Y^{i}_{\mathrm{signal}}$ denotes the signal yield measured in the $i$th bin, $\epsilon_{i}$ is the corresponding signal-reconstruction efficiency, ${{\cal B}(\Upsilon(4S)\to B^{+}B^{-})=(51.4\pm 0.6)\%}$, and $N_{B\bar{B}}=(771.6\pm 10.6)\times 10^{6}$ is the number of $B\bar{B}$-pairs produced for the complete Belle dataset. We determine the signal reconstruction efficiency from MC simulation, with corrections for differences between data and simulated detector performance. The factor of four in the denominator averages the observed branching fraction across the four channels: ${B^{+}\to\pi^{+}\pi^{-}e^{+}\nu_{e}}$, ${B^{-}\to\pi^{+}\pi^{-}e^{-}\bar{\nu}_{e}}$, ${B^{+}\to\pi^{+}\pi^{-}\mu^{+}\nu_{\mu}}$, and ${B^{-}\to\pi^{+}\pi^{-}\mu^{-}\bar{\nu}_{\mu}}$.

\begin{figure*}[htpb]
\centering
\subfloat{
\includegraphics[width=0.5\textwidth ]{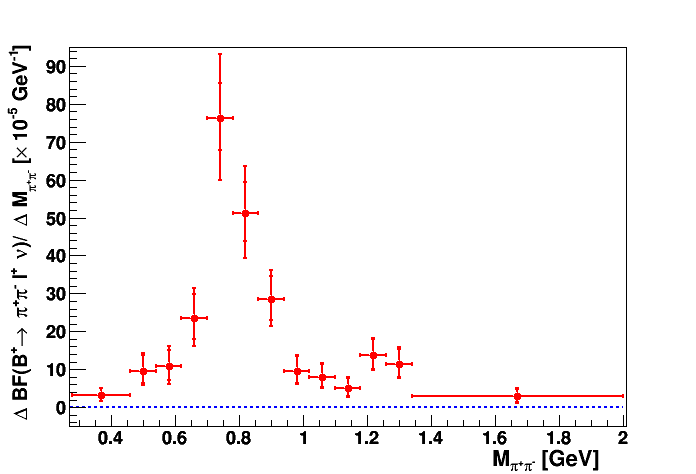}
}
\subfloat{
\includegraphics[width=0.5\textwidth ]{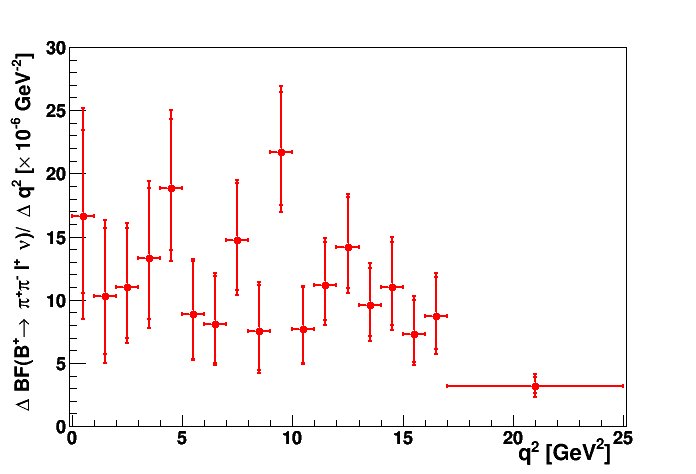}
}
\caption{Partial branching fractions for the decay $B^{+}\to \pi^{+}\pi^{-}\ell^{+}\nu_{\ell}$ in bins of: (left) the $\pi^{+}\pi^{-}$ invariant mass according to the results in the 1D$(M_{\pi\pi})$ configuration, and (right) the momentum-transfer squared according to the results in the 1D$(q^{2})$ configuration. As there is no upper limit in the $\pi^{+}\pi^{-}$ invariant mass, we use a cut-off at 2~GeV for the left figure.}
\label{fig:BR_in_bins}
\end{figure*} 

The values of the input parameters for Eq.~\ref{eq:BF}, as well as the partial branching fractions for each bin, are presented in Table~\ref{tab:BR_results} for the 1D$(M_{\pi\pi})$ and 2D configurations, and in Table~\ref{tab:BR_resultsq2} for the 1D$(q^{2})$ configuration. Adding the partial branching fractions, the total branching fraction for each configuration results in 

\begin{equation}
\begin{aligned}
 &{\cal B}(B^{+}\to \pi^{+}\pi^{-}\ell^{+}\nu_{\ell})[\mathrm{1D}(M_{\pi\pi})]\\
 &= (22.3^{+2.0}_{-1.8} (\mathrm{stat}) \pm 4.0(\mathrm{syst}) )\times 10^{-5},
\end{aligned}
\label{eq:BR_Bl4}
\end{equation}

\begin{equation}
\begin{aligned}
&{\cal B}(B^{+}\to \pi^{+}\pi^{-}\ell^{+}\nu_{\ell})[\mathrm{1D}(q^{2})]\\
&= (22.7 ^{+1.9}_{-1.6} (\mathrm{stat}) \pm 3.5(\mathrm{syst}) )\times 10^{-5},
\end{aligned}
\label{eq:BR_Bl4q2}
\end{equation}

\begin{equation}
	\begin{aligned}
		&{\cal B}(B^{+}\to \pi^{+}\pi^{-}\ell^{+}\nu_{\ell})[\mathrm{2D}]\\
		&= (23.0^{+2.0}_{-1.8} (\mathrm{stat}) \pm 3.0(\mathrm{syst}) )\times 10^{-5},
	\end{aligned}
\end{equation}
where the quoted uncertainties are the statistical and systematic uncertainties, respectively. As the $[\mathrm{1D}(q^{2})]$ result lies between the $[\mathrm{1D}(M_{\pi\pi})]$ and $[\mathrm{2D}]$ results, with the difference in central values being negligible as compared to the quoted systematic uncertainty, we take the $[\mathrm{1D}(q^{2})]$ measurement as our final result:

\begin{equation}
\begin{aligned}
&{\cal B}(B^{+}\to \pi^{+}\pi^{-}\ell^{+}\nu_{\ell})\\
&= (22.7 ^{+1.9}_{-1.6} (\mathrm{stat}) \pm 3.5(\mathrm{syst}) )\times 10^{-5}.
\end{aligned}
\label{eq:ReportedBR}
\end{equation}

In the three configurations, our measurement is dominated by systematic uncertainties. The most significant source of systematic uncertainty comes from signal modeling. The value given in Eq.~\ref{eq:ReportedBR} is the first reported measurement of the  branching fraction for the $B^{+}\to \pi^{+}\pi^{-}\ell^{+}\nu_{\ell}$ decay. A correlation matrix between the measurements using the 18 bins in $q^{2}$ and the 13 bins in $M_{\pi\pi}$ is provided in Table~\ref{tab:Corr_q2_mpipi_SR}.

Figure~\ref{fig:BR_in_bins} shows the dependence of the partial branching fractions on the $\pi^{+}\pi^{-}$ invariant mass and the squared momentum transfer. Although a detailed analysis of the resonant and nonresonant composition of the dipion mass spectrum is beyond the scope of this paper due to the limited statistics, we can observe a dominant peak associated with the $\rho^{0}$ meson and a small bump around the mass window for the $f_{2}(1270)$ meson. In a previous analysis~\cite{Sibidanov}, using the same data set, the contribution of the $\rho^{0}$ meson corresponds to a branching ratio of ${{\cal B}(B^{+}\to \rho^{0}\ell^{+}\nu_{\ell})= [18.3 \pm 1.0(\mathrm{stat}) \pm 1.0(\mathrm{syst}) ]\times 10^{-5}}$. The excess observed in the dipion mass spectrum motivates the search and measurement of other exclusive charmless semileptonic $B$ decays with masses above 1 GeV in the next generation of $B$ factories, such as Belle II~\cite{B2TIP}. Our measurement of $B^{+}\to \pi^{+}\pi^{-}\ell^{+}\nu_{\ell}$ should improve the modeling of $B$ semileptonic decays and thus increase the precision with which $|V_{ub}|$ can be measured.





\section{ACKNOWLEDGEMENTS}
We thank the KEKB group for the excellent operation of the
accelerator; the KEK cryogenics group for the efficient
operation of the solenoid; and the KEK computer group, and the Pacific Northwest National
Laboratory (PNNL) Environmental Molecular Sciences Laboratory (EMSL)
computing group for strong computing support; and the National
Institute of Informatics, and Science Information NETwork 5 (SINET5) for
valuable network support.  We acknowledge support from
the Ministry of Education, Culture, Sports, Science, and
Technology (MEXT) of Japan, the Japan Society for the 
Promotion of Science (JSPS), and the Tau-Lepton Physics 
Research Center of Nagoya University; 
the Australian Research Council including grants
DP180102629, 
DP170102389, 
DP170102204, 
DP150103061, 
FT130100303; 
Austrian Federal Ministry of Education, Science and Research (FWF) and
FWF Austrian Science Fund No.~P~31361-N36;
the National Natural Science Foundation of China under Contracts
No.~11435013,  
No.~11475187,  
No.~11521505,  
No.~11575017,  
No.~11675166,  
No.~11705209;  
Key Research Program of Frontier Sciences, Chinese Academy of Sciences (CAS), Grant No.~QYZDJ-SSW-SLH011; 
the  CAS Center for Excellence in Particle Physics (CCEPP); 
the Shanghai Pujiang Program under Grant No.~18PJ1401000;  
the Shanghai Science and Technology Committee (STCSM) under Grant No.~19ZR1403000; 
the Ministry of Education, Youth and Sports of the Czech
Republic under Contract No.~LTT17020;
Horizon 2020 ERC Advanced Grant No.~884719 and ERC Starting Grant No.~947006 ``InterLeptons'' (European Union);
the Carl Zeiss Foundation, the Deutsche Forschungsgemeinschaft, the
Excellence Cluster Universe, and the VolkswagenStiftung;
the Department of Atomic Energy (Project Identification No. RTI 4002) and the Department of Science and Technology of India; 
the Istituto Nazionale di Fisica Nucleare of Italy; 
National Research Foundation (NRF) of Korea Grant
Nos.~2016R1\-D1A1B\-01010135, 2016R1\-D1A1B\-02012900, 2018R1\-A2B\-3003643,
2018R1\-A6A1A\-06024970, 2018R1\-D1A1B\-07047294, 2019K1\-A3A7A\-09033840,
2019R1\-I1A3A\-01058933;
Radiation Science Research Institute, Foreign Large-size Research Facility Application Supporting project, the Global Science Experimental Data Hub Center of the Korea Institute of Science and Technology Information and KREONET/GLORIAD;
the Polish Ministry of Science and Higher Education and 
the National Science Center;
the Ministry of Science and Higher Education of the Russian Federation, Agreement 14.W03.31.0026, 
and the HSE University Basic Research Program, Moscow; 
University of Tabuk research grants
S-1440-0321, S-0256-1438, and S-0280-1439 (Saudi Arabia);
the Slovenian Research Agency Grant Nos. J1-9124 and P1-0135;
Ikerbasque, Basque Foundation for Science, Spain;
the Swiss National Science Foundation; 
the Ministry of Education and the Ministry of Science and Technology of Taiwan;
and the United States Department of Energy and the National Science Foundation.

%

\appendix
\counterwithin{table}{section}
\section{Supplementary tables}

\clearpage

\begin{turnpage}
\begin{table}[tpb]
\caption{Reconstruction efficiency ($\times 10^{-4}$) in $M_{\pi\pi}$ bins for the phase space $B^{+}\to \pi^{+}\pi^{-}\ell^{+}\nu_{\ell}$ and for  $B^{+}\to X\ell^{+}\nu_{\ell}$, with the intermediate resonance $X$ simulated using different MC generators. The bin-number convention is defined in Table~\ref{tab:fit1D}.}
\label{tab:SigModEff1D}
\begin{center}
\bgroup
\def\arraystretch{2}
{\footnotesize
\begin{tabular}{cccccccccccc}
\toprule
Bin	& \makecell[c]{PHSP\\ $\pi^{+}\pi^{-}$}  & \makecell[c]{ISGW2 \\ $f_{0}(500)$} & \makecell[c]{PHSP \\ $f_{0}(500)$} & \makecell[c]{ISGW2 \\ $\rho^{0}$} & \makecell[c]{LCSR~\cite{Ball98} \\ $\rho^{0}$} & \makecell[c]{PHSP \\ $\rho^{0}$} & \makecell[c]{ISGW2 \\ $f_{2}(1270)$} & \makecell[c]{PHSP \\ $f_{2}(1270)$} & \makecell[c]{LCSR~\cite{f2ffLCSR} \\ $f_{2}(1270)$} & \makecell[c]{ISGW2 \\ $\rho(1450)^{0}$} & \makecell[c]{PHSP \\ $\rho(1450)^{0}$} \\
\midrule
1	&	7.92	$\pm$	0.66	&	8.06	$\pm$	0.21	&	8.17	$\pm$	0.21	&				&				&				&				&				&				&				&				\\
2	&	8.20	$\pm$	0.77	&	8.27	$\pm$	0.29	&	8.18	$\pm$	0.29	&	10.05	$\pm$	0.70	&	9.74	$\pm$	0.69	&	9.27	$\pm$	0.68	&				&				&				&				&				\\
3	&	7.75	$\pm$	0.68	&	8.39	$\pm$	0.28	&	7.96	$\pm$	0.27	&	9.43	$\pm$	0.46	&	9.40	$\pm$	0.46	&	8.89	$\pm$	0.45	&				&				&				&				&				\\
4	&	7.82	$\pm$	0.64	&	8.21	$\pm$	0.30	&	8.28	$\pm$	0.30	&	9.01	$\pm$	0.28	&	9.08	$\pm$	0.28	&	8.64	$\pm$	0.27	&				&				&				&				&				\\
5	&	9.32	$\pm$	0.66	&	8.10	$\pm$	0.36	&	8.33	$\pm$	0.37	&	9.08	$\pm$	0.17	&	8.54	$\pm$	0.17	&	8.20	$\pm$	0.16	&				&				&				&				&				\\
6	&	7.76	$\pm$	0.58	&	8.61	$\pm$	0.48	&	8.31	$\pm$	0.47	&	8.93	$\pm$	0.17	&	8.28	$\pm$	0.17	&	8.06	$\pm$	0.18	&				&				&				&				&				\\
7	&	8.18	$\pm$	0.57	&				&				&	9.34	$\pm$	0.32	&	8.91	$\pm$	0.31	&	8.70	$\pm$	0.31	&				&				&				&				&				\\
8	&	8.47	$\pm$	0.57	&				&				&	9.57	$\pm$	0.49	&	8.81	$\pm$	0.47	&	9.19	$\pm$	0.48	&	10.12	$\pm$	0.96	&	9.63	$\pm$	0.94	&	8.32	$\pm$	0.87	&	8.73	$\pm$	0.96	&	9.67	$\pm$	1.01	\\
9	&	8.79	$\pm$	0.56	&				&				&	9.35	$\pm$	0.67	&	10.02	$\pm$	0.69	&	10.17	$\pm$	0.70	&	9.14	$\pm$	0.64	&	9.71	$\pm$	0.66	&	7.74	$\pm$	0.59	&	11.70	$\pm$	0.97	&	8.13	$\pm$	0.80	\\
10	&	8.98	$\pm$	0.56	&				&				&				&				&				&	8.9	$\pm$	0.41	&	9.45	$\pm$	0.42	&	8.67	$\pm$	0.41	&	11.91	$\pm$	0.83	&	8.93	$\pm$	0.72	\\
11	&	9.04	$\pm$	0.55	&				&				&				&				&				&	8.91	$\pm$	0.27	&	8.49	$\pm$	0.26	&	8.23	$\pm$	0.26	&	9.55	$\pm$	0.63	&	8.63	$\pm$	0.60	\\
12	&	8.2	$\pm$	0.52	&				&				&				&				&				&	8.40	$\pm$	0.24	&	8.55	$\pm$	0.25	&	8.23	$\pm$	0.24	&	9.99	$\pm$	0.54	&	9.49	$\pm$	0.52	\\
13	&	7.45	$\pm$	0.10	&				&				&				&				&				&	8.74	$\pm$	0.23	&	8.56	$\pm$	0.22	&	9.02	$\pm$	0.23	&	9.69	$\pm$	0.19	&	9.05	$\pm$	0.19	\\
\bottomrule																					
\end{tabular}}
\egroup
\end{center}
\end{table}
\end{turnpage}

\clearpage

\begin{turnpage}
\begin{table}[tpb]
\caption{Reconstruction efficiency ($\times 10^{-4}$) in $q^{2}$ bins for the phase space $B^{+}\to \pi^{+}\pi^{-}\ell^{+}\nu_{\ell}$ and for  $B^{+}\to X\ell^{+}\nu_{\ell}$, with the intermediate resonance $X$ simulated using different MC generators. The bin-number convention is defined in Table~\ref{tab:fit1Dq2}.}
\label{tab:SigModEff1Dq2}
\begin{center}
\bgroup
\def\arraystretch{2}
{\footnotesize
\begin{tabular}{cccccccccccc}
\toprule
Bin	& \makecell[c]{PHSP\\ $\pi^{+}\pi^{-}$}  & \makecell[c]{ISGW2 \\ $f_{0}(500)$} & \makecell[c]{PHSP \\ $f_{0}(500)$} & \makecell[c]{ISGW2 \\ $\rho^{0}$} & \makecell[c]{LCSR~\cite{Ball98} \\ $\rho^{0}$} & \makecell[c]{PHSP \\ $\rho^{0}$} & \makecell[c]{ISGW2 \\ $f_{2}(1270)$} & \makecell[c]{PHSP \\ $f_{2}(1270)$} & \makecell[c]{LCSR~\cite{f2ffLCSR} \\ $f_{2}(1270)$} & \makecell[c]{ISGW2 \\ $\rho(1450)^{0}$} & \makecell[c]{PHSP \\ $\rho(1450)^{0}$} \\
\midrule
1	&	6.27	$\pm$	0.19	&	8.10	$\pm$	0.40	&	7.94	$\pm$	0.40	&	7.95	$\pm$	0.83	&	8.76	$\pm$	0.48	&	7.63	$\pm$	0.31	&	8.46	$\pm$	0.43	&	8.13	$\pm$	0.39	&	8.55	$\pm$	0.28	&	7.53	$\pm$	0.63	&	7.91	$\pm$	0.44	\\
2	&	6.97	$\pm$	0.22	&	7.28	$\pm$	0.38	&	7.55	$\pm$	0.39	&	10.14	$\pm$	0.75	&	8.84	$\pm$	0.46	&	7.81	$\pm$	0.32	&	8.75	$\pm$	0.42	&	7.73	$\pm$	0.38	&	8.61	$\pm$	0.30	&	8.62	$\pm$	0.62	&	8.31	$\pm$	0.46	\\
3	&	7.45	$\pm$	0.25	&	8.17	$\pm$	0.41	&	7.78	$\pm$	0.4	&	9.15	$\pm$	0.61	&	8.26	$\pm$	0.43	&	7.67	$\pm$	0.32	&	8.82	$\pm$	0.42	&	7.95	$\pm$	0.40	&	8.22	$\pm$	0.32	&	9.55	$\pm$	0.61	&	9.41	$\pm$	0.50	\\
4	&	7.56	$\pm$	0.27	&	8.39	$\pm$	0.43	&	8.26	$\pm$	0.43	&	9.17	$\pm$	0.55	&	9.06	$\pm$	0.44	&	7.94	$\pm$	0.34	&	8.89	$\pm$	0.42	&	8.15	$\pm$	0.42	&	8.30	$\pm$	0.36	&	9.72	$\pm$	0.59	&	9.87	$\pm$	0.53	\\
5	&	8.13	$\pm$	0.31	&	8.15	$\pm$	0.43	&	8.25	$\pm$	0.43	&	9.51	$\pm$	0.51	&	8.56	$\pm$	0.42	&	8.06	$\pm$	0.35	&	8.82	$\pm$	0.42	&	9.16	$\pm$	0.45	&	8.26	$\pm$	0.40	&	10.88	$\pm$	0.61	&	9.24	$\pm$	0.53	\\
6	&	8.60	$\pm$	0.35	&	8.85	$\pm$	0.46	&	8.22	$\pm$	0.44	&	10.13	$\pm$	0.49	&	8.35	$\pm$	0.41	&	9.06	$\pm$	0.38	&	8.62	$\pm$	0.42	&	9.02	$\pm$	0.47	&	8.95	$\pm$	0.47	&	10.21	$\pm$	0.57	&	8.52	$\pm$	0.53	\\
7	&	8.43	$\pm$	0.38	&	8.20	$\pm$	0.46	&	8.84	$\pm$	0.47	&	9.84	$\pm$	0.46	&	8.37	$\pm$	0.40	&	8.44	$\pm$	0.37	&	9.07	$\pm$	0.45	&	9.44	$\pm$	0.49	&	8.43	$\pm$	0.52	&	11.48	$\pm$	0.60	&	9.12	$\pm$	0.57	\\
8	&	9.17	$\pm$	0.44	&	8.24	$\pm$	0.47	&	8.82	$\pm$	0.49	&	9.40	$\pm$	0.43	&	9.31	$\pm$	0.42	&	8.60	$\pm$	0.39	&	9.14	$\pm$	0.47	&	9.15	$\pm$	0.51	&	9.06	$\pm$	0.63	&	10.88	$\pm$	0.59	&	10.63	$\pm$	0.64	\\
9	&	8.03	$\pm$	0.45	&	8.58	$\pm$	0.49	&	9.02	$\pm$	0.51	&	8.63	$\pm$	0.39	&	8.75	$\pm$	0.40	&	8.50	$\pm$	0.40	&	8.37	$\pm$	0.47	&	9.83	$\pm$	0.55	&	10.15	$\pm$	0.80	&				&				\\
10	&	8.96	$\pm$	0.53	&	9.75	$\pm$	0.54	&	8.71	$\pm$	0.51	&	9.50	$\pm$	0.40	&	9.44	$\pm$	0.41	&	8.91	$\pm$	0.42	&	8.68	$\pm$	0.52	&	9.62	$\pm$	0.57	&	9.84	$\pm$	0.97	&				&				\\
11	&	9.52	$\pm$	0.60	&	9.41	$\pm$	0.55	&	8.84	$\pm$	0.53	&	9.62	$\pm$	0.39	&	9.14	$\pm$	0.40	&	9.06	$\pm$	0.44	&	9.01	$\pm$	0.58	&	8.66	$\pm$	0.57	&	10.07	$\pm$	1.25	&				&				\\
12	&	9.14	$\pm$	0.66	&	9.13	$\pm$	0.56	&	8.44	$\pm$	0.54	&	8.96	$\pm$	0.37	&	9.10	$\pm$	0.40	&	9.04	$\pm$	0.46	&	9.44	$\pm$	0.67	&	10.29	$\pm$	0.67	&	7.56	$\pm$	1.46	&				&				\\
13	&	8.62	$\pm$	0.72	&	8.44	$\pm$	0.56	&	8.19	$\pm$	0.55	&	9.58	$\pm$	0.38	&	8.50	$\pm$	0.39	&	8.84	$\pm$	0.47	&				&				&				&				&				\\
14	&	10.14	$\pm$	0.88	&	8.66	$\pm$	0.59	&	8.68	$\pm$	0.59	&	9.65	$\pm$	0.38	&	9.06	$\pm$	0.41	&	9.52	$\pm$	0.51	&				&				&				&				&				\\
15	&	8.65	$\pm$	0.93	&	8.00	$\pm$	0.60	&	8.86	$\pm$	0.63	&	8.92	$\pm$	0.37	&	9.01	$\pm$	0.42	&	8.91	$\pm$	0.52	&				&				&				&				&				\\
16	&	9.31	$\pm$	1.12	&	8.33	$\pm$	0.65	&	8.24	$\pm$	0.64	&	8.58	$\pm$	0.37	&	9.10	$\pm$	0.43	&	8.82	$\pm$	0.56	&				&				&				&				&				\\
17	&	8.94	$\pm$	1.28	&	8.3	$\pm$	0.69	&	8.64	$\pm$	0.70	&	9.12	$\pm$	0.40	&	8.43	$\pm$	0.43	&	8.54	$\pm$	0.59	&				&				&				&				&				\\
18	&	7.85	$\pm$	0.88	&	7.07	$\pm$	0.36	&	7.76	$\pm$	0.38	&	8.56	$\pm$	0.27	&	8.60	$\pm$	0.29	&	9.09	$\pm$	0.43	&				&				&				&				&				\\
\bottomrule																					
\end{tabular}}
\egroup
\end{center}
\end{table}
\end{turnpage}

\clearpage

\begin{turnpage}
\begin{table}[tpb]
\caption{Reconstruction efficiency ($\times 10^{-4}$) in $M_{\pi\pi}$ and $q^{2}$ bins for the phase space $B^{+}\to \pi^{+}\pi^{-}\ell^{+}\nu_{\ell}$ and for  $B^{+}\to X\ell^{+}\nu_{\ell}$, with the intermediate resonance $X$ simulated using different MC generators. The bin-number convention is defined in Table~\ref{tab:fit2D}.}
\label{tab:SigModEff2D}
\begin{center}
\bgroup
\def\arraystretch{2}
{\footnotesize
\begin{tabular}{cccccccccccc}
\toprule
Bin	& \makecell[c]{PHSP\\ $\pi^{+}\pi^{-}$}  & \makecell[c]{ISGW2 \\ $f_{0}(500)$} & \makecell[c]{PHSP \\ $f_{0}(500)$} & \makecell[c]{ISGW2 \\ $\rho^{0}$} & \makecell[c]{LCSR~\cite{Ball98} \\ $\rho^{0}$} & \makecell[c]{PHSP \\ $\rho^{0}$} & \makecell[c]{ISGW2 \\ $f_{2}(1270)$} & \makecell[c]{PHSP \\ $f_{2}(1270)$} & \makecell[c]{LCSR~\cite{f2ffLCSR} \\ $f_{2}(1270)$} & \makecell[c]{ISGW2 \\ $\rho(1450)^{0}$} & \makecell[c]{PHSP \\ $\rho(1450)^{0}$} \\
\midrule
1	&	7.39	$\pm$	0.57	&	8.02	$\pm$	0.21	&	7.88	$\pm$	0.21	&	9.65	$\pm$	0.72	&	9.19	$\pm$	0.64	&	8.92	$\pm$	0.52	&				&				&				&				&				\\
2	&	8.45	$\pm$	0.63	&	8.46	$\pm$	0.22	&	8.35	$\pm$	0.22	&	10.30	$\pm$	0.49	&	9.97	$\pm$	0.50	&	9.80	$\pm$	0.58	&				&				&				&				&				\\
3	&	8.61	$\pm$	0.60	&	8.43	$\pm$	0.35	&	8.32	$\pm$	0.34	&	8.78	$\pm$	0.37	&	8.49	$\pm$	0.25	&	7.73	$\pm$	0.18	&				&				&				&				&				\\
4	&	8.35	$\pm$	0.63	&	8.14	$\pm$	0.37	&	8.21	$\pm$	0.37	&	9.61	$\pm$	0.27	&	8.32	$\pm$	0.23	&	8.24	$\pm$	0.21	&				&				&				&				&				\\
5	&	7.98	$\pm$	0.48	&	8.01	$\pm$	0.28	&	8.56	$\pm$	0.29	&	8.88	$\pm$	0.13	&	8.61	$\pm$	0.14	&	8.68	$\pm$	0.16	&				&				&				&				&				\\
6	&	7.20	$\pm$	0.46	&	7.79	$\pm$	0.76	&	8.78	$\pm$	0.81	&	10.68	$\pm$	1.10	&	8.66	$\pm$	0.63	&	7.64	$\pm$	0.46	&	8.79	$\pm$	0.47	&	8.84	$\pm$	0.48	&	8.05	$\pm$	0.34	&	9.65	$\pm$	0.96	&	8.39	$\pm$	0.71	\\
7	&	9.07	$\pm$	0.56	&	8.30	$\pm$	0.85	&	9.46	$\pm$	0.91	&	9.84	$\pm$	0.65	&	9.47	$\pm$	0.59	&	9.20	$\pm$	0.55	&	9.26	$\pm$	0.51	&	10.12	$\pm$	0.56	&	8.34	$\pm$	0.51	&	11.18	$\pm$	0.94	&	8.72	$\pm$	0.80	\\
8	&	9.78	$\pm$	0.50	&	10.64	$\pm$	0.81	&	9.28	$\pm$	0.76	&	9.65	$\pm$	0.35	&	10.05	$\pm$	0.40	&	10.04	$\pm$	0.48	&	9.26	$\pm$	0.51	&	9.12	$\pm$	0.47	&	9.30	$\pm$	0.91	&	11.30	$\pm$	0.67	&	9.40	$\pm$	0.72	\\
9	&	7.82	$\pm$	0.43	&				&				&				&				&				&	8.54	$\pm$	0.26	&	7.71	$\pm$	0.24	&	8.23	$\pm$	0.19	&	8.79	$\pm$	0.51	&	8.72	$\pm$	0.38	\\
10	&	8.45	$\pm$	0.49	&				&				&				&				&				&	8.51	$\pm$	0.27	&	8.82	$\pm$	0.29	&	8.40	$\pm$	0.29	&	9.86	$\pm$	0.45	&	8.52	$\pm$	0.42	\\
11	&	8.97	$\pm$	0.49	&				&				&				&				&				&	8.51	$\pm$	0.29	&	9.24	$\pm$	0.29	&	9.60	$\pm$	0.62	&	10.14	$\pm$	0.38	&	10.00	$\pm$	0.46	\\
12	&	6.77	$\pm$	0.12	&				&				&				&				&				&	9.08	$\pm$	0.54	&	7.89	$\pm$	0.49	&	9.56	$\pm$	0.43	&	8.46	$\pm$	0.43	&	8.58	$\pm$	0.34	\\
13	&	8.55	$\pm$	0.20	&				&				&				&				&				&	9.53	$\pm$	0.51	&	9.23	$\pm$	0.52	&	9.65	$\pm$	0.78	&	10.17	$\pm$	0.30	&	9.55	$\pm$	0.34	\\
\bottomrule																					
\end{tabular}}
\egroup
\end{center}
\end{table}
\end{turnpage}

\clearpage

\begin{turnpage}
\begin{table}[tpb]
\caption{Relative systematic uncertainties in percent for the fits in $M_{\pi\pi}$ bins. The bin-number convention is defined in Table~\ref{tab:fit1D}.}
\label{tab:syst1D}
\begin{center}
\bgroup
\def\arraystretch{1.7}
{\small
\begin{tabular}{p{2.7cm}ccccccccccccc}
\toprule
Source	&	Bin 1	&	Bin 2	&	Bin 3	&	Bin 4	&	Bin 5	&	Bin 6	&	Bin 7	&	Bin 8	&	Bin 9	&	Bin  10	&	Bin 11	&	Bin 12	&	Bin 13	\\
\midrule																											
FF ${B \to D^{(*)}\ell\nu}$	&	0.12	&	0.04	&	0.10	&	0.10	&	0.03	&	0.02	&	0.06	&	0.07	&	0.12	&	0.07	&	0.10	&	0.05	&	0.65	\\
FF ${B \to D^{**}\ell\nu}$	&	0.79	&	0.27	&	0.37	&	0.52	&	0.29	&	0.08	&	0.13	&	0.38	&	0.47	&	0.21	&	0.23	&	0.06	&	1.40	\\
Shapes ${B \to X_{u}\ell\nu}$	&	2.11	&	1.39	&	0.49	&	1.10	&	0.17	&	0.37	&	0.39	&	0.56	&	0.13	&	1.52	&	0.06	&	0.06	&	0.11	\\
${{\cal B}(B \to D^{(*)}\ell\nu)}$	&	0.37	&	0.01	&	0.34	&	0.19	&	1.70	&	0.16	&	0.05	&	0.14	&	0.23	&	0.14	&	0.31	&	0.12	&	0.41	\\
${{\cal B}(B \to D^{**}\ell\nu)}$	&	0.12	&	0.08	&	0.38	&	0.21	&	0.05	&	0.08	&	0.14	&	0.12	&	0.14	&	0.09	&	0.14	&	0.20	&	0.39	\\
${{\cal B}(B \to X_{u}\ell\nu)}$	&	6.54	&	3.02	&	4.14	&	3.44	&	0.92	&	1.74	&	5.13	&	3.48	&	0.55	&	0.41	&	0.43	&	0.31	&	0.94	\\
Continuum	&	0.58	&	0.39	&	0.09	&	0.24	&	0.08	&	0.09	&	0.30	&	0.99	&	0.09	&	0.60	&	0.20	&	0.34	&	1.85	\\
Rare	&	0.80	&	1.04	&	0.62	&	0.65	&	0.15	&	0.25	&	0.41	&	0.61	&	0.53	&	1.49	&	0.65	&	0.48	&	4.80	\\
Sec. Leptons	&	0.05	&	0.13	&	0.00	&	0.19	&	0.02	&	0.01	&	0.14	&	0.14	&	0.16	&	0.01	&	0.00	&	0.17	&	1.14	\\
Fake leptons	&	0.76	&	0.07	&	0.11	&	0.24	&	0.04	&	0.06	&	0.03	&	0.90	&	0.11	&	0.23	&	0.41	&	0.06	&	2.00	\\
$\ell$ID	&	1.85	&	1.90	&	1.90	&	1.87	&	1.93	&	1.90	&	1.93	&	1.83	&	1.89	&	1.85	&	1.89	&	1.89	&	2.02	\\
$\pi$ID	&	0.98	&	0.98	&	0.95	&	1.00	&	0.98	&	1.00	&	1.01	&	0.99	&	1.00	&	1.01	&	1.04	&	1.02	&	1.18	\\
FSR	&	0.00	&	0.15	&	0.20	&	0.28	&	0.37	&	0.11	&	0.27	&	0.37	&	1.04	&	1.46	&	1.38	&	0.58	&	0.53	\\
Signal model	&	2.56	&	14.4	&	24.7	&	18.2	&	15.9	&	15.1	&	10.4	&	11.4	&	15.4	&	14.9	&	5.86	&	12.2	&	21.6	\\
Fit procedure	&	3.01	&	1.42	&	1.43	&	0.19	&	0.33	&	0.27	&	1.40	&	1.09	&	1.82	&	2.35	&	1.15	&	1.07	&	2.70	\\
Nom. Eff. Stats. 	&	8.33	&	9.39	&	8.77	&	8.18	&	7.08	&	7.47	&	6.97	&	6.73	&	6.37	&	6.24	&	6.08	&	6.34	&	1.34	\\
BDT selection	&	0.84	&	1.16	&	0.00	&	0.32	&	0.77	&	1.64	&	1.42	&	2.30	&	1.22	&	0.68	&	1.73	&	4.01	&	2.18	\\
No. $B\bar{B}$ pairs	&	1.40	&	1.40	&	1.40	&	1.40	&	1.40	&	1.40	&	1.40	&	1.40	&	1.40	&	1.40	&	1.40	&	1.40	&	1.40	\\
${{\cal B}(\Upsilon(4S)\to B^{+}B^{-})}$	&	1.17	&	1.17	&	1.17	&	1.17	&	1.17	&	1.17	&	1.17	&	1.17	&	1.17	&	1.17	&	1.17	&	1.17	&	1.17	\\
Tracking efficiency	&	1.05	&	1.05	&	1.05	&	1.05	&	1.05	&	1.05	&	1.05	&	1.05	&	1.05	&	1.05	&	1.05	&	1.05	&	1.05	\\
Tagging efficiency	&	4.20	&	4.20	&	4.20	&	4.20	&	4.20	&	4.20	&	4.20	&	4.20	&	4.20	&	4.20	&	4.20	&	4.20	&	4.20	\\
Nonresonant ${B \to X_{u}\ell\nu}$	&	9.86	&	9.00	&	3.77	&	5.58	&	1.99	&	3.17	&	5.74	&	3.92	&	3.37	&	3.51	&	2.55	&	1.69	&	2.14	\\
\midrule																											
Total	&	16.1	&	20.4	&	27.3	&	21.7	&	18.4	&	18.1	&	15.7	&	15.4	&	18.0	&	17.7	&	10.6	&	15.4	&	23.4	\\
\bottomrule																											
\end{tabular}}
\egroup
\end{center}
\end{table}
\end{turnpage}

\clearpage

\begin{turnpage}
\begin{table}[tpb]
\caption{Relative systematic uncertainties in percent for the fits in $q^{2}$ bins. The bin-number convention is defined in Table~\ref{tab:fit1Dq2}.}
\label{tab:systq2}
\begin{center}
\bgroup
\def\arraystretch{1.5}
{\small
\begin{tabular}{p{2.7cm}cccccccccccccccccc}
\toprule
Source	&	Bin 1	&	Bin 2	&	Bin 3	&	Bin 4	&	Bin 5	&	Bin 6	&	Bin 7	&	Bin 8	&	Bin 9	&	Bin  10	&	Bin 11	&	Bin 12	&	Bin 13	&	Bin  14	&	Bin  15	&	Bin  16	&	Bin  17	&	Bin  18	\\
\midrule
FF ${B \to D^{(*)}\ell\nu}$	&	1.19	&	0.85	&	0.40	&	0.40	&	0.20	&	0.45	&	0.56	&	0.71	&	1.08	&	0.20	&	1.07	&	0.10	&	0.04	&	0.07	&	0.05	&	0.02	&	0.15	&	0.03	\\
FF ${B \to D^{**}\ell\nu}$	&	1.93	&	2.36	&	1.27	&	1.84	&	1.05	&	1.09	&	1.12	&	0.73	&	0.72	&	0.29	&	0.66	&	0.10	&	0.07	&	0.20	&	0.15	&	0.21	&	0.22	&	0.08	\\
Shapes ${B \to X_{u}\ell\nu}$	&	0.20	&	0.36	&	0.42	&	0.33	&	0.23	&	0.32	&	0.25	&	0.26	&	0.11	&	0.08	&	0.21	&	0.22	&	0.37	&	0.35	&	1.32	&	1.30	&	0.40	&	1.89	\\
${{\cal B}(B \to D^{(*)}\ell\nu)}$	&	1.92	&	3.10	&	2.05	&	2.35	&	1.41	&	2.63	&	1.17	&	1.11	&	1.79	&	0.24	&	0.91	&	0.06	&	0.04	&	0.08	&	0.09	&	0.05	&	0.24	&	0.07	\\
${{\cal B}(B \to D^{**}\ell\nu)}$	&	0.66	&	0.78	&	0.42	&	0.61	&	0.31	&	0.37	&	0.39	&	0.22	&	0.18	&	0.08	&	0.17	&	0.02	&	0.02	&	0.04	&	0.05	&	0.02	&	0.08	&	0.02	\\
${{\cal B}(B \to X_{u}\ell\nu)}$	&	0.68	&	0.71	&	0.15	&	1.07	&	0.83	&	0.67	&	0.66	&	0.53	&	0.53	&	0.42	&	1.85	&	1.06	&	1.37	&	1.20	&	3.81	&	2.30	&	0.95	&	1.88	\\
Continuum	&	1.70	&	0.93	&	0.34	&	0.43	&	0.13	&	0.45	&	0.32	&	0.41	&	0.54	&	0.02	&	0.19	&	0.03	&	0.28	&	0.07	&	0.01	&	0.27	&	0.40	&	0.14	\\
Rare	&	7.44	&	1.47	&	0.73	&	0.97	&	0.60	&	0.30	&	0.31	&	0.46	&	1.09	&	0.34	&	0.44	&	0.29	&	0.30	&	0.20	&	0.39	&	0.01	&	0.25	&	0.16	\\
Sec. Leptons	&	0.29	&	0.14	&	0.16	&	0.20	&	0.01	&	0.32	&	0.03	&	0.01	&	0.11	&	0.02	&	0.00	&	0.01	&	0.00	&	0.01	&	0.00	&	0.01	&	0.07	&	0.03	\\
Fake leptons	&	2.56	&	1.05	&	0.20	&	0.17	&	0.11	&	0.15	&	0.04	&	0.23	&	0.09	&	0.04	&	0.20	&	0.02	&	0.16	&	0.01	&	0.40	&	0.08	&	0.55	&	0.27	\\
$\ell$ID	&	1.99	&	2.03	&	2.08	&	2.02	&	2.04	&	1.98	&	1.96	&	1.91	&	1.85	&	1.87	&	1.82	&	1.83	&	1.74	&	1.83	&	1.83	&	1.89	&	1.89	&	1.89	\\
$\pi$ID	&	1.23	&	1.21	&	1.19	&	1.17	&	1.15	&	1.12	&	1.11	&	1.07	&	1.06	&	1.03	&	0.97	&	0.96	&	0.93	&	0.84	&	0.81	&	0.81	&	0.81	&	0.68	\\
FSR	&	0.19	&	0.50	&	0.06	&	0.58	&	0.05	&	0.06	&	1.05	&	0.65	&	2.11	&	0.03	&	0.36	&	0.43	&	0.24	&	0.46	&	0.73	&	0.92	&	0.35	&	0.35	\\
Signal model	&	29.8	&	23.0	&	16.6	&	18.2	&	13.7	&	8.79	&	13.7	&	8.76	&	14.1	&	6.24	&	5.44	&	8.15	&	5.69	&	10.9	&	4.44	&	8.21	&	4.88	&	10.2	\\
Nom. Eff. Stats. 	&	3.03	&	3.16	&	3.36	&	3.57	&	3.81	&	4.07	&	4.51	&	4.80	&	5.60	&	5.92	&	6.30	&	7.22	&	8.35	&	8.68	&	10.8	&	12.0	&	14.3	&	11.2	\\
Fit procedure	&	0.27	&	1.79	&	1.46	&	2.09	&	0.99	&	1.27	&	2.43	&	1.18	&	2.36	&	1.08	&	3.67	&	0.68	&	1.29	&	1.00	&	2.50	&	2.53	&	1.44	&	1.19	\\
BDT selection	&	2.31	&	1.88	&	1.55	&	2.22	&	1.00	&	1.65	&	0.88	&	2.19	&	1.08	&	0.76	&	1.07	&	0.64	&	0.21	&	0.32	&	0.74	&	0.21	&	0.32	&	2.15	\\
No. $B\bar{B}$ pairs	&	1.40	&	1.40	&	1.40	&	1.40	&	1.40	&	1.40	&	1.40	&	1.40	&	1.40	&	1.40	&	1.40	&	1.40	&	1.40	&	1.40	&	1.40	&	1.40	&	1.40	&	1.40	\\
${{\cal B}(\Upsilon(4S)\to B^{+}B^{-})}$	&	1.17	&	1.17	&	1.17	&	1.17	&	1.17	&	1.17	&	1.17	&	1.17	&	1.17	&	1.17	&	1.17	&	1.17	&	1.17	&	2.17	&	3.17	&	4.17	&	5.17	&	6.17	\\
Tracking efficiency	&	1.05	&	1.05	&	1.05	&	1.05	&	1.05	&	1.05	&	1.05	&	1.05	&	1.05	&	1.05	&	1.05	&	1.05	&	1.05	&	1.05	&	1.05	&	1.05	&	1.05	&	1.05	\\
Tagging efficiency	&	4.20	&	4.20	&	4.20	&	4.20	&	4.20	&	4.20	&	4.20	&	4.20	&	4.20	&	4.20	&	4.20	&	4.20	&	4.20	&	4.20	&	4.20	&	4.20	&	4.20	&	4.20	\\
Nonresonant ${B \to X_{u}\ell\nu}$	&	0.42	&	1.75	&	1.54	&	1.25	&	0.82	&	1.64	&	0.93	&	1.87	&	5.21	&	2.60	&	2.59	&	3.07	&	4.12	&	5.19	&	6.62	&	4.63	&	4.07	&	5.26	\\
\midrule								
Total	&	31.7	&	24.5	&	18.1	&	19.9	&	15.4	&	11.7	&	15.7	&	11.8	&	17.4	&	10.5	&	11.1	&	12.5	&	12.2	&	15.9	&	15.5	&	17.0	&	17.3	&	18.3	\\
\bottomrule
\end{tabular}}
\egroup
\end{center}
\end{table}
\end{turnpage}

\clearpage

\begin{turnpage}
\begin{table}[tpb]
\caption{Relative systematic uncertainties in percent for the fits in $M_{\pi\pi}$ and $q^{2}$ bins. The bin-number convention is defined in Table~\ref{tab:fit2D}.}
\label{tab:syst2D}
\begin{center}
\bgroup
\def\arraystretch{1.5}
{\small
\begin{tabular}{p{2.7cm}ccccccccccccccc}
\toprule
Source	&	Bin 1	&	Bin 2	&	Bin 3	&	Bin 4	&	Bin 5	&	Bin 6	&	Bin 7	&	Bin 8	&	Bin 9	&	Bin  10	&	Bin 11	&	Bin 12	&	Bin 13	\\
\midrule
FF ${B \to D^{(*)}\ell\nu}$	&	0.18	&	0.07	&	0.06	&	0.03	&	0.05	&	0.09	&	0.06	&	0.08	&	0.10	&	0.14	&	0.10	&	2.39	&	1.13	\\
FF ${B \to D^{**}\ell\nu}$	&	1.34	&	0.18	&	0.21	&	0.16	&	0.10	&	0.51	&	0.69	&	0.17	&	0.23	&	0.38	&	0.20	&	1.59	&	0.76	\\
Shapes ${B \to X_{u}\ell\nu}$	&	2.24	&	2.49	&	0.56	&	0.63	&	0.57	&	0.62	&	0.26	&	0.27	&	0.01	&	0.12	&	0.05	&	0.48	&	0.09	\\
${{\cal B}(B \to D^{(*)}\ell\nu)}$	&	0.37	&	0.01	&	0.34	&	0.19	&	1.70	&	0.16	&	0.05	&	0.14	&	0.23	&	0.14	&	0.31	&	0.12	&	0.41	\\
${{\cal B}(B \to D^{**}\ell\nu)}$	&	0.12	&	0.08	&	0.38	&	0.21	&	0.05	&	0.08	&	0.14	&	0.12	&	0.14	&	0.09	&	0.14	&	0.20	&	0.39	\\
${{\cal B}(B \to X_{u}\ell\nu)}$	&	6.54	&	3.02	&	4.14	&	3.44	&	0.92	&	1.74	&	5.13	&	3.48	&	0.55	&	0.41	&	0.43	&	0.31	&	0.94	\\
Continuum	&	0.53	&	0.25	&	0.51	&	0.02	&	0.07	&	0.82	&	0.44	&	0.27	&	1.00	&	0.03	&	0.19	&	2.93	&	2.76	\\
Rare	&	1.04	&	0.47	&	0.95	&	0.32	&	0.14	&	2.10	&	0.33	&	0.23	&	2.20	&	0.45	&	0.60	&	8.47	&	1.78	\\
Sec. Leptons	&	0.17	&	0.00	&	0.05	&	0.04	&	0.02	&	0.04	&	0.58	&	0.07	&	0.18	&	0.03	&	0.02	&	1.24	&	0.15	\\
Fake leptons	&	0.25	&	0.68	&	0.15	&	0.01	&	0.07	&	0.72	&	0.01	&	0.29	&	0.35	&	0.15	&	0.13	&	1.32	&	1.02	\\
$\ell$ID	&	1.89	&	1.85	&	1.86	&	2.00	&	1.90	&	1.83	&	1.97	&	1.83	&	1.89	&	1.97	&	1.83	&	2.08	&	1.94	\\
$\pi$ID	&	1.12	&	0.81	&	1.14	&	1.04	&	0.85	&	1.10	&	1.05	&	0.90	&	1.08	&	1.04	&	0.97	&	1.23	&	1.15	\\
FSR	&	0.14	&	0.00	&	0.06	&	0.22	&	0.09	&	0.24	&	0.13	&	0.16	&	0.03	&	0.99	&	0.48	&	1.49	&	1.29	\\
Signal model	&	18.2	&	12.7	&	4.46	&	6.90	&	6.98	&	26.9	&	13.4	&	11.4	&	8.95	&	7.73	&	8.79	&	29.9	&	13.1	\\
Nom. Eff. Stats. 	&	7.71	&	7.46	&	6.97	&	7.54	&	6.02	&	6.39	&	6.17	&	5.11	&	5.50	&	5.80	&	5.46	&	1.77	&	2.34	\\
Fit procedure	&	1.88	&	0.93	&	0.85	&	0.82	&	0.21	&	1.48	&	2.03	&	0.57	&	2.41	&	1.65	&	0.96	&	6.89	&	1.35	\\
BDT selection	&	1.49	&	0.21	&	0.99	&	0.65	&	0.86	&	1.66	&	1.33	&	1.11	&	3.60	&	2.52	&	2.61	&	2.51	&	1.54	\\
No. $B\bar{B}$ pairs	&	1.40	&	1.40	&	1.40	&	1.40	&	1.40	&	1.40	&	1.40	&	1.40	&	1.40	&	1.40	&	1.40	&	1.40	&	1.40	\\
${{\cal B}(\Upsilon(4S)\to B^{+}B^{-})}$	&	1.17	&	1.17	&	1.17	&	1.17	&	1.17	&	1.17	&	1.17	&	1.17	&	1.17	&	1.17	&	1.17	&	1.17	&	1.17	\\
Tracking efficiency	&	1.05	&	1.05	&	1.05	&	1.05	&	1.05	&	1.05	&	1.05	&	1.05	&	1.05	&	1.05	&	1.05	&	1.05	&	1.05	\\
Tagging efficiency	&	4.20	&	4.20	&	4.20	&	4.20	&	4.20	&	4.20	&	4.20	&	4.20	&	4.20	&	4.20	&	4.20	&	4.20	&	4.20	\\
Nonresonant ${B \to X_{u}\ell\nu}$	&	0.42	&	3.80	&	1.02	&	2.59	&	2.50	&	1.25	&	1.09	&	1.45	&	0.00	&	0.00	&	0.00	&	0.00	&	0.00	\\
\midrule																											
Total	&	21.8	&	16.6	&	10.8	&	12.3	&	11.1	&	28.4	&	16.7	&	14.1	&	12.7	&	11.4	&	11.9	&	32.8	&	15.0	\\
\bottomrule																											
\end{tabular}}
\egroup
\end{center}
\end{table}
\end{turnpage}

\setlength{\tabcolsep}{3pt} 
\begin{table*}[tpb]
\caption{Systematic uncertainty correlation matrix of the $B^{+}\to \pi^{+}\pi^{-}\ell^{+}\nu_{\ell}$ measurement in bins of the dipion mass, 1D$(M_{\pi\pi})$ configuration. The binning convention is defined in Table~\ref{tab:fit1D}. }
\label{tab:SystCorr1D}
\begin{center}
\bgroup
\def\arraystretch{1.1}
{\small
\begin{tabular}{p{1cm}ccccccccccccc}
\toprule 
 Bin &  1 &  2 &  3 &  4 &  5 &  6 &  7 &  8 &  9 &  10 &  11 &  12 &  13  \\
\midrule  
1 &   1  &0.787  &0.529  &0.635  &0.523  &0.595  &0.777  &0.722  &0.562  &0.573  &0.735  &0.547  &0.327  \\
2 &     &1  &0.918  &0.966  &0.919  &0.948  &0.990  &0.967  &0.936  &0.934  &0.907  &0.902  &0.749  \\
3 &     &  &1  &0.986  &0.986  &0.981  &0.919  &0.944  &0.979  &0.967  &0.824  &0.934  &0.871  \\
4 &     &  &  &1  &0.981  &0.989  &0.964  &0.971  &0.980  &0.971  &0.874  &0.938  &0.837  \\
5 &     &  &  &  &1  &0.994  &0.934  &0.953  &0.991  &0.979  &0.882  &0.969  &0.859  \\
6 &     &  &  &  &  &1  &0.961  &0.976  &0.989  &0.976  &0.905  &0.975  &0.842  \\
7 &     &  &  &  &  &  &1  &0.981  &0.951  &0.945  &0.946  &0.933  &0.761  \\
8 &     &  &  &  &  &  &  &1  &0.959  &0.949  &0.931  &0.956  &0.792  \\
9 &     &  &  &  &  &  &  &  &1  &0.992  &0.898  &0.972  &0.870  \\
10 &     &  &  &  &  &  &  &  &  &1  &0.896  &0.958  &0.860  \\
11 &     &  &  &  &  &  &  &  &  &  &1  &0.925  &0.681  \\
12 &     &  &  &  &  &  &  &  &  &  &  &1  &0.831  \\
13 &     &  &  &  &  &  &  &  &  &  &  &  &1   \\

\bottomrule
\end{tabular}}
\egroup
\end{center}
\end{table*}

\begin{table*}[tpb]
\caption{Systematic uncertainty correlation matrix of the $B^{+}\to \pi^{+}\pi^{-}\ell^{+}\nu_{\ell}$ measurement in bins of the momentum transfer square, 1D$(q^{2})$ configuration. The binning convention is defined in Table~\ref{tab:fit1Dq2}.}
\label{tab:SystCorr1Dq2}
\begin{center}
\bgroup
\def\arraystretch{1.1}
{\small
\begin{tabular}{p{1cm}cccccccccccccccccc}
\toprule 
Bin  &  1 &  2 &  3 &  4 &  5 &  6 &  7 &  8 &  9 &  10 &  11 &  12 &  13  &  14 &  15 &  16 &  17 &  18  \\
\midrule  \\
1 &   1  &0.961  &0.949  &0.950  &0.936  &0.850  &0.920  &0.840  &0.867  &0.711  &0.622  &0.746  &0.584  &0.763  &0.419  &0.598  &0.423  &0.686  \\
2 &     &1  &0.991  &0.992  &0.978  &0.918  &0.969  &0.901  &0.929  &0.779  &0.709  &0.806  &0.656  &0.823  &0.499  &0.668  &0.488  &0.748  \\
3 &     &  &1  &0.995  &0.993  &0.948  &0.985  &0.937  &0.949  &0.835  &0.761  &0.854  &0.716  &0.864  &0.558  &0.721  &0.553  &0.795  \\
4 &     &  &  &1  &0.989  &0.944  &0.985  &0.932  &0.944  &0.821  &0.761  &0.843  &0.703  &0.852  &0.553  &0.714  &0.541  &0.787  \\
5 &     &  &  &  &1  &0.961  &0.991  &0.954  &0.948  &0.870  &0.797  &0.888  &0.760  &0.888  &0.605  &0.763  &0.607  &0.826  \\
6 &     &  &  &  &  &1  &0.962  &0.984  &0.951  &0.930  &0.891  &0.926  &0.843  &0.914  &0.722  &0.832  &0.714  &0.880  \\
7 &     &  &  &  &  &  &1  &0.962  &0.963  &0.885  &0.836  &0.901  &0.785  &0.901  &0.641  &0.796  &0.642  &0.844  \\
8 &     &  &  &  &  &  &  &1  &0.960  &0.957  &0.911  &0.957  &0.881  &0.944  &0.765  &0.872  &0.762  &0.920  \\
9 &     &  &  &  &  &  &  &  &1  &0.912  &0.865  &0.929  &0.846  &0.948  &0.739  &0.847  &0.703  &0.894  \\
10 &     &  &  &  &  &  &  &  &  &1  &0.952  &0.991  &0.971  &0.974  &0.883  &0.952  &0.890  &0.964  \\
11 &     &  &  &  &  &  &  &  &  &  &1  &0.933  &0.946  &0.914  &0.911  &0.943  &0.878  &0.924  \\
12 &     &  &  &  &  &  &  &  &  &  &  &1  &0.969  &0.991  &0.882  &0.962  &0.888  &0.979  \\
13 &     &  &  &  &  &  &  &  &  &  &  &  &1  &0.959  &0.962  &0.987  &0.956  &0.973  \\
14 &     &  &  &  &  &  &  &  &  &  &  &  &  &1  &0.878  &0.956  &0.868  &0.979  \\
15 &     &  &  &  &  &  &  &  &  &  &  &  &  &  &1  &0.959  &0.945  &0.925  \\
16 &     &  &  &  &  &  &  &  &  &  &  &  &  &  &  &1  &0.961  &0.980  \\
17 &     &  &  &  &  &  &  &  &  &  &  &  &  &  &  &  &1  &0.920  \\
18 &     &  &  &  &  &  &  &  &  &  &  &  &  &  &  &  &  &1   \\

\bottomrule
\end{tabular}}
\egroup
\end{center}
\end{table*}

\begin{table*}[tpb]
\caption{Systematic uncertainty correlation matrix of the $B^{+}\to \pi^{+}\pi^{-}\ell^{+}\nu_{\ell}$ measurement in bins of the dipion mass and the momentum transfer square, 2D configuration. The binning convention is defined in Table~\ref{tab:fit2D}.}
\label{tab:SystCorr2D}
\begin{center}
\bgroup
\def\arraystretch{1.1}
{\small
\begin{tabular}{p{1cm}ccccccccccccc}
\toprule 
Bin  &  1 &  2 &  3 &  4 &  5 &  6 &  7 &  8 &  9 &  10 &  11 &  12 &  13  \\
\midrule 
1 &   1  &0.940  &0.767  &0.861  &0.881  &0.907  &0.936  &0.934  &0.843  &0.856  &0.877  &0.806  &0.844  \\
2 &     &1  &0.867  &0.950  &0.965  &0.906  &0.958  &0.968  &0.874  &0.899  &0.914  &0.769  &0.832  \\
3 &     &  &1  &0.969  &0.950  &0.693  &0.846  &0.856  &0.876  &0.925  &0.902  &0.545  &0.683  \\
4 &     &  &  &1  &0.992  &0.790  &0.911  &0.922  &0.883  &0.931  &0.921  &0.636  &0.746  \\
5 &     &  &  &  &1  &0.829  &0.932  &0.951  &0.897  &0.938  &0.938  &0.677  &0.789  \\
6 &     &  &  &  &  &1  &0.962  &0.956  &0.867  &0.860  &0.900  &0.902  &0.909  \\
7 &     &  &  &  &  &  &1  &0.992  &0.939  &0.952  &0.969  &0.858  &0.904  \\
8 &     &  &  &  &  &  &  &1  &0.931  &0.946  &0.969  &0.837  &0.904  \\
9 &     &  &  &  &  &  &  &  &1  &0.969  &0.972  &0.781  &0.856  \\
10 &     &  &  &  &  &  &  &  &  &1  &0.991  &0.758  &0.852  \\
11 &     &  &  &  &  &  &  &  &  &  &1  &0.788  &0.874  \\
12 &     &  &  &  &  &  &  &  &  &  &  &1  &0.911  \\
13 &     &  &  &  &  &  &  &  &  &  &  &  &1   \\
\bottomrule
\end{tabular}}
\egroup
\end{center}
\end{table*}

\begin{table*}[tpb]
\caption{Approximated correlation matrix between the 18 bins in $q^{2}$ and 13 bins in $M_{\pi\pi}$ derived from data by requiring $M_{\textrm{miss}}^{2}<0.5$ GeV$^{2}$. }
\label{tab:Corr_q2_mpipi_SR}
\begin{center}
\bgroup
\def\arraystretch{1.1}
{\small
\begin{tabular}{c|ccccccccccccc}
\toprule
Bin  &  1 &  2 &  3 &  4 &  5 & 6 & 7 & 8 & 9 & 10 & 11& 12& 13      \\
\midrule  
1 & 0.000 &  0.000 &  0.091 &  0.000 &  0.067 &  0.085 &  0.028 &  0.048 &  0.000 &  0.057 &  0.036 &  0.182 &  0.281 \\
2 & 0.000 &  0.000 &  0.050 &  0.038 &  0.091 &  0.046 &  0.000 &  0.052 &  0.052 &  0.000 &  0.039 &  0.050 &  0.283 \\
3 & 0.000 &  0.000 &  0.058 &  0.044 &  0.021 &  0.027 &  0.035 &  0.000 &  0.000 &  0.000 &  0.091 &  0.000 &  0.276 \\
4 & 0.050 &  0.000 &  0.050 &  0.038 &  0.054 &  0.046 &  0.030 &  0.052 &  0.052 &  0.000 &  0.000 &  0.000 &  0.307 \\
5 & 0.045 &  0.091 &  0.045 &  0.069 &  0.033 &  0.107 &  0.083 &  0.048 &  0.095 &  0.000 &  0.000 &  0.000 &  0.271 \\
6 & 0.171 &  0.057 &  0.000 &  0.000 &  0.083 &  0.053 &  0.000 &  0.000 &  0.060 &  0.000 &  0.045 &  0.000 &  0.217 \\
7 & 0.000 &  0.000 &  0.000 &  0.000 &  0.022 &  0.055 &  0.107 &  0.000 &  0.062 &  0.000 &  0.185 &  0.000 &  0.211 \\
8 & 0.000 &  0.051 &  0.051 &  0.039 &  0.112 &  0.048 &  0.093 &  0.000 &  0.000 &  0.000 &  0.120 &  0.000 &  0.218 \\
9 & 0.000 &  0.000 &  0.000 &  0.111 &  0.080 &  0.069 &  0.044 &  0.000 &  0.000 &  0.092 &  0.000 &  0.000 &  0.139 \\
10& 0.000 &  0.000 &  0.000 &  0.000 &  0.096 &  0.074 &  0.095 &  0.110 &  0.000 &  0.197 &  0.164 &  0.210 &  0.112 \\
11& 0.000 &  0.000 &  0.000 &  0.150 &  0.145 &  0.062 &  0.000 &  0.000 &  0.000 &  0.165 &  0.103 &  0.000 &  0.094 \\
12& 0.000 &  0.078 &  0.078 &  0.000 &  0.171 &  0.110 &  0.000 &  0.082 &  0.082 &  0.000 &  0.000 &  0.078 &  0.019 \\
13& 0.000 &  0.208 &  0.000 &  0.053 &  0.152 &  0.162 &  0.168 &  0.000 &  0.000 &  0.000 &  0.000 &  0.000 &  0.000 \\
14& 0.000 &  0.161 &  0.000 &  0.061 &  0.030 &  0.189 &  0.098 &  0.085 &  0.085 &  0.000 &  0.000 &  0.081 &  0.000 \\
15& 0.000 &  0.000 &  0.000 &  0.000 &  0.177 &  0.076 &  0.195 &  0.085 &  0.085 &  0.000 &  0.000 &  0.000 &  0.000 \\
16& 0.000 &  0.000 &  0.091 &  0.069 &  0.100 &  0.085 &  0.165 &  0.000 &  0.095 &  0.000 &  0.000 &  0.000 &  0.000 \\
17& 0.167 &  0.084 &  0.000 &  0.000 &  0.214 &  0.078 &  0.000 &  0.088 &  0.000 &  0.000 &  0.000 &  0.000 &  0.000 \\
18& 0.224 &  0.000 &  0.112 &  0.213 &  0.267 &  0.105 &  0.034 &  0.000 &  0.000 &  0.000 &  0.000 &  0.000 &  0.000 \\
\bottomrule
\end{tabular}}
\egroup
\end{center}
\end{table*}


\begin{thebibliography}{120}
\bibitem{natural-units}
  The use of natural units ($\hbar=c=1$) is assumed throughout the paper.
\end{thebibliography}
\end{document}